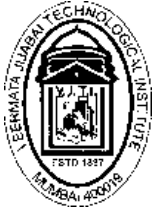

# VEERMATA JIJABAI TECHNOLOGICAL INSTITUTE
[Central Technological Institute, Maharashtra State ]
Matunga, Mumbai-400019
[Autonomous Institute]

# Electrical Engineering Department

### Report of Examiners for Ph.D Defence Viva Voce

| | |
|---|---|
| Name of the Research Topic | : **Synchronization Strategies for Multi-agent Networked Control Systems** |
| Name of the Candidate | : **Mr. Pratik K. Bajaria** |
| Venue | : Online mode, Zoom platform |
| Date | : Wednesday, July 29, 2020 |
| Time | : 05.40 pm Onwards |
| Examination Committee: | **Prof. Dr. Madhu Belur** |
| | Professor, Department of Electrical Engineering, IIT, Bombay. |
| | **Prof. Dr. D. N. Raut** |
| | Chairman, Examination Committee, |
| | VJTI, Mumbai. |
| | **Prof. Dr. N. M. Singh** |
| | Supervisor, |
| | Electrical Engg. Dept., VJTI, Mumbai. |

**Recommendation of Examination Committee:**

The PhD defense of the candidate, Mr. Pratik K. Bajaria was conducted using Zoom online platform because of the prevailing Covid-19 situation. The candidate has addressed all the questions raised by the thesis examiners and has defended his thesis well. He also answered the questions posed by the examiners and chairman satisfactorily. The work presented is supported by journal papers and international conference paper publications. The examination committee recommends the award of PhD degree to Mr. Pratik K. Bajaria.

| | | |
|---|---|---|
| *[signature]* | *[signature]* | *[signature]* |
| **Prof. Dr. Madhu Belur** <br> **External Examiner** | **Prof. Dr. D.N. Raut** <br> **Chairman Examination Committee** | **Prof. Dr. N. M. Singh** <br> **Supervisor** |

# Synchronization Strategies for Multi-agent Networked Control Systems

By

Pratik K. Bajaria

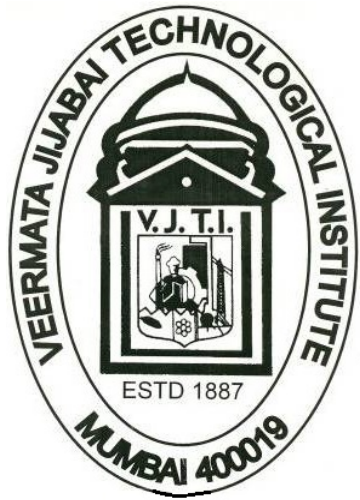

Department of Electrical Engineering
Veermata Jijabai Technological Institute
University of Mumbai

A thesis submitted for the degree of
Doctor of Philosophy
July, 2020

**Synchronization Strategies for Multi-agent Networked Control Systems**

for the degree of
PhD in Electrical Engineering

Submitted by

**PRATIK K. BAJARIA**
**(ROLL NO. 159030005)**

Under the Guidance of

**DR. N. M. SINGH**

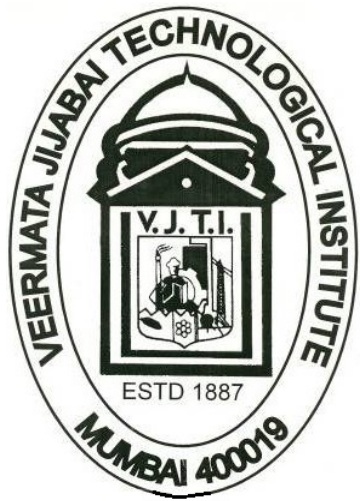

Department of Electrical Engineering
VEERMATA JIJABAI TECHNOLOGICAL INSTITUTE, MUMBAI-400 019
(Autonomous Institute Affiliated to University of Mumbai)

2015-2020

# Certificate

This is to certify that **Mr. Pratik K. Bajaria**, a student of PhD in Electrical Engineering has completed the thesis and his research work on, **"Synchronization Strategies for Multi-agent Networked Control Systems"** to our satisfaction.

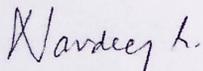

Dr. N. M. Singh
Supervisor

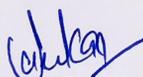

Dr. F. S. Kazi
Head, Electrical Engineering Department

Director
VJTI, Mumbai

# Approval

The thesis, **"Synchronization Strategies for Multi-agent Networked Control Systems"** submitted by **Mr. Pratik K. Bajaria**, is found to be satisfactory and is approved for the degree of Doctor of Philosophy in Electrical Engineering from Veermata Jijabai Technological Institute affiliated to University of Mumbai.

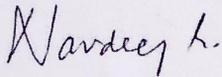

Dr. N. M. Singh
(Supervisor)

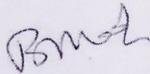       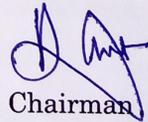

Madhu N. Belur
Examiner
(EE Dept, IITBombay)                                      Chairman

# ABSTRACT

With the advent of $21^{st}$ century and increasing advancements in the field of technology and connectivity, inter-networking in real-time has achieved great importance. Distributed control and multi-agent paradigm has groped rapidly with history of big time failures of centralized systems in the past. The concepts of synchronization and network control systems have been used extensively in the near past to map, analyze and solve defined set of objectives. In this thesis, a diverse set of applications from power flow point of view are taken into consideration and modelled/analyzed using synchronization as the central theme. These systems are proposed (or assumed) to be network connected and its control has been devised accordingly. It has been shown how some examples from nature can help recreate similar dynamics synthetically and help achieve system objectives. Few of the applications of the smart world have been ascribed in the thesis and distributed control of these seen from a multi-agent perspective have been devised in order to better design and operate such systems in real-time. Synchronization happens to be the heart of all the networks, as all the agents work in tandem in order to provide to a common objective as well abide by the defined constraints.

Micro grids are system of distributed renewable/non-renewable energy sources connected together to provide to a local load and in some cases distribute it across other micro grids (if in excess) or get paid back by dumping it on the central grid. In the course of achieving demand-supply balance, various distributed source of energies have to be balanced (in terms of its system parameters) so as to make it economically viable/feasible. In this work, set of converters being sourced from various distributed sources have been considered as multiple nodes of a complex network providing to a common or set of distributed loads. A problem of frequency synchronization, thereby maintaining active power sharing has been considered. Micro grids are modelled as Lure' systems in order to analyse system stability using incremental passivity fundamentals. A novel control is proposed in order to achieve system objectives in an uncertain environment and various network defects are simulated in order to understand its effectiveness in reference to existing controllers in the literature.

In order to further study the effects of power system instabilities, I try to articulate sustainable strategies by exploiting ancillary services to provide grid stability. I show for the first time how Kuramoto models can become handy to solve regulation problems in a power grid. A set-up of thermostatically controlled loads (TCLs - or heating and cooling units) in residential sector has been considered. A novel Boolean phase oscillator model based de-synchronization of TCLs has been proposed. For the first time in a power network, it is shown how effective de-synchronization of electrical components can achieve power system stability. Two sets of laws using Kuramoto based phase oscillator model and distributed averaging has been conceived. Results (both software and hardware) have been simulated for homogeneous, heterogeneous and population of TCLs using a computer program. Dynamic dispatch of TCLs shows a great promise in providing meaningful ancillary services to the central grid. A load following scenario has been simulated in order to verify the effectiveness of the proposed model to a utility driven architecture and the outcomes are verified across literature.

Power systems deal with issues like cascade failures, blackouts etc., and thereby require in depth analysis of occurrence/inception of such phenomena. One of the very well-studied concepts is that of low frequency oscillations in power grid; particularly, interarea oscillations that lead to system failures. In the second part of the thesis, a power network is taken into consideration. The power grid is seen as a network of connected generators that oscillate against each other to provide to a common load. Swing equations commonly referred to in power systems are remodelled into second order Kuramoto form. A novel nonlinear model for analyzing interarea oscillations that could help power system designers to construct a robust grid free from these oscillations without relying on small-signal methods available in the literature has been proposed. A mapping between the well-studied Kuramoto paradigm and power systems is derived and should facilitate better understanding of the complex dynamics of a power network. A standard four generator power system with all-to-all connectivity is considered and results obtained from the proposed model are verified. Further, bifurcation analysis of the proposed model has been carried out and the results so obtained are visualized as some of the blackouts scenarios registered in power networks. 'Chimera' states or condition of partial synchrony has been achieved using the developed theory and mapped to blackouts with islanding scenarios.

As an inflection to the work, I set up a platform for development of new distributed and fast acting control strategies for conventional as well as futuristic control systems existing/non-existing in the literature. As a major contribution of this dissertation, by combining the ideas from physics and power systems, I open up various interesting phenomena already existing in either fields to be explored for one another.

# Dedication and acknowledgements

As it goes, anything you do in your life requires perseverance, hard work and support from some or the other source. For me that source was my family; hence I would like to dedicate this thesis to my mom, my dad and my sister, who have always been by my side in thick and thins. I wish everyone gets such a supportive family so as to bloom in a constraints-free environment. I also want this thesis to serve as an inspiration (if it deserves) to those who think that inventions and discoveries are impossible in $21^{st}$ century. I hope my doctoral dissertation serves as a perspective to existent applications as well as new theories.

As the acknowledgement goes, primarily I would like to recognize the efforts of Prof.N.M.Singh for taming a crazy individual like me into something this meaningful. His constant projections to drive me on a technical oblivion turned me into a confident and presentable personality. It is worth mentioning though that his constant endeavours to turn me into a sensible character by keeping me socially active helped me a lot. Being one of the last doctoral student, I feel prideful on having such supervision. I would like to mention one of the quotes that he mentioned when I got a crucial publication,

*"You may lose a battle, but you must win the war!"*

I would also like to quote Prof.F.S.Kazi, Mrs.A.A.Sharma and Prof.S.R.Wagh who helped me offline on administrative and psychological fronts. Prof.S.D.Varwandkar taught me about power systems from scratch and discussions with him allowed to clear fog off the sky.

Also, I would sincerely like to mention few of my friends at (or not at) the university campus who kept me constantly motivated. In particular, Mr.Darshak Sanghavi, Dr.Prashant Bhopale, always had something in the store when I felt low or diffident. Mr.Amol Yerudkar, who would turn a doc soon before this thesis gets published online; he was a constant push to get my work done in time. Not to mention, so many friends that emerged from the post-graduation batches whom I saw pass by. Additionally, I would like to quote passionate efforts of an undergraduate apprentice Mr.Kshitij Singh to shape some of the outcomes of this thesis and who emerged as a student-turned-great-friend.

A 4 years of journey wouldn't have been possible with a great institutional reputation and facilities that were provided by my institute. I oblige to every single institutional facility provided by Veermata Jijabai Technological Institute, India, where I could blossom like a flower and pace up like a Panther.

# LIST OF TABLES

# LIST OF FIGURES

# LIST OF ABBREVIATIONS/ACRONYMS

CHAPTER 1

**INTRODUCTION**

**Table of Contents**

Synchronization refers broadly to cohesive patterns that emerge in several natural and man-made systems, due to coordinated actions by design or through spontaneity [1–4]. These systems can be broadly classified into the ones that synchronize to a reference trajectory or try to achieve a coherent frequency/phase of oscillations. To re-frame, the latter case is no different from the former one, with reference trajectory generated from one of the agents participating in the system. Formally, oscillations are repetitive variations in time of a parameter measured from a physical system about a central value (equilibrium) or between two or more states. Oscillatory systems from heart beats and birds flapping their wings, to firing of neurons [5] and brain rhythms [6]; power grids and computer networks [7], to de-synchronization in air-conditioners [8] and artificial neural networks [9], and many more achieve synchronization through phase/frequency along time. Oscillatory systems are ubiquitous and so is synchronization and there's no harm to say that periodicity is the nature of all the systems existing in universe.

## 1.1 Synchronization in Nature

In the sequel, I present some of the interesting examples existing in nature that has helped me understand synchronization theory in a better way. For instance, it has been observed that huge congregations of Fireflies flash in unison by using visionary signals as a negative feedback to tune its own frequency of oscillations and thereby achieve synchronization [10–15]. In human body, a cluster of about more than a thousand cells called the sinoatrial (SA) node, functions together to generate electrical pulses at a frequency commonly referred to as the heart rate. The firing of pacemaker comprise of clusters of completely synchronized and de-synchronized cells so as to achieve desired objective [16–21].

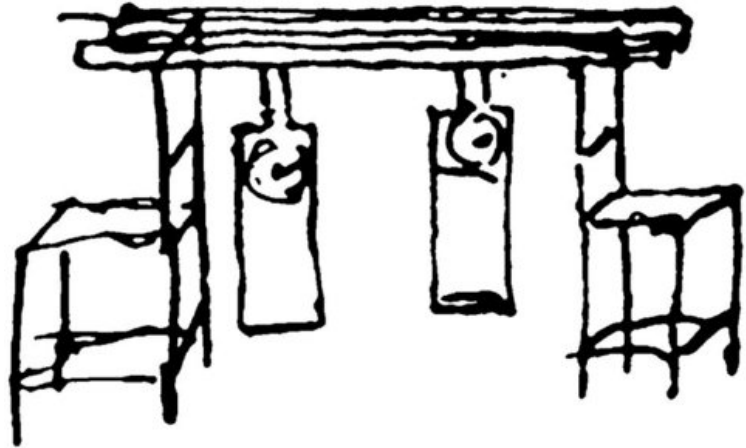

Figure 1.1: The original diagram scribbled by Christiaan Huygens in his letter to Royal Society of London [30].

It was no later than the seminal work of Nobert Weiner, when it was shown that brain waves exhibit synchronization too. His model was corrected by Art Winfree, by adding interaction terms and framing a nonlinear structure through set of ordinary differential equations. With this discovery mathematicians could forge a connection between nonlinear dynamics and statistical mechanics [22–27]. Though effective, these models could handle small systems with handful variables, but had problems with large constellations. To fix this, Yoshiki Kuramoto refined Winfree's model using highly symmetric rule integrating frequency congregations [28]. Steven Strogatz, a research scholar at Harvard university happened to observe sleep-wake cycles in humans as a result of spontaneous internal de-synchronization [29].

One such observation that changed the way system of coupled pendulums to be looked at was made by Christian Huygens. To his attention, when the clocks were in sympathy (refer Figure (1.1); i.e., when the swing of pendulums were equal in magnitude but opposite in direction), the equal and opposite forces cancelled each other leaving the chair stand still. Huygens had just discovered the use of negative feedback to achieve stabilization and inspired field of physics/mathematics/engineering for years [4, 30–33]. Various applications like Josephson junction, comprising a pair of superconductors coupled by a weak link allowing indefinite flow of current without applying external voltage have been based on Huygens observations [34]. Synchronization forms the basic of modern day physics too. For instance, quantum coherence has given lasers abiding to Bose-Einstein statistics synchronizing all the photons in the same quantum state acting like a gigantic wave of light [35–37].

Synchronization has also helped solve various man-made fiascos. The Millenium bridge is one such incidence, where the Londoners couldn't hold there excitement and crowded the newly opened bridge chaotically. The bridge showed huge oscillations side to side horrifying people, which was later investigated using Winfree, Kuramoto and Huygens observations [38, 39]. It can be noted that, synchronization is ubiquitous not only in systems with order but also in a chaotic frame. In the seminal work of Louis Pecora, he proved synchronized chaos breaking conventional idea of rhythmicity and bought understanding of synchronicity through loops, cycles and repetitions [40]. Further in the age of interconnected systems, sociologists relate occurrence of riots, propagation of fads and diffusion of innovations to be very closely related to the concept of synchronization [41]. In the succeeding sections I investigate science of a connected age to extend/propose the ideas to existent/emerging applications.

## 1.2 Synchronization in Interconnected Networks

From previous discussions, it can be inferred that right from flashing of fireflies, to sleep-wake cycles, 'The Huygens' Experiment' and complex networks, synchronization theory has shown a great promise to help understand various laws of nature and implement them to create artifacts with great precision and accuracy. Some of the examples stated above offer a great sort of motivation to look around synchronization not as a theory but from a law that has potential to drive huge masses towards a common objective.

It can be noted though, that one of the important factors that decide rate of synchronization is degree of coupling between the participating agents. One of the ways of visualizing these agents and its dynamics is via structuring them in a complex network (a network of multiple nodes connected together and sharing information). With the introduction of internet-of-things (IoT) and devices having its own decision making powers, complex networks have emerged as the next big thing. Conventional power grids have turned smart, taxi booking have turned from huge loss of time and efforts to a single click, buying clothes and ordering food all have turned into an effortless job with the introduction of networking. In a complex network (say congregate of fireflies for instance), every node corresponds to an agent contributing to the effective outcome (e.g., oscillations) of the system. This network can be wired or wireless and can experience many hindrances during operation. This gives rise to the concept of closing a conventional feedback control network using a communication channel, usually termed as a 'Networked Control Systems (NCS)'.

[42] provides an extensive understanding of NCS and its components as well as its articulation as a complex network and information flow. System stability is examined using passivity (using Lyapunov functions that map energy flow of a system to state information) fundamentals and robustness of the network in presence of external disturbances and uncertainties have been discussed. Various control schemes for robustness, switched feedback, fuzzy systems with time-delay, network boundary control, adaptive laws and positive real transfer function based control design has been devised. It has been shown

how every problem related to complex networks or consensus or IoT connected systems can be formulated collectively into a NCS problem. Well, after framing a problem into synchronization framework using NCS, the next problem arises of that related to selecting the right scheme of control. Various hierarchical schemes like centralized (those being controlled using a central controller), distributed (system with distributed nodes and control hotspots but actions dependent on information shared with each other) and decentralized (system with local control points but not sharing information with each other) have been formalized to control a huge multi-agent network of systems.

Researchers have adopted a distributed framework (owing to similarities persisting in nature) over centralized one in order to reduce the effects of malicious intrusions and increase the robustness of the network. Various examples from the past have shown effectiveness of a distributed model over centralised one. [43] shows various techniques by which distributed control over a multi-node network can be achieved for variety of applications. Consensus for second and higher order dynamical equations has been achieved for multi-agent set-up. A variety of control scenarios like communication link uncertainties, time delay, consensus filtering, Quasi-consensus, etc., have been taken into account. A distributed adaptive control with capabilities of recalculating control parameters in real-time in reference to the deflection in objective function has been formulated.

In this work, I make use of various ideas from the systems existing in nature to gather inspiration and device algorithms to solve problems related to power transmission networks or power grids. Some of the research works that highly inspired the outcome of this thesis and fathom the potentials of synchronization as a theory are mentioned as follows. In [44], it is shown how a distributed hierarchical control can be implemented to achieve frequency and voltage regulation in a micro grid. The framework is simple yet effective and allows room for betterment in terms of time scale separation and fastness of the strategy. On similar lines, a passivity based approach has been proposed to achieve group coordination objectives and thereby some practical applications have been suggested in [45]. Further, an adaptive strategy to achieve consensus on a complex network has been devised in [46] and forms the basis to develop control that can be improved in terms of speed of action. These papers provide fundamental ideas about controlling a network of oscillators not only effectively but in an organised manner. These research work formed basis to develop some fast acting controllers for frequency synchronization in a micro grid.

Following the work already proposed in the direction of power flow, [47] provides an interesting insights to existence of partial synchronization states in group coordination theory. [47] gives a deep understand of existence of clusters of synchronized and de-synchronized states simultaneously. These states normally referred to as 'chimera' in oscillator theory, opens up various phenomena in power systems like cascade failures or blackouts with islanding to be studied in a similar frame. On the other hand, [48] showed/proved conventional built-up of positive coupling in Kuramoto oscillators to be trivial. The research in [48] explains existence of positive and negative coupling in a Kuramoto framework and discusses its applications. These two papers turned handy and helped the research in direction of mapping different power system scenarios to a novel Kuramoto framework to be introduced.

While exploring the Kuramoto framework and beauty of its synchronization behavior via symmetry, I exploited model characteristics to achieve controlled de-synchronization. In various systems like those related to Thermostatically Controlled Load (TCL), synchronization is an undesirable state and de-synchronization is a necessity. These ideas were made evident to me by interesting insights provided by [49], where utility controlled load following techniques have been proposed. Although, it must be noted though that [49] uses an ensemble of TCLs rather than using coupled ones. Digitization of Kuramoto oscillators given in [50] inspired me to define direction for controlled de-synchronization in TCLs.

In reference to the various strategies and concepts mentioned in previous sections and an effort to connect the idea of synchronization to multi-agent networked control systems, the major contribution of the thesis can be stated as follows,

1. Frequency is one of the most significant factors in power flow. Any change in load of the power network gets reflected in the effective frequency of the grid. In reference to an alternating current (AC) micro grid with distributed energy sources, converters require to synchronize at a single frequency, thereby maintaining effective power sharing according to their capacities. In such cases, it is necessary to implement a hierarchical control to abide by the system objectives. Also, it would be of great advantage if a model can not only give necessary insight into micro grid dynamics, but also ease stability analysis. Firstly, I restructure micro grid dynamics into its equivalent Lure' form (an interconnection of linear and nonlinear model). Then, a novel adaptive control law for frequency synchronization thereby effective active power sharing in an uncertain environment is devised. Finally, the system stability is analysed using incremental passivity fundamentals and effective lower bound on the passivity function has been calculated.

2. TCLs form a major part of residential loads and are ubiquitous in any modern day infrastructure. These loads when managed by utility can provide significant Ancillary Service (AS) to the central power grid. It may be noted that these TCLs can be constructed as set of discrete oscillators oscillating in harmony. Also, synchronization of switching states of TCLs is a highly undesired state and must be de-synchronized for all times. Using the conventional Kuramoto model and a novel mathematical tweak, TCLs are de-synchronized and the effective power aggregate consumed is observed to be in accordance to the reference utility signal defined by the operator. Further, a distributed averaging protocol is introduced to overcome few computational complexities in the Kuramoto model. In order to strengthen the idea, a set of light emitting diodes (LEDs) have been considered to emulate as TCLs and hardware based results are acquired. It has been shown how TCLs can prove to be an excellent ASs to the grid by imitating load following scenarios.

3. Next, the Kuramoto model is extended to a power grid. In power systems, swing

equations help understand the dynamics of power flow and thereby define system stability. A novel conformist-contrarian Kuramoto (CC-Kuramoto, explained later) model has been proposed using the idea of coupled pendulums. This model simplifies the analysis of effects of intra and inter area oscillations commonly referred to in power systems. The results obtained by the model are verified against those available in the literature and implications discussed. Further, bifurcation analysis of the proposed model has been carried out and its analogy to a power grid has been discussed. It has been shown, how a simple nonlinear structure can help study 'chimera' behaviours (simultaneous existence of synchrony and de-synchrony) arising in power network. These behaviors have been mapped to blackout type scenarios in power grid.

The thesis is organized as follows. First, frequency synchronization problem in micro grid has been studied and a novel control law has been introduced for regulation in an uncertain environment in Chapter 2. Next, a load following scenario using controlled de-synchronization of TCLs has been achieved using phase oscillator model and a distributed averaging approach in Chapter 3. Further, in Chapter 4 a power grid is considered and is studied using a novel Kuramoto model thereby providing deeper insights into the system dynamics. Finally, I conclude the thesis on a futuristic note to extend the ideas proposed.

## 1.3 Thesis at a Glance

A brief flow chart describing the work performed in this thesis has been given below.

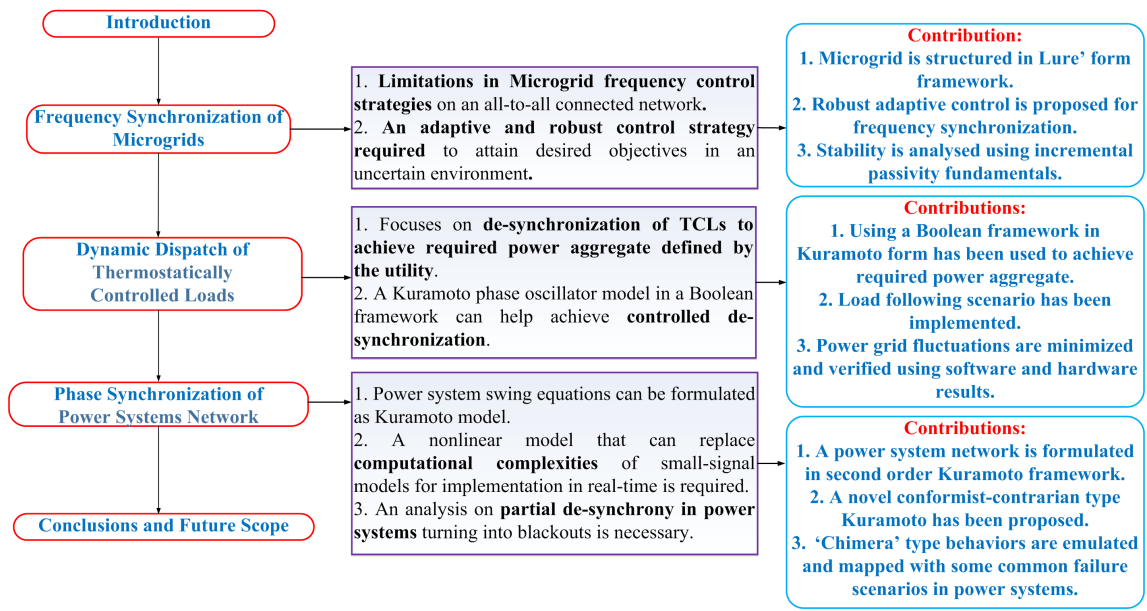

Figure 1.2: Thesis at a glance.

## 1.4 Summary and Inferences

The major takeaways from this chapter can be summarized as follows:

1. Synchronization is ubiquitous and its existence in nature allows mankind to solve complex problems existing in achieving synthetic intelligence. These ideas can be adopted from nature and can be replicated to achieve/impart intelligence to the machines/algorithms existent in the market to make them more efficient and accurate.

2. Internet allows these algorithms to be extended on a global scale and can be studied and implemented under the canopy of NCS. With the cause-effect already studied in the literature, NCS can become a liability to any organization working towards a significant objective.

3. Centralized control has been replaced by distributed laws with its reliability in terms of controlling huge complex networks and hence requires powerful tools to drive them. Distributed space has been studied extensively in the literature and provides effective tools for a control system designer to control a huge network.

4. Further, the contributions of the thesis have been listed and shown how the ideas inspired from nature can help achieve desired objectives to various applications.

CHAPTER 2

**FREQUENCY SYNCHRONIZATION OF MICRO GRIDS**

**Table of Contents**

In order to manage ever increasing future energy demands and avoid system failures, a flexible energy system with a distributed and customizable architecture is expected to reshape future low level distribution networks [51, 52]. Micro grids is one such system that adheres to similar philosophy of zero net energy. A direct current (DC) or alternating current (AC) or hybrid micro grid, comprises of small networked generating sources providing local set of loads, thereby consuming little from the central grid. Specifically, AC micro grids have gained importance in the near past (due to existing AC power grid), since they allow the integration of renewable energy sources [53, 54] allowing fewer points of failure and a more reliable and uninterrupted power distribution. Technically, AC micro grids require parameters like operating frequencies, line voltages, active and reactive power sharing as well as power factors to be controlled in real time in order to compensate changing load demands. Out of the mentioned parameters, frequency is considered to be one of the significant element that reflects power network stability. As known, frequency is a direct indicative of the load being served and hence high deflections from the nominal value can be considered to be undesirable.

Also, in case of a distributed network of source-loads, it is important for all the converters to provide frequency at a desired reference thereby maintain active power sharing.

In this work, an AC micro grid with frequency synchronization and regulation alongwith power sharing on a multi-agent framework is studied and analyzed. This work tackles the problem of frequency synchronization via adaptation in presence of external disturbances. To achieve frequency synchronization in a micro grid, typically a hierarchical approach is employed [44, 55], whereby a micro grid is controlled in stages. A primary controller stationed inside the inverter abides by its inherent droop characteristics [56], whereas a secondary controller is used to compensate for any deviations caused by the unmatched inter-component behaviours (i.e., power sharing in relation of their respective capacities). In the literature, a centralised averaging proportional integral (CAPI) control action has been proposed as an advanced version of a centralised control scheme at secondary level that can be used to attain the frequency synchronization objective [57]. On the other hand, Distributed Averaging Proportional Integral Control (DAPI) [44] is a distributed scheme used to solve the problem of frequency synchronization by means of relative measurement errors. Both being comparatively effective, these are slow acting schemes and require time to adapt to fast acting load changes. To mitigate this issue, various strategies in literature employing similar structural/hierarchical architectures have been proposed to control distributed micro grids [57, 58].

In course of moving towards smart grid solutions (which are envisioned as networked micro grids), hierarchical control theory has moved from wired to wireless communication scheme. Although, being beneficial in terms of flexibility, there are various other drawbacks that make these schemes difficult to implement and are weakly reliable. Precisely, time delays, packet drops, intrusion of data noises and system failures are major issues to be addressed. Hence, a robust control scheme should be devised for the set of uncertainties that are implied by these communication channels to assure an uninterrupted/reliable supply of energy. [59, 60] define various strategies to achieve synchronization over diffusively coupled systems and issues related to NCS.

In addition to slow acting behaviours, multiple time scales and uncertainties in the network, issues related to instabilities introduced by the nonlinear mathematical characteristics of micro grids are pretty evident in a micro grid. Hence, it is necessary to analyze stability of the effective system. In this work, micro grid system dynamics are redesigned into its equivalent interconnected linear and nonlinear subsystems known as Lure' forms [61] in order to analyze system stability using passivity fundamentals. Benefits of Lure' systems over a network are well studied and exploited in [62], where a linear controller-observer scheme is devised to compensate for channel uncertainties. [46] shows various applications including those related to mechanical fly-ball power generators to be re-casted in Lure' form structures and evaluate its benefits.

[63] achieves objective of robust consensus over networked Lure' systems using output synchronization technique similar to the one proposed in [46], but without evaluating the impact of external disturbances. The coupling gains are allowed to grow unbounded in the event of perturbations that effect synchronization. [46] though effective needs to be extended to maintain levels of synchrony, which is achieved in [44]. As shown in [44], synchronization

is maintained over primary and secondary levels, but compromises on the rate of action and happens to be a slow acting control in the events of fast switching loads and also hasn't been evaluated for external disturbances or network uncertainties. On the other hand, [64] evaluates the impact of external disturbances and network uncertainties in Lure' form, but doesn't maintain levels of synchrony. It must be noted though that levels of synchrony helps distribution of control in a network and hence would be better if hierarchy as well as synchrony can be achieved simultaneously.

The major contribution of this work is along the lines of the issues discussed above. Firstly, Lure' form structures are used to redesign system dynamics into its linear-nonlinear form, in the presence of external disturbances and network uncertainties. This allows to analyze system stability using incremental passivity fundamentals and a positive lower bound in the event of external disturbances and network uncertainties can be calculated. Second, in order to achieve faster synchronization, in the event of rapid switching loads and maintain proportional (in ratio of their respective capacities) load sharing in presence of external disturbances, a robust adaptive control law has been proposed. Thirdly, when connected to a wireless communication channel, network coupling gains have been optimised in order to regulate control action in case of no further load switching. Finally, the effects of several communication network issues have been simulated and its impact on the system stability have been investigated and compared with the available controllers in the literature to evaluate its benefits.

This chapter has been organized as follows. Initially, frequency synchronization in micro grids has been pitched as a motivating problem in section 2.1. Then, mathematical model for the same has been articulated in section 2.2. A novel robust control strategy has been proposed in section 2.3, thereby providing stability analysis using incremental passivity ideas. Finally, a test case has been considered to verify the concept in section 2.4.

## 2.1 Motivating Application and Problem Statement

A micro grid is a system of renewable/non-renewable energy sources providing to a local load, restricted to a geography and sometimes communicating to the central power grid. Subjectively, a micro grid is termed as islanded if it doesn't have a provision to connect to the central power grid and so on. In this work, I consider an AC micro grid for its popularity due to ease of getting directly interfaced with the central power grid. In an islanded micro grid, several distributed energy sources supply a common load. Various renewable sources like solar or photovoltaics (PV), wind, etc., get interfaced together using an AC-AC, DC-AC converters. Although, the frequencies generated by these must be in tandem and so must be there phases too. With no reference points, (like those available in grid-connected systems) synchronization of every connected generating source becomes a primary concern. A generic schematic diagram of a set of inverters supplying a common load in a micro grid frame is shown in Figure 2.1.

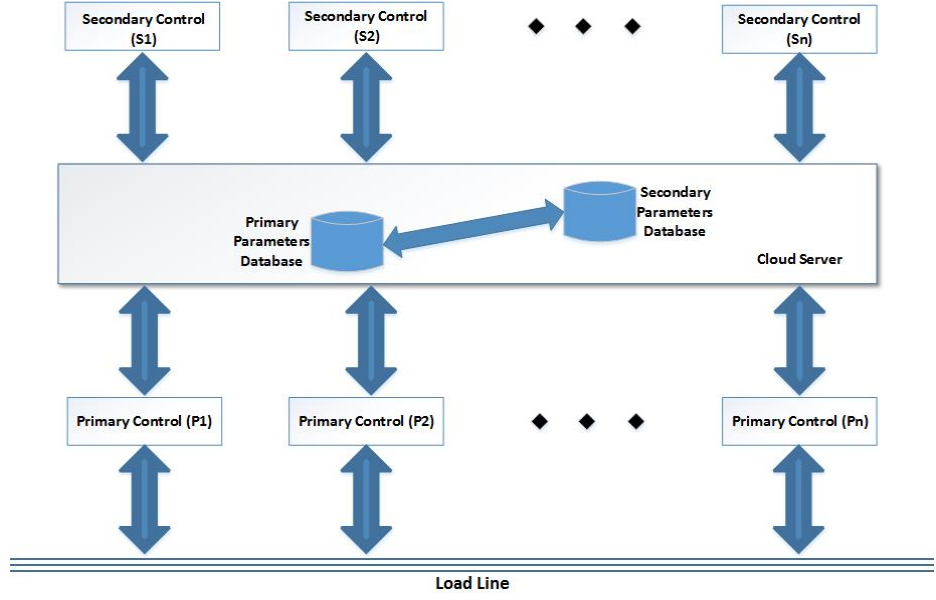

Figure 2.1: Schematic diagram of a distributed control of micro grid.

### 2.1.1 Droop Control

In a distributed setup of coupled inverters, the droop laws [56] define effective frequencies of individual inverters. Inverters adjust themselves so as to share load power proportional to their capacities and drop frequencies accordingly. Droop laws decide the inherent drop in frequencies in reference to the load being served to. The dynamics of each inverter with droop laws can be modelled as follows,

$$\dot{\delta}_i = \omega_i \tag{2.1a}$$
$$\omega_i = \omega_0 - m_{p_i}(P_i^m - P_i^*) \tag{2.1b}$$
$$V_i = V_0 - m_{q_i}(Q_i^m - Q_i^*) \tag{2.1c}$$
$$\tau_p \dot{P}_i^m = P_i - P_i^m \tag{2.1d}$$
$$\tau_p \dot{Q}_i^m = Q_i - Q_i^m \tag{2.1e}$$

where $\delta_i \in [-\pi/2, \pi/2]$ radians is the power angle with respect to frequency $\omega_i$, $[\omega_i, \omega_0] \in \mathbb{R}$ are the operating frequencies at any given node and the nominal network frequency respectively. $[V_i, V_0] \in \mathbb{R}$ are the operating voltages at a given node and the nominal network voltages respectively, $\tau_p$ is the filter time constant, $P_i^m \in \mathbb{R}$ is the measured active power, $Q_i^m \in \mathbb{R}$ is the measured reactive power, $P_i, Q_i$ are the overall active and reactive power respectively, given by,

$$\begin{aligned} P_i &= \sum_{j=1}^{N} \frac{E_i E_j}{X_{ij}} sin(\delta_i - \delta_j) \\ Q_i &= \frac{E_i^2}{X_i} - \sum_{j=1}^{N} \frac{E_i E_j}{X_{ij}} cos(\delta_i - \delta_j) \end{aligned} \tag{2.2}$$

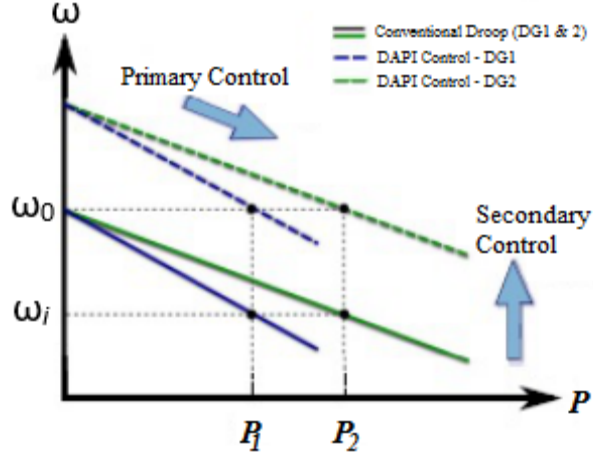

Figure 2.2: DAPI control scheme showing active power sharing by ordinate shifting [44].

where $E_i, E_j$ are the voltage magnitudes and $X_{ij}$ is the reactance of transmission line, $X_i = 1/\sum_{j=1}^{N} X_{ij}^{-1}$; $m_{p_i}$ and $m_{q_i}$ are droop coefficients of $\omega$ and $V$, respectively. $P_i^*$ and $Q_i^*$ are desired equilibrium constants. Differentiating (2.1b) (respectively (2.1c)), substituting $\dot{P}_i^m$ (respectively $\dot{Q}_i^m$) from (2.1d) (respectively (2.1e)) and further substituting $P_i^m$ (respectively $Q_i^m$) from (2.1d) (respectively (2.1e)) the dynamics of $i^{th}$ inverter can be re-framed as,

$$\dot{\delta}_i = \omega_i \tag{2.3a}$$
$$\tau_{p_i}\dot{\omega}_i = \omega_0 - \omega_i - m_{p_i}(P_i - P_i^*) \tag{2.3b}$$
$$\tau_{p_i}\dot{V}_i = V_0 - V_i - m_{q_i}(Q_i - Q_i^*) \tag{2.3c}$$

The laws mentioned in (2.3) generate this deviations in the operating frequency from the set point (namely $\omega_0$), for all values of $P_i$ (respectively $Q_i$) except the trivial equilibrium $P_i = P_i^*$ (respectively $Q_i = Q_i^*$) and hence, an additional control action to restore these parameters to their nominal value is required. In the next subsection, I define a secondary controller on a hierarchical topology for the same.

### 2.1.2 Secondary Control

In order to achieve frequency synchronization and to track set point frequency, DAPI has been proposed in the literature [44]. A micro grid with dynamics (2.3b) can be controlled for frequency synchronization as under (refer Figure 2.1 and Figure 2.2),

$$\omega_i = \omega_0 - m_i P_i + \Omega_i \tag{2.4a}$$
$$k_i \frac{d\Omega_i}{dt} = -(\omega_i - \omega_0) - \sum_{j=1}^{N} c_{ij}(\Omega_i - \Omega_j) \tag{2.4b}$$

where the parameters are as defined in the section 2.1.1 above, with $\Omega_i$ being the control input, $c_{ij}$ diffusive coupling gains and $k_i$ the time constant of the controller dynamics. As shown in Figure 2.2, as load varies the converters try to maintain active power sharing by correcting respective frequencies. These deviations can be compensated using a secondary control $\Omega_i$ so as to shift it back to the set point. Although, the rate at which they synchronise is dependent on time constant $k_i$ and network coupling constant $c_{ij}$. Fast changes in power accounts for fast actions to be taken in control input $\Omega_i$, which is dependent again on $k_i$ and $c_{ij}$. Assuming that value of $k_i$ remains constant (being system characteristics), coupling weights can be adapted as per the deviations in frequency in order to adjust real-time rate of synchronization.

### 2.1.3 Control over Wireless Networks

Micro grid can be viewed as a cyber-physical system (CPS) integrating physical inputs from the field with software capabilities of centralized sophisticated control and communication networks. Such kind of framework helps analyze and control micro grids in an effective and organised manner. Hence, a micro grid model must include the impact of communication networks and other CPS components for efficiency, reliability and stability of the overall system. For instance, the data to be controlled (i.e., the operating frequency in the given case) can be communicated over a network to a centralised cloud server. On the server, real-time measurement database are shared across the nodes as per the control protocol. The effective control inputs so calculated using (2.4b) are then communicated to the respective source inverters to take necessary actions. Communicating the data over a network channel may face several network related issues such as time delays, packet drops, system failures and malicious intrusions. In such scenarios, it is worth evaluating the performance of the control paradigm so as to assure system robustness and performance.

### 2.1.4 Problem Formulation

As mentioned in previous sections, a droop control strategy helps achieve active power sharing thereby generating deviations in set point parameters. These deviations could be quite evident in the case of sudden load change and could be tested for Gaussian distributed loads too. To cater to the former problem and its effects on operating frequencies secondary controllers as in [44] can be implemented. Although, it has been observed that these are slow acting controllers and experience time scale issues. Also, robustness to external disturbances and network uncertainties still remains an open problem. Hence, problem statement can be formulated as to design a controller such that,

1. Hierarchical synchronization (i.e., synchronization of primary control as well as secondary control parameters), in presence of external disturbances and network uncertainties can be achieved,

2. Improvise speed of control action for fast switching loads,

3. Optimise coupling weights so as to reduce stress on power spectrum of communication channel.

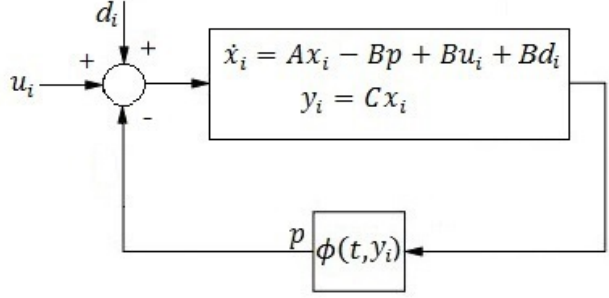

Figure 2.3: A nonlinear system represented in Lure' form.

## 2.2 Mathematical Background

### 2.2.1 Lure' Forms in Presence of External Disturbances

A nonlinear system represented by its corresponding interconnected linear and nonlinear parts in order to render passivity analysis feasible is termed as Lure' forms manipulation in control theory. Systems in Lure' forms are given according to (2.5), with $A, B, C$ defining the system characteristics of the linear part and $\phi(t, y)$ the static nonlinearity. The feedback control loop of a Lure' system can be represented as shown in Figure 2.3,

$$\begin{aligned} \dot{x}_i &= Ax_i - B\phi_i(t, y_i) + Bu_i + Bd_i \\ y_i &= Cx_i \end{aligned} \quad (2.5)$$

where $x_i \in \mathbb{R}^n, u_i, y_i, d_i \in \mathbb{R}^m, i = 1, \ldots, N$; $d_i$ satisfies $\| d_i \|_2 \leq \theta_i$, (Definition 2.1), with $\theta_i > 0$ being a constant bound. Linear characteristics of system are given by $A \in \mathbb{R}^{n \times n}$ and is Hurwitz (i.e., eigenvalues$(A) \leq 0$), $B \in \mathbb{R}^{n \times m}$ and $C \in \mathbb{R}^{m \times n}$. $(A, B)$ is controllable and $(A, C)$ is observable. $\phi_i(t, y) : \mathbb{R}^+ \times \mathbb{R}^m \to \mathbb{R}^m$ is a structured monotonously increasing sector bounded nonlinearity (Definition 2.2). Compartmentalization of a system as in (2.5) helps analyze a nonlinear system segregated into its linear and nonlinear interconnected counterparts. It must be noted though that this manipulation is to analyze the system in a passive frame (i.e., outputs<inputs) and hence derive its stability (bounded inputs bounded outputs) instantaneously.

**Definition 2.1.** $\|d_i\|_2 = \left(|d_1|^2 + |d_1|^2 + \cdots + |d_n|^2\right)^{1/2}$

**Definition 2.2.** [61], [63] A nonlinearity of the form $\phi_i(t, y_i)$ is said to be monotonously increasing sector bounded in the sector $[a, b] \in \mathbb{R}$ if,

$$\begin{aligned} \phi_1(t, y_1(t)) &< \phi_2(t, y_2(t)), \quad y_1(t) < y_2(t); \quad t \in \mathbb{R}^+, \\ q &\leq \frac{\phi_2(t, y_2(t)) - \phi_1(t, y_1(t))}{y_2(t) - y_1(t)} \leq r, \quad \forall r > q \geq 0; \quad r, q \in [a, b]. \end{aligned} \quad (2.6)$$

CHAPTER 2. FREQUENCY SYNCHRONIZATION OF MICRO GRIDS### 2.2.2 Output-feedback Passivity

**Definition 2.3.** [61] A function $f(x)$ is positive definite, iff $f(x) > 0, \forall x \neq 0$. A weaker condition is when $f(x) \geq 0, \forall x \neq 0$, then $f(x)$ is called as positive semidefinite function.

**Definition 2.4.** [61] A system of the form (2.5) without disturbance (i.e., $d_i = 0$), is output-feedback passive, if a continuously differentiable positive semidefinite function $S(x(t))$ exists (Definition 2.3), called the storage function given by,

$$\frac{dS(x(t))}{dt} \leq u^T y - y^T \rho(y) \tag{2.7}$$

for some function $\rho(y)$, and is output feedback strictly passive for $y^T \rho(y) > 0, \forall y \neq 0$ (refer Definition 6.3 from [61]).

**Definition 2.5.** [65] A system with nonlinearity of the form, $z_i = \phi_i(t, y_i)$ is incrementally passive if the function $\phi_i(t, \cdot)$ satisfies,

$$(y_j - y_i)^T (\phi_j(t, y_j) - \phi_i(t, y_i)) \geq 0 \tag{2.8}$$

for all, $y_i, y_j \in \mathbb{R}$ and $i, j \in \mathbb{Z}^+$.

**Definition 2.6.** [61] A proper rational transfer function $Z(s)$ is called strictly positive real if,

1. $Z(s)$ is Hurwitz, i.e., all the poles of $Z(s)$ are in $\mathbb{R}_{\leq 0} = \{x \in \mathbb{R} : x \leq 0\}$,

2. $Z(j\omega) + Z^T(-j\omega) > 0, \forall \omega \in \mathbb{R}$, and

3. $Z(\infty) + Z^T(\infty) \geq 0$; there exists set of singular values $\sigma_i(\omega)$ and $\sigma_0 > 0, \omega_0 > 0$, such that
$$\omega^2 \sigma_{min}[Z(j\omega) + Z^T(-j\omega)] \geq \sigma_0, \forall |\omega| \geq \omega_0. \tag{2.9}$$

**Lemma 2.1.** *Lure' systems of the form given in (2.5), with $PB = C^T$ are strictly output feedback incrementally passive (SOFIP).*

**Proof.** Let $S$ be a storage function of $N$ interconnected systems ($S : \mathbb{R}^N \to \mathbb{R}$) given by $S(\xi_i) = \frac{1}{2}\xi_i^T P \xi_i$, $P$ being a positive definite symmetric matrix and $L, W$ chosen to be $0_{N \times N}$, where $\xi_i = \sum_{j=1}^{N}(x_i - x_j) = [\xi_1, \xi_2, \ldots, \xi_N] \in \mathbb{R}^N$ is the parameter of synchronization. Now, differentiating $S$ with respect to time,

$$\begin{aligned}\dot{S}(\xi_i) &= \dot{\xi}_i^T P \xi_i \\ &= \left[\sum_{j=1}^{N}(\dot{x}_i - \dot{x}_j)\right]^T P \xi_i \\ &= \left[A\sum_{j=1}^{N}(x_i - x_j) - B\sum_{j=1}^{N}(\phi_i(t, y_i) - \phi_j(t, y_j)) + B\sum_{j=1}^{N}(u_i - u_j) + B\sum_{j=1}^{N}(d_i - d_j)\right]^T P \xi_i\end{aligned} \tag{2.10}$$

Now let $\sum_{j=1}^{N}(u_i - u_j) = \delta u_i$, similarly $\sum_{j=1}^{N}(\phi_i(t, y_i) - \phi_j(t, y_j)) = \delta \phi_i$ and $\sum_{j=1}^{N}(d_i - d_j) = \delta d_i$, hence following can be deduced,

$$\dot{S}(\xi_i) = \xi_i^T A^T P \xi_i + \delta \phi_i^T B^T P \xi_i + \delta u_i^T B^T P \xi_i + \delta d_i^T B^T P \xi_i \tag{2.11}$$

Using the Definition 2.6 and Lemma 6.3 from [61], $\xi_i^T P A \xi_i \leq 0$, knowing $\delta y_i = C \xi_i$ thereby substituting $PB = C^T$, (2.11) can be re-formulated as below,

$$\begin{aligned}\dot{S}(\xi_i) &\leq -\delta \phi_i^T C \xi_i + \delta u_i^T C \xi_i + \delta d_i^T C \xi_i \\ &\leq -\delta \phi_i^T \delta y_i + \delta u_i^T \delta y_i + \delta d_i^T \delta y_i \\ &\leq -(\delta \phi_i - \delta d_i)^T \delta y_i + \delta u_i^T \delta y_i \\ &\leq -\delta \Phi_i^T \delta y_i + \delta u_i^T \delta y_i\end{aligned} \tag{2.12}$$

Now, taking disturbance to be zero and using the Definition 2.4 and Definition 2.8 renders $S(\xi_i)$ SOFIP. ∎

### 2.2.3 Network Connectivity

Consider an interconnected dynamical system comprising of $N$ nodes defined by (2.5). Let $G$ be an undirected graph with $M$ set of links connected via $N$ set of nodes. To represent orientation of links in graph $G$, consider an incidence matrix $E$ with links $l = 1, \ldots, M$ given by,

$$E(i, j) = \begin{cases} 1, & \text{if information transfer is directed from } i \text{ to } j \\ -1, & \text{if information transfer is directed from } j \text{ to } i \\ 0, & \text{if they are unconnected} \end{cases}$$

Let, $K$ be the matrix of link weights. Then the graph Laplacian of $G$ can be given by,

$$L(t) = EK(t)E^T \tag{2.13}$$

**Definition 2.7.** The Kronecker product of any two matrices $P \in \mathbb{R}^{m_1 \times n_1}$ and $Q \in \mathbb{R}^{m_2 \times n_2}$ is defined as

$$P \otimes Q := \begin{bmatrix} p_{11}Q & \cdots & p_{1n}Q \\ \vdots & \ddots & \vdots \\ p_{m1}Q & \cdots & p_{mn}Q \end{bmatrix} \in \mathbb{R}^{m_1 m_2 \times n_1 n_2}$$

## 2.3 Main Result: Robust Adaptive Synchronization over a Network

In this section, a novel control law to achieve robust synchronization over an all-to-all connected network is proposed. The control law is termed as Robust Adaptive Distributed Averaging Proportional Integral Control (RADAPI) protocol, in reference to the objective being served and an increment to existing DAPI law. In Theorem 2.1, RADAPI is proposed thereby stability analysis is provided using incremental passivity fundamentals.

**Theorem 2.1.** *Let the state-space dynamics of a system with N agents be given by (2.5), with parameters as defined previous sections. Let $u_i$ be defined as under,*

$$\dot{u}_i = -y_i + y_{ref} - \sum_{j=1}^{N} c_{ij}(u_i - u_j)$$
$$\dot{c}_{ij} = \left[ -\Delta_{ij} c_{ij} + (u_i - u_j)^T \Gamma_{ij}(u_i - u_j) \right] \quad (2.14)$$

*where $\Gamma_{ij}, c_{ij}, \Delta_{ij} \in \mathbb{R}^{m \times m}$ are diagonal control design parameter matrix, coupling weights and control regulation gain of of G respectively. The deviations in the value of $u_i - u_j$ define the coupling weights $c_{ij}$ in real-time. Then, for the control law in (2.14) and for any $\Delta_{ij} > 0$, the following holds,*

1. $\lim_{t \to \infty} \sum_{j=1}^{N} (y_i - y_j) = 0; \forall i, j = 1, \ldots, N,$

2. $\lim_{t \to \infty} \sum_{j=1}^{N} (u_i - u_j) = 0$ *and thereby* $\lim_{t \to \infty} \sum_{j=1}^{N} (\phi_i(t, y_i) - \phi_j(t, y_j)) = 0, \forall i, j = 1, \ldots, N,$

3. $\lim_{t \to \infty} c_{ij} = c_*; c_{ij} = c_{ij_0} e^{-\Delta_{ij} t} + c_*,$ *where $c_*$ is the lower bound on the value of coupling weight $c_{ij}, \forall i, j = 1, \ldots, N$ and $c_{ij_0}$ is the initial value of $c_{ij}$,*

4. *The closed loop system is incrementally passive, i.e. $\xi_i \to 0$, as $t \to \infty$; where, $\xi_i = \sum_{j=1}^{N} (x_i - x_j)$ and*

5. *the variables $\xi_i$ and $c_{ij}$ exponentially converge with lower bound given by $\Theta$ as under,*

$$\Theta = \left\{ \xi_i, c_{ij} : \dot{Z} \leq -\frac{1}{2} \sum_{i=1}^{N} \sum_{j=1}^{N} \frac{1}{\Gamma_{ij}} \Delta_{ij} \hat{c}_{ij}^2 - \frac{1}{2} \left( \delta \tilde{\Phi}_i^T \delta \tilde{y}_i + c_* \lambda_2 \delta \tilde{u}_i^T \delta \tilde{u}_i \right) \right\} \quad (2.15)$$

*where, $\tilde{\xi}_i = \xi_i - \bar{\xi}$, and $\tilde{u}_i, \tilde{y}_i, \tilde{d}_i; \bar{\xi} = \frac{1}{N} \sum_{j=1}^{N} \xi_j$, analogously $\bar{u}, \bar{y}, \bar{d}$. Z is the incremental Lyapunov function ($Z : \mathbb{R}^n \to \mathbb{R}$) given by,*

$$Z = \frac{1}{4} \tilde{\xi}_i^T P \tilde{\xi}_i + \frac{1}{4 \Gamma_{ij}} \sum_{i=1}^{N} \sum_{j=1}^{N} \hat{c}_{ij}^2 + \frac{1}{4} \delta \tilde{u}_i^T \delta \tilde{u}_i \quad (2.16)$$

**Proof.** Author starts the proof by assuming incremental Lyapunov function Z as given in the statement of the theorem. Let $Z = V + W + U$, where $V = \frac{1}{4} \tilde{\xi}_i^T P \tilde{\xi}_i$, $W = \frac{1}{4\Gamma_{ij}} \sum_{i=1}^{N} \sum_{j=1}^{N} \hat{c}_{ij}^2$ and $U = \frac{1}{4} \delta \tilde{u}_i^T \delta \tilde{u}_i$. Then, differentiating Z with respect to time will give,

$$\dot{Z} = \dot{V} + \dot{W} + \dot{U} \quad (2.17)$$

Now, $\tilde{\xi}_i = \xi_i - \bar{\xi}$ thereby following can be derived,

$$\begin{aligned}
\bar{\xi} &= \frac{1}{N}\sum_{j=1}^{N}\xi_j = \frac{1}{N}\sum_{j=1}^{N}\sum_{i=1}^{N}(x_j - x_i) \\
&= \frac{1}{N}\sum_{j=1}^{N}(x_j\sum_{i=1}^{N}(1) - \sum_{i=1}^{N}x_i) \\
&= \frac{1}{N}\sum_{j=1}^{N}(Nx_j - N\bar{x}) \\
&= \frac{1}{N}(N\sum_{j=1}^{N}x_j - N\bar{x}\sum_{j=1}^{N}(1)) \\
&= \frac{1}{N}(N^2\bar{x} - N^2\bar{x}) = 0.
\end{aligned} \quad (2.18)$$

thus, $\tilde{\xi}_i = \xi_i$. Now, using few steps from Lemma 2.1, let there be a storage function $V(\tilde{\xi}_i) = \frac{1}{4}\tilde{\xi}_i^T P \tilde{\xi}_i$, then by making use of (2.10),

$$\begin{aligned}
\dot{V}(\tilde{\xi}_i) &= \frac{1}{2}\left[A\sum_{j=1}^{N}(x_i - x_j) - B\sum_{j=1}^{N}(\phi_i(t,y_i) - \phi_j(t,y_j)) + B\sum_{j=1}^{N}(u_i - u_j) + B\sum_{j=1}^{N}(d_i - d_j)\right]^T P\tilde{\xi}_i \\
&= \frac{1}{2}\left[A\sum_{j=1}^{N}((x_i - \bar{x}) - (x_j - \bar{x})) - B\sum_{j=1}^{N}((\phi_i(t,y_i) - \bar{\phi}) - (\phi_j(t,y_j) - \bar{\phi})) \right. \\
&\qquad \left. + B\sum_{j=1}^{N}((u_i - \bar{u}) - (u_j - \bar{u})) + B\sum_{j=1}^{N}((d_i - \bar{d}) - (d_j - \bar{d}))\right]^T P\tilde{\xi}_i \\
&= \frac{1}{2}\left[A\tilde{\xi}_i - B\delta\tilde{\phi}_i + B\delta\tilde{u}_i + B\delta\tilde{d}_i\right]^T P\tilde{\xi}_i
\end{aligned} \quad (2.19)$$

and following same steps used in Lemma 2.1,

$$\dot{V} \leq \frac{1}{2}\left[-\delta\tilde{\Phi}_i^T \delta\tilde{y}_i + \delta\tilde{u}_i^T \delta\tilde{y}_i\right] \quad (2.20)$$

An assumption that introduction of disturbance only reduces the sector of nonlinearity, i.e., $d_i = N(0,\gamma_i) \in \mathbb{R}^m$ being a Gaussian distributed random variable thereby $\Phi_i(t,y_i(t))$ retains the properties mentioned in Definition (2.6) with a smaller set $[\hat{a},\hat{b}] \in [a,b]$ is made. Hence, $V$ is SOFIP.

Let $W = \frac{1}{4}\sum_{i=1}^{N}\sum_{j=1}^{N}\frac{1}{\Gamma_{ij}}\hat{c}_{ij}^2$, where $\hat{c}_{ij} = c_{ij} - c_{ij}^*$

$$c_{ij}^* = \begin{cases} c_*, & \text{if } i \text{ and } j \text{ are neighbours in } G \\ 0, & \text{otherwise} \end{cases}$$

$c_*$ is a constant left to the control designer. Thus, differentiating $W$ and substituting $c_{ij}$

given by (2.14),

$$\begin{aligned}
\dot{W} &= \frac{1}{2}\sum_{i=1}^{N}\sum_{j=1}^{N}\frac{1}{\Gamma_{ij}}\hat{c}_{ij}\dot{\hat{c}}_{ij} \\
&= \frac{1}{2}\sum_{i=1}^{N}\sum_{j=1}^{N}\frac{1}{\Gamma_{ij}}\left[-\Delta_{ij}\hat{c}_{ij}^2 + \hat{c}_{ij}(\tilde{u}_i - \tilde{u}_j)^T\Gamma_{ij}(\tilde{u}_i - \tilde{u}_j)\right] \\
&= \frac{1}{2}\sum_{i=1}^{N}\sum_{j=1}^{N}\frac{1}{\Gamma_{ij}}\left[-\Delta_{ij}\hat{c}_{ij}^2 + \hat{c}_{ij}\delta\tilde{u}_i^T\Gamma_{ij}\delta\tilde{u}_i\right] \\
&= \frac{1}{2}\sum_{i=1}^{N}\sum_{j=1}^{N}\frac{1}{\Gamma_{ij}}\left[-\Delta_{ij}\tilde{c}_{ij}^2\right] + \frac{1}{2}\sum_{i=1}^{N}\left[c_{ij}\delta\tilde{u}_i^T\delta\tilde{u}_i - c_{ij}^*\delta\tilde{u}_i^T\delta\tilde{u}_i\right] \\
&= \frac{1}{2}\sum_{i=1}^{N}\sum_{j=1}^{N}\frac{1}{\Gamma_{ij}}\left[-\Delta_{ij}\tilde{c}_{ij}^2\right] - \frac{1}{2}\left[\delta\tilde{u}_i^T(L(G)\otimes I_p)\delta\tilde{u}_i + c^*\lambda_2\delta\tilde{u}_i^T\delta\tilde{u}_i\right]
\end{aligned} \quad (2.21)$$

where $c^* > 0$, $-(L(G)\otimes I_p)\delta\tilde{u}_i = \sum_{i=1}^{N}\sum_{j=1}^{N}c_{ij}(u_j - u_i) = \sum_{i=1}^{N}c_{ij}\delta\tilde{u}_i$ (refer Definition 2.7) and $\lambda_2 > 0$ [46] being first positive eigenvalue of $L(G)$, implying $\dot{W} \leq 0, \forall c_{ij}$.

Differentiating $\delta\tilde{u}_i$ with respect to time,

$$\begin{aligned}
\delta\dot{\tilde{u}}_i = \delta\dot{u}_i &= \sum_{j=1}^{N}(\dot{u}_i - \dot{u}_j) \\
&= \sum_{j=1}^{N}\left[\left(-y_i + y_{ref} - \sum_{k=1}^{N}c_{ik}(u_i - u_k)\right) - \left(-y_j + y_{ref} - \sum_{k=1}^{N}c_{ik}(u_j - u_k)\right)\right] \\
&= -\delta\tilde{y}_i - \sum_{i=1}^{N}\sum_{j=1}^{N}c_{ij}(\tilde{u}_i - \tilde{u}_j)
\end{aligned} \quad (2.22)$$

Using $U = \frac{1}{4}\delta\tilde{u}_i^T\delta\tilde{u}_i$, differentiating with respect to time and substituting for $\delta\dot{\tilde{u}}_i$ from (2.22),

$$\begin{aligned}
\dot{U} &= \frac{1}{2}\left\{-\delta\tilde{y}_i - \sum_{i=1}^{N}\sum_{j=1}^{N}c_{ij}(\tilde{u}_i - \tilde{u}_j)\right\}^T\delta\tilde{u}_i \\
&= \frac{1}{2}\left\{-\delta\tilde{y}_i^T\delta\tilde{u}_i + \delta\tilde{u}_i^T(L(G)\otimes I_p)\delta\tilde{u}_i\right\}
\end{aligned} \quad (2.23)$$

Now from (2.20), (2.21) and (2.23), $\dot{Z}$ can be formed using (2.17) as follows,

$$\begin{aligned}
\dot{Z} &\leq \dot{V} + \dot{W} + \dot{U} \\
&\leq -\frac{1}{2}\left\{\sum_{i=1}^{N}\sum_{j=1}^{N}\frac{1}{\Gamma_{ij}}\Delta_{ij}\hat{c}_{ij}^2 + \left(\delta\tilde{\Phi}_i^T\delta\tilde{y}_i + c_*\lambda_2\delta\tilde{u}_i^T\delta\tilde{u}_i\right)\right\} \leq 0
\end{aligned} \quad (2.24)$$

By integrating on both sides, $\tilde{u}_i \in L_2$, further boundedness implies, $\xi_i \to \dot{\tilde{y}}_i$ is bounded and Barbalat's lemma (Theorem 8.4 from [61]) guarantees,

$$\lim_{t\to\infty}\xi_i = \lim_{t\to\infty}\sum_{j=1}^{N}(y_i - y_j) = \lim_{t\to\infty}\sum_{j=1}^{N}(\phi_i - \phi_j) = \lim_{t\to\infty}\sum_{j=1}^{N}(u_i - u_j) = 0 \quad (2.25)$$

Integrating, $\dot{c}_{ij}$ from (2.14), when (2.25) holds true gives,

$$c_{ij} = c_{ij_0}e^{-\Delta_{ij}t} + c_* \to \lim_{t\to\infty}c_{ij} = c_*. \quad (2.26)$$

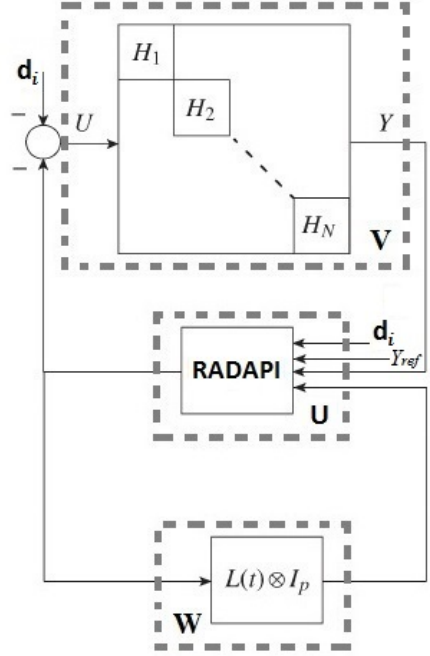

Figure 2.4: An interconnection of storage functions $U, V$ and $W$ representing input-output flows of respective blocks.

∎

## 2.4 Case Study: Hierarchical Control of Micro grids

To validate the proposed idea, an islanded micro grid comprising of two inverters serving a switching load is considered i.e., $P_{Load} \in [0.33, 0.67]$pu or per unit, as shown in Figure 2.5(a). Primary droop control induces deviations in frequencies, which are regulated to set point value using RADAPI control defined in (2.14),

$$\begin{aligned}
\omega_i &= \omega_0 - m_i P_i + \Omega_i + d_i \\
\dot{\Omega}_i &= \omega_0 - \omega_i - \sum_{j=1}^{N} c_{ij}(\Omega_i - \Omega_j) \\
\dot{c}_{ij} &= \left[ -\Delta_{ij} c_{ij} + (\Omega_i - \Omega_j)^T \Gamma_{ij}(\Omega_i - \Omega_j) \right]
\end{aligned} \quad (2.27)$$

where $d_i = N(0, 1e^{-4})$ is considered to be Gaussian, $\Gamma_{ij} = \Gamma_{ji} > 0$ is the controller response gain left to the designer, to be chosen as per the required system characteristics. In reference to (2.14), $y_i = \omega_i, u_i = \Omega_i$ and set of inverters seen from a multi-agent point-of-view, are controlled by RADAPI law updating the rate of synchronization at every instant thereby eliminating frequency deviations. This helps achieve synchronization amongst inverters as well as attain a set point trajectory of $\omega_0$, even in an uncertain environment.

Using the micro grid system dynamics in (2.27), nonlinearity being the power flow component with sinusoidal characteristics. Using the Lure' form restructuring, the overall system dynamics can be rewritten as follows,

$$\begin{bmatrix} \dot{\delta_i} \\ \dot{\omega_i} \end{bmatrix} = \begin{bmatrix} O_{N \times N} & I_{N \times N} \\ O_{N \times N} & -I_{N \times N} \end{bmatrix} \begin{bmatrix} \delta_i \\ \omega_i \end{bmatrix} + \begin{bmatrix} O_{N \times 1} \\ I_{N \times 1} \end{bmatrix} (m_{p_i}(P_i - P_i^*)) + \begin{bmatrix} O_{N \times 1} \\ \omega_0 I_{N \times 1} \end{bmatrix}. \quad (2.28)$$

where $(m_{p_i}(P_i - P_i^*)) \in [-\pi/2, \pi/2]$ is monotonically increasing and bounded.

### 2.4.1 Simulation Outcomes

In this section, RADAPI is considered against DAPI control to evaluate its efficiency. In order to quantify amount of advantages gained following metrics are defined.

#### 2.4.1.1 Metrics of Evaluation

There are 3 major metrics that are used to evaluate the performance of controllers in consideration (RADAPI vs DAPI).

1. *Output Frequency* - The major criteria of acceptance is synchronization of output frequencies of inverters and their regulation towards set point trajectory, first criteria of evaluation would be to identify which control allows frequencies to synchronise the fastest.

2. *Control Effort* - Second criteria of evaluation is amount of control efforts required to stabilise the system, as well as time taken for them to synchronise amongst each other.

3. *Active Power Sharing* - The final criteria of concern would be to make sure controller doesn't lose power sharing in the course of achieving set point trajectory. Also, in order to quantify the fastness of RADAPI over DAPI a fourth metric is defined as the ratio of the two subtracted over 1,

$$\text{Net Gain}(\%) = \frac{(\text{DAPI Metric} - \text{RADAPI Metric})}{\text{DAPI Metric}} \times 100. \quad (2.29)$$

#### 2.4.1.2 Results

Sudden load variations as shown in Figure 2.5(a) are considered and simulations carried out using MATLAB/Simulink. Subsequent figures illustrate specific cases of synchronization for defined parameters. For simplicity, the injected uncertainty has been kept constant for all the participating inverters and is chosen to be $N(\mu, \sigma) = N(0, 0.0001)$ with $\Delta_{ij} = \Delta = 0.0005$.

#### 2.4.1.3 Nominal Operation

In this case, nominal operation of RADAPI (2.14) is compared with that of DAPI as defined in (2.4). As seen from Figure 2.5-2.7, RADAPI is evaluated against DAPI as per the metrics defined before. It is observed that RADAPI helps achieve faster synchronization and thus compares against some commonly used distributed averaging strategies [44].

In order to further exploit the benefits of RADAPI, a set of test case scenarios have been considered, to validate effectiveness of the control strategy in challenging scenarios. Various test cases with several failure modes, have been performed and tested. For reference, it is considered that fault occurs anywhere in the time frame of 30-60 seconds, RADAPI tries to restore synchronization as and when the network links are functional again.

Table 2.1: Nominal operation of micro grid.

|  | **RADAPI** | **DAPI** | **Net Gain** |
|---|---|---|---|
| **Metric 1** | 20 seconds | 30 seconds | 33 % |
| **Metric 2** | 20 seconds | 30 seconds | 33 % |
| **Metric 3** | 20 seconds | 30 seconds | 33 % |

### 2.4.2 Time Delay

Next, a random time delay of 250milliseconds is induced in the output of the micro grid defined in (2.27). The results are quantified using the metrics defined in section 2.4.1.1 and are tabulated in Table 2.2. Also, as observed in Figure 2.6, RADAPI is seen effectively faster than DAPI scheme.

Table 2.2: Effects in output network channel with time delay of 250 milliseconds.

|  | **RADAPI** | **DAPI** | **Net Gain** |
|---|---|---|---|
| **Metric 1** | 18 seconds | 40 seconds | 55 % |
| **Metric 2** | 18 seconds | 40 seconds | 55 % |
| **Metric 3** | 18 seconds | 40 seconds | 55 % |

### 2.4.3 Malicious Data

Next, a malicious data or an uncertainty of certain order has been induced in the output of the micro grid defined in (2.27). The results are quantified using the metrics defined in section 2.4.1.1 and are tabulated in Table 2.3. Also, as observed in Figure 2.7, RADAPI is seen effectively faster than DAPI scheme.

Table 2.3: Effects of malicious data in output network channel.

|  | **RADAPI** | **DAPI** | **Net Gain** |
|---|---|---|---|
| **Metric 1** | 0 seconds | 3 seconds | - |
| **Metric 2** | 0 seconds | 3 seconds | - |
| **Metric 3** | 20 seconds | 35 seconds | 43 % |

#### 2.4.3.1 Special Case - Effect of change in $\Delta_{ij}$

$\Delta_{ij}$ is a critical parameter, ensuring both boundedness and larger region of attraction. The effect of changes in its value on the system parameters has been considered here. As shown in Figure 2.7, with $\Delta_{ij} = \Delta = 0.0005$, the gain drops at a lower rate once the threshold has been attained. Although, when $\Delta_{ij} = \Delta$ is chosen to be 0.5, the drop is faster and it loses synchronization due to absence of coupling. In order to prevent it from loosing consensus, a minor modification is made to RADAPI, where the controller is forced to attain minimum gain of 1. This allows the network to maintain synchronization, and emulate behaviour of an event based strategy. Outputs are analyzed per metric basis in Table 2.4.

Table 2.4: Effect of change in $\Delta_{ij}$ on system dynamics.

|  | **RADAPI** | **DAPI** | **Net Gain** |
| --- | --- | --- | --- |
| **Metric 1** | 25 seconds | 30 seconds | 17 % |
| **Metric 2** | 25 seconds | 30 seconds | 17 % |
| **Metric 3** | 25 seconds | 30 seconds | 17 % |

## 2.5 Summary and Inferences

To summarize,

1. In this work, a novel robust adaptive synchronization scheme (named RADAPI) has been proposed to maintain nominal operation of a nonlinear system in an uncertain environment.

2. An adaptive strategy has been derived where the deviations in the system parameters due to presence of disturbances/mismatches have been solved.

3. The stability of the closed loop system has been theoretically proved using Lure' form restructuring and region of attraction for the same has been computed. Quantification of effectiveness of the proposed algorithm helps define its competencies.

4. The proposed idea is validated using a problem of proportional load sharing and frequency synchronization in an islanded AC micro grid.

5. Results have been verified using computer simulations and the stability of the system analyzed using incremental passivity fundamentals.

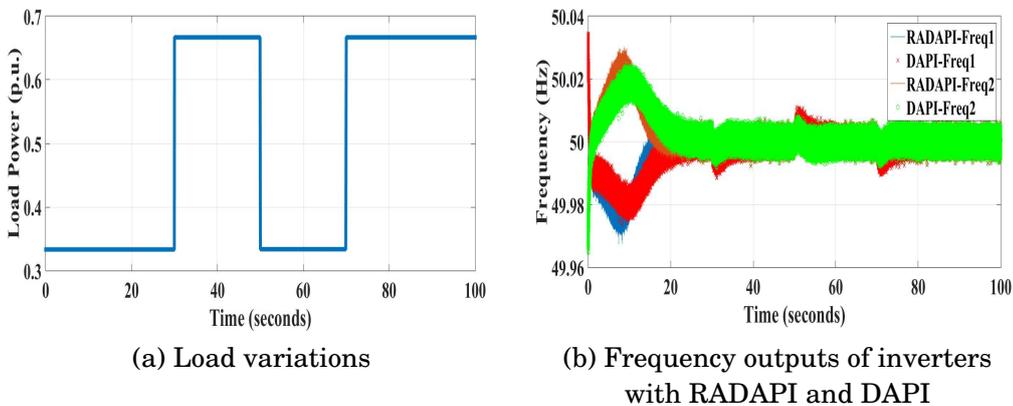

(a) Load variations

(b) Frequency outputs of inverters with RADAPI and DAPI

Figure 2.5: Nominal operation of RADAPI versus DAPI.

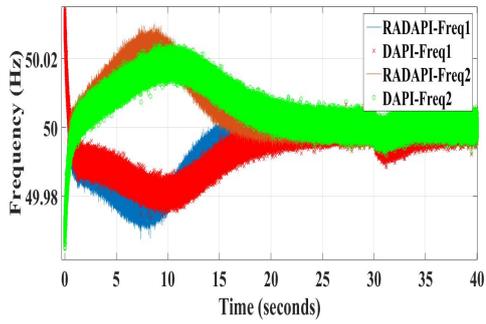

(a) Magnified image of Figure (2.5)(b), showing RADAPI benefits over DAPI

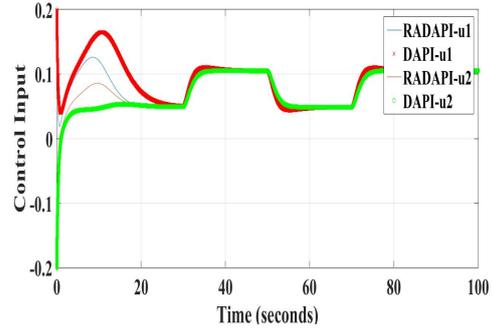

(b) Secondary control inputs

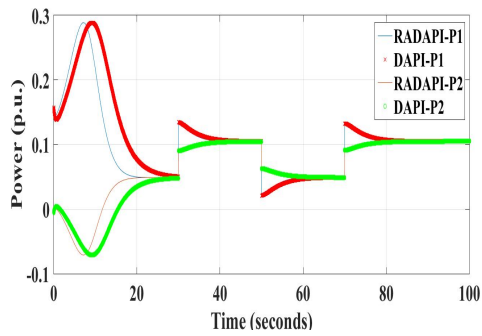

(c) Power sharing of inverters with RADAPI and DAPI

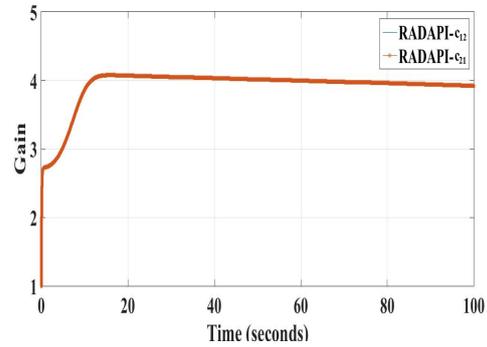

(d) Coupling gains with and without RADAPI

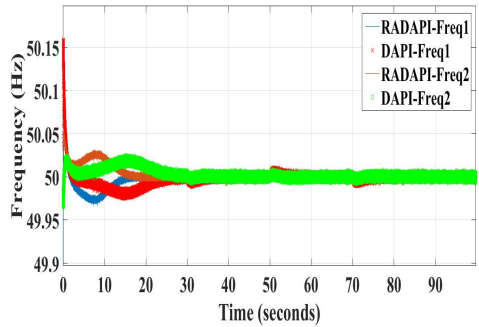

(e) Time delay impact on output frequencies

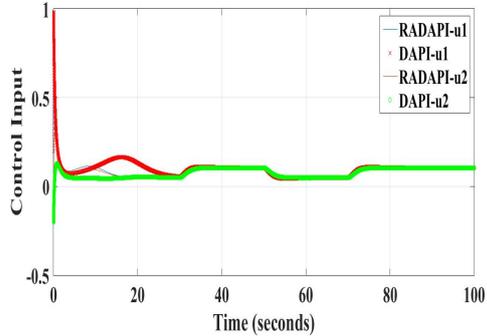

(f) Time delay impact on controller inputs

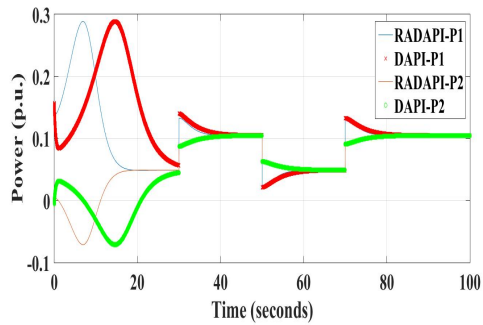

(g) Time delay impact on power sharing

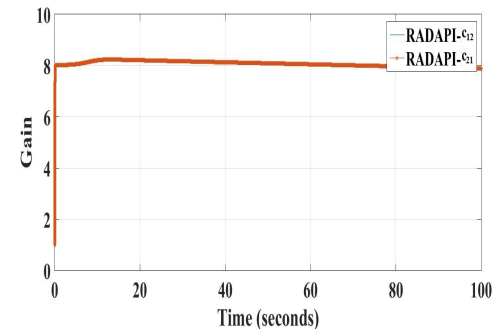

(h) Time delay impact on adaptive gains

Figure 2.6: Nominal operation and time delay response of RADAPI.

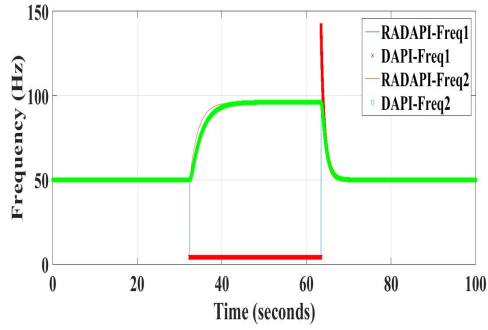

(a) Malicious data impact on output frequencies

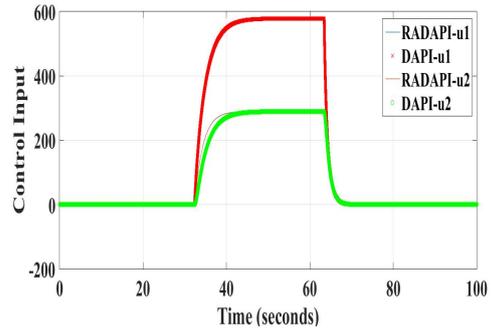

(b) Malicious data impact on controller inputs

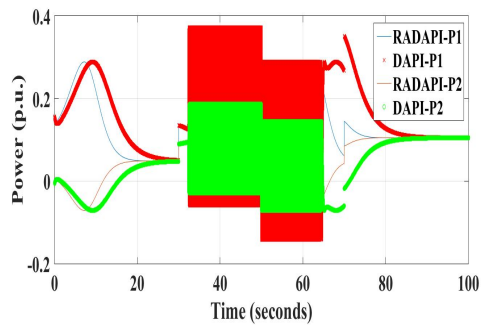

(c) Malicious data impact on power sharing

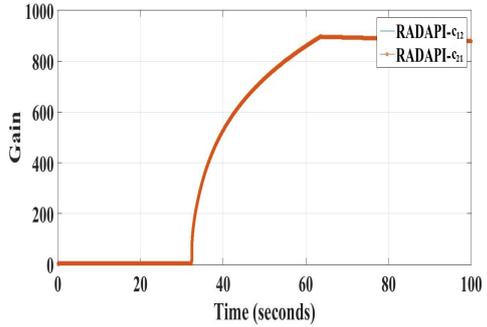

(d) Malicious data impact on adaptive gains

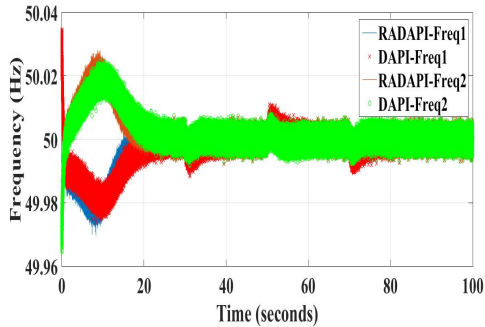

(e) Impact of tuning robustness parameter ($\Delta_{ij}$) on output frequencies

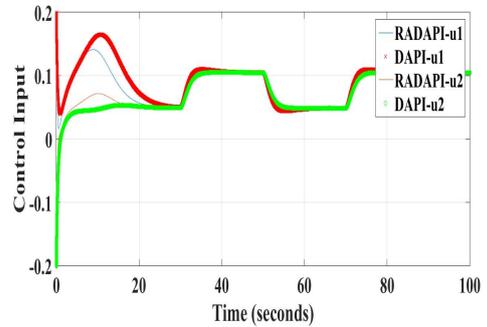

(f) Impact of tuning robustness parameter ($\Delta_{ij}$) on controller inputs

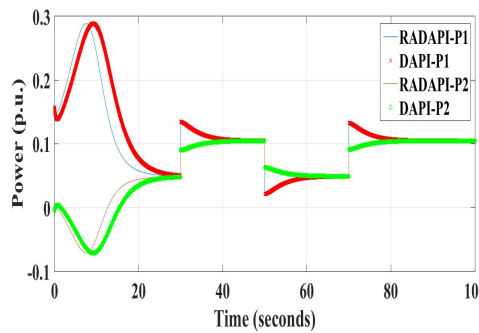

(g) Impact of tuning robustness parameter ($\Delta_{ij}$) on power sharing

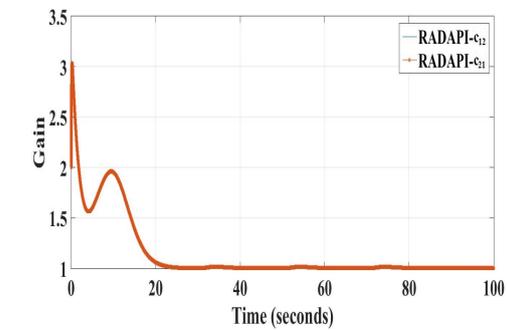

(h) Impact of tuning robustness parameter ($\Delta_{ij}$) on adaptive gains

Figure 2.7: Malicious data and robustness tuning parameter response of RADAPI.

# Controlled De-Synchronization of Thermostatically Controlled Loads

**Table of Contents**

Sustainability is the process of maintaining a balanced environment subject to ecological changes; wherein exploitation of resources, the technical know-how, the inclination of investments, the organization/planning of technological development and institutional changes are all in harmony thereby help enhance present as well as future potential to meet human requirements and aspirations. With the advent of civilization, in order to serve electrical and transportation needs, fossil fuel reserves have been over-utilized. Excessive use of fossil fuels and increasing human settlements, has brought imbalance to the environment. Thus, researchers have moved towards building settlements that can serve human requirements, thereby mean no harm to the environment utilizing least natural resources; hence are sustainable. Conventional power grid architecture of unidirectional power flow has been modified to a bidirectional one whilst promoting penetration of Ancillary Service (AS).

ASs are the supportive programs for dispatch of electric power from seller to purchaser, through the use of smart control techniques thereby maintaining the operations of an interconnected transmission system reliable [66]. Some common ASs are load following, load dispatch services, energy imbalance services, reactive power voltage regulation, loss compensation and system protection services. Load following is referred to the service program that monitors real time consumption vis-a-vis the scheduled loads so as to consume what was estimated. Load dispatch services are real time adjustments in case load following doesn't tally. These real time extracts are supported by system protective services which serve as energy reserves compensating for any discrepancies. The load dispatch services are guided by loss compensation schemes, allowing to regulate total loss allowable. All these actions are taken under energy imbalance programs using system control services. Perturbations in voltage are mitigated online using reactive power voltage regulation service. As observed, these services help keep balance between demand and supply power flows, with majority of them falling on load management end to optimize increasing demand to supply ratios.

ASs have gained attention with increasing penetration of renewable energy sources (RESs) in the course of replacing fossil fuels. Although, RESs are abundantly available in nature, they suffer from intermittencies. These intermittencies are significant enough to render a power grid unstable and hence require additional regulating circuits to avoid direct injection [67–70]. One of the ways of injection is to store it in a battery and then provide it to the grid using a battery management system (BMS) with a stable discharge rate. Although reliable, this strategy bears a high capital expenditure (CAPEX) for installation and hence is commercially less feasible [71–74]. Therefore, research has moved in direction of devising methods to regulate the loads with techno-commercial feasibility. Demand Response (DR) programs have been developed in order to ally with several handshaking criteria and thereby maintain power grid stability.

DR programs allow consumers to adjust their electricity consumption in response to energy prices or incentive payments. ASs form a major part of these DR programs and an optimal use of these services can come to advantage in terms of adjusting loads as per utility demands. As per a survey in U.S. [75], it has been observed that the cooling and heating loads account for the largest annual residential electricity consumption. The scenario remains the same globally, thereby requiring optimization of these loads in order to attain required power aggregation and reduce overall power grid loading. Fundamentally, all heating/cooling loads (Thermostatically Controlled Load (TCL)) comprise of a temperature transducer, a thermostat and other assemblies like compressor, condenser, etc. It can be noted though, that TCLs follow a scheme whereby they switch between ON and OFF states to attain required set-point temperatures. ON states consume full capacity power of the TCL and OFF states imply dormancy. Thus, a clever adjustment of these operation states relatively can help achieve required system objectives as well as can be controlled to provide load following AS.

In [76, 77], field experiments are conducted using household domestic refrigerators to gather, analyze advance or delay of operation time of TCLs to quantify flexibility of achieving required aggregate power. A novel parameter randomization control scheme of TCLs by compartmentalization of compressor cycles so as to attain effective de-synchronization is shown in [74]. Automatic generation control (AGC - a demand-supply balancing technique) tracking technique is employed in [78] using three different protocols. On similar lines, flexible TCL operating cycles are achieved using a two-level scheduling method in an intra-day electricity market in [79]. All the above mentioned schemes though effective, has been observed to deteriorate overall system performance with increase in heterogeneity of system parameters. On a practical frame, [80, 81] proposes a multi-objective model predictive control (MPC) for heating in residential areas using heat pumps (or using heat as energy carrier) to attain similar objective. Again, the scheme seems difficult to be implemented for a large congregation of TCLs due to many parameters being used.

In literature, aggregated models for TCLs are widely studied due to its ability to carry macro level information of total power consumed by TCLs, and are framed using discrete state-space dynamics. Using aggregated models a centralised control scheme has been proposed in [81, 82], whereby the problem of frequency regulation has been solved using controlled dispatch of TCLs. A Markovian probabilistic framework is used in [83] to achieve similar objectives. An event-triggered control of TCLs on a hierarchical architecture is utilized over a network for optimization of communication cost in [84]. A load shedding strategy is implemented in [85] to achieve desired aggregated power. Concepts from optimization theory related to tracking device, convex polytopes and binary multi-swarm, particle optimization are used in [86, 87] and [88]. In [89] batch reinforcement learning has been implemented to achieve stated DR objectives. In order to make TCLs available in day-ahead markets, a model free estimation technique using Monte Carlo methods has been proposed. In [90], a Nano grid solution to store thermal energy in TCLs for later use is experimented. It has been shown, how regulation of maximum power point tracking (MPPT) of a photovoltaic (PV) panel can help avoid wastage of thermal energy. A partial differential equation (PDE) based Fokker-Plank model is proposed in [91]. All of the above-mentioned methods are either probabilistic, discretized (affecting stability, controllability and other properties) or involve considerable computation cost.

Complexity of population of TCLs increases with the number of TCLs, and with the reference aggregate power set by the utility. It is observed that the undesired synchronization of TCL switching signals by control policies have the tendency to limit the time scale as well as capacity of ASs that a set of or a population of TCLs can provide. In such cases, it is important to de-synchronize/randomize the operation times of these switching devices, as well as induce sequential transition in order to avoid any damage to the device. One of the ways of solving the latter issue is by sequentially changing the set point temperature. A sequential transitioning safe protocol has been developed in [92, 93] to minimize the unwanted power oscillations. In [92], a variable deadband strategy has been employed until the temperature hits one of the transition points to provide power regulation. [92] uses similar strategy by injecting a delay of $M$-minutes and thereby achieving required objectives. [94], [95] deploys state stacking technique based on priority measures, which are defined as the distance from the switching boundaries to reduction in aggregate power. Briefly, the TCLs belonging to the ON stack nearing the switching boundary turn ON first, for increase in aggregate power; similarly, TCLs nearest to the OFF stack are signalled to turn OFF first and so on. A phase response curve (PRC) based technique has been proposed in [49] that helps control an ensemble of TCLs using an exogenous control input to manoeuvre OFF times of switching signals. This in turn modulates duty cycles as well as frequencies and an additional delay/advance parameter helps achieve system objectives. Although, it has been observed that it impacts user experience by changing effective set point temperatures. The major highlight of the work being, a one-to-one analogy of switching dynamics of TCLs to oscillator dynamics, a phase dependent model realized on similar lines helps achieve desired objectives. A continuum model is proposed in [96] to understand, analyze and mitigate potential risks of a decentralized response provider using synchronization statistics of TCLs. Again, a common observation to the stated literature being involvement of high computation cost and challenges for practical implementation, an effective solution to incorporate these issues would be advantageous.

The work proposed in this study, primarily focuses on the standard idea of controlling OFF times of TCLs using a novel Kuramoto phase oscillator model in Boolean form. It has been shown, how a simple phase delay/advance parameter can be used to achieve required system objectives, i.e., reduction in aggregate power consumed in reference to utility defined load requirements thereby maintain user experience. A novel coupled framework of TCLs has been developed with synchronization dynamics employing a bottom-up approach against majority of the work in the literature. The main motivation behind articulating such framework is to introduce multi-agent interactions amongst TCLs thereby regulate total power consumed even for cases when utility demand signal fails to communicate. To the best of my knowledge, the proposed model is one of its kind and has the potential to bring in various interesting phenomena that are usually studied using Kuramoto framework. It has been observed that, the computational cost reduces by a significant amount compared to [49] for the Kuramoto models have a pre-defined PRC. On the other hand, the proposed model employs basic tools like fast Fourier transforms (FFT) to calculate/incorporate switching statistics in real-time. It must be noted though that, unlike Kuramoto models normally employed for coupled synchronization of oscillators, here I make use of controlled de-synchronization using time-delays. Also, in order to allow practical implementation of the proposed idea the model is realized using Boolean fundamentals.

In succeeding sections, I extend the phase oscillator framework to distributed averaging model. An extensive study on the possibilities of practical implementation of the phase oscillator model has been discussed. On the basis of findings, a distributed averaging protocol to achieve similar (and in a better way) objectives as obtained in Kuramoto frame has been proposed. Further, the model outcomes have been extended to understand the implications on the central power grid. Reduction in oscillations imposed by switching loads on the central power grid has been quantified using computer simulations. It has been shown, how phase manipulations can be used to advantage thereby stabilize power system oscillations. To support the idea, a hardware-based practical implementation using analogous switching device has been performed. Hardware-based outcomes in conjunction to computer simulations provide validity to the proposed idea. It has been shown, how controlled de-synchronization can help achieve load following as well as help reduce power fluctuations on the central power grid.

The chapter has been organized as follows. In section 3.1 I sets up preliminaries in order to provide a platform to the proposed contributions. Section 3.2 explains and proposes a novel mathematical model using preliminaries developed in section 3.1. Section 3.2, provides the underlying motivation using the idea of synchronizing pendulums. In section 3.3, the proposed idea has been validated using parameters available in the literature. Next, in section 3.4, a distributed protocol is discussed, whereby a novel model for control of TCLs has been proposed. In section 3.5, hardware outcomes using LEDs as acting TCLs which simulate similar characteristics has been shown, thereby validating the proposed idea. A competitive benchmark is provided in section 3.6 where the proposed model is compared against those available in the literature and provide its competencies using peer-to-peer reviews.

## 3.1 Preliminaries

### 3.1.1 Mathematical Model of TCLs

Consider the behavior of an air-conditioning (AC) unit, wherein a thermostat-micro-controller combination controls switching action of the entire unit. The TCL switches ON when the temperature of a closed system (i.e., a system that doesn't allow mass transfer but permits exchange of energy in the form of heat) attains the higher threshold limit (predefined by the TCL unit) and turns OFF when it hits the lower temperature threshold. The internal temperature $T(t)$ of such closed system with an AC unit can be modelled mathematically integrating thermostat parameters and the switching characteristics $(s(t) \in \{0, 1\})$. A hybrid state model is widely studied in literature, in order to understand/analyze TCL dynamics [97] and can be stated as follows,

$$\dot{T}(t) = -\frac{1}{RC}[T(t) - T_a + s(t)PR]$$

$$s(t) = \begin{cases} 0 & \text{if } T(t) < T_{min} \\ 1 & \text{if } T(t) > T_{max} \\ s(t) & \text{otherwise,} \end{cases} \qquad (3.1)$$

where $T_a$ is the ambient temperature and $P$ is the average TCL ON state energy transfer rate; $C$ is the thermal capacitance and $R$ is the thermal resistance of the closed system under consideration. The minimum and maximum threshold temperature of the TCL are $T_{min} = T_s - \delta/2$ and $T_{max} = T_s + \delta/2$, respectively. $T_s$ is the thermostat set point temperature and $\delta$ represents the dead band. The total power aggregate can be computed using sum of power consumed by individual TCLs multiplied by their respective performance ratios. The parameters listed in Table 3.1 [49] are used to replicate the system dynamics of a TCL (and further for all succeeding simulations in this work).

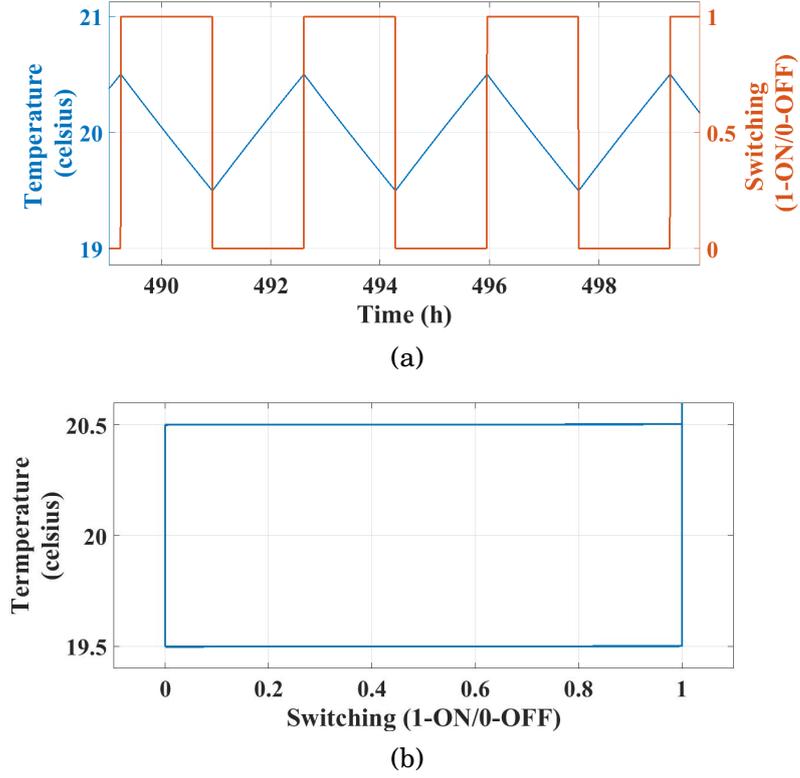

Figure 3.1: Dynamics of TCL using hybrid state model. (a) Behavior of internal temperature $T$ and switching signal $s$ plotted against time. (b) Phase portrait of TCL; internal temperature $T$ plotted against switching signal $s$.

As observed from Figure 3.1, TCL switching is oscillatory with its upper and lower temperature threshold limits set by the operator within the designed dead band. $R$ and $C$ can be seen as design parameters that define how fast or slow these extreme thresholds can be achieved. The effective temperature drop ($PR$) as well as thermal conductivity parameters ($R,C$) allow tuning of the duty cycle; for instance, $[2T_a - (T_{max} + T_{min})] = PR \longrightarrow 50\%$ duty cycle and so on.

Table 3.1: List of Parameters

| Parameter | Meaning | Value | Unit |
|:---:|:---:|:---:|:---:|
| $T_a$ | ambient temperature | 32 | $°C$ |
| $\delta$ | thermostat deadband | 0.5 | $°C$ |
| R | thermal resistance | 2 | $°C/kW$ |
| C | thermal capacitance | 10 | $kWh/°C$ |
| P | energy transfer rate | 14 | kW |
| K | coupling strength | 0.267 | MHz |

### 3.1.2 Kuramoto based Phase Oscillator Model for a System of Coupled TCLs

Kuramoto models [98] are widely studied in order to understand the synchronization dynamics of a system of '$N$' oscillators coupled by a link. It can be noted though, that the TCL switching signals are binary (i.e., $s(t) \in [1,0]$) in contrast to the Kuramoto model coupled by interaction terms of continuous type. Thus, I make use of some mathematical manipulations to achieve a Kuramoto model with digital interaction terms. An analogous phase oscillator model of Kuramoto type to understand the effective system dynamics of a coupled population of TCLs has been derived.

A standard first order Kuramoto equation of $N$ coupled oscillators can be given as follows,

$$\dot{\phi}_i = \omega_i + \frac{\epsilon}{N} \sum_{j=1, j \neq i}^{N} \left(sin(\phi_j - \phi_i)\right), \tag{3.2}$$

where $\phi_i$ is the phase of the $i$-th oscillator in radians, $\omega_i$ is the natural frequency and $\epsilon$ is a matrix of coupling constants. The overall interconnections of a system of coupled TCLs can be shown as in Figure 3.2, where a temperature sensor (or a transducer) measures $(\phi_i, s_i)$ the switching parameters of respective units. These switching parameters are then fed back as interaction terms in (3.2) and helps achieve synchronization via information transfer. For TCLs, these interacting terms must first be converted into its digital equivalents due to its inherent characteristics. Making use of the fact that $s_i(t) = f(\phi_i(t))$, switching signal $s_i(t)$ can be mapped to Heaviside function through oscillatory dynamics $sin(\phi_i(t))$, i.e., $sin(\phi_i(t)) \in [-1,1] \longrightarrow \Theta[sin(\phi_i(t))] \in [0,1]$. Hence, by using Heaviside map and a standard continuous-time distributed averaging 'consensus' type protocol [99] $\left(\dot{\phi}_i = \sum_{j=1, j\neq i}^{N} (s_j - s_i) = \sum_{j=1, j\neq i}^{N} [\Theta[sin(\phi_j)] - \Theta[sin(\phi_i)]]\right)$, (3.2) can be rewritten as,

$$\dot{\phi}_i = \omega_i + K \sum_{j=1, j \neq i}^{N} \left[\Theta\left[sin(\phi_j)\right] - \Theta\left[sin(\phi_i)\right]\right], \tag{3.3}$$

where $K = \epsilon/N$ and the Heaviside function $\Theta[sin(\cdot)]$ is given by,

$$\Theta[sin(\cdot)] = \begin{cases} 0 & \text{if } sin(\cdot) < 0 \\ 1 & \text{if } sin(\cdot) \geq 0. \end{cases} \tag{3.4}$$

In addition, a modulus function makes sure that the interaction terms are always positive (reasoned out in section 3.1.3). Thus, (3.3) can be modified as under,

$$\dot{\phi}_i = \omega_i + K \sum_{j=1, j \neq i}^{N} \left|\Theta\left[sin(\phi_j)\right] - \Theta\left[sin(\phi_i)\right]\right|, \tag{3.5}$$

which can be considered as a binary Kuramoto model for phase (switching) synchronization of $N$ coupled digital oscillators. Further, pertaining to the conventional Kuramoto dynamics the TCL's switching would converge in phase as per (3.5). In order to induce separation and achieve effective controlled de-synchronization an additional term $\alpha_{ij}$ being the time

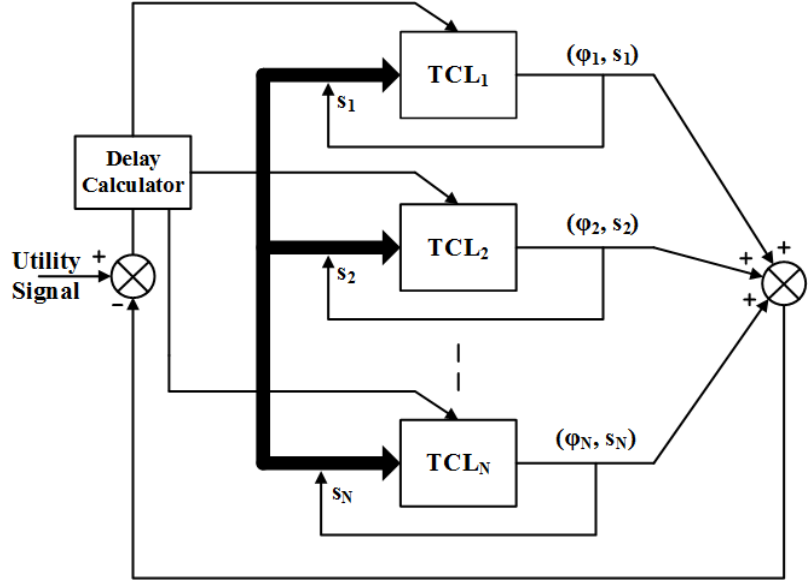

Figure 3.2: A system of coupled TCLs driven by a negative feedback in order to track utility defined signal.

delay/advance signal can be introduced. This can be seen in (3.6) below,

$$\dot{\phi}_i = \omega_i + K \sum_{j=1, j \neq i}^{N} \left| \Theta\left[sin(\phi_j)\right] - \Theta\left[sin(\phi_i + \alpha_{ij})\right] \right|. \qquad (3.6)$$

Variation in the time delay $\alpha_{ij}$ can be a controlling parameter used to achieve required power aggregate defined by the utility in real-time. As shown in Figure 3.2, utility operator defines desired demand reference, which can be subtracted from the aggregated power of a system of TCLs forming a closed loop feedback. $\alpha_{ij}$ can be varied accordingly, in order to minimize the tracking error, thereby achieving required power aggregate. A delay calculator (DelayC, defined in section 3.3.5), computes effective time delay $\alpha_{ij}$ to be induced in reference to required aggregate power consumption. Further, it can be easily shown that $\dot{\phi} = \omega$ dynamics work for controlling independent TCLs too.

### 3.1.3 Boolean Phase Oscillators

Next, I explain how a system of coupled phase oscillators derived in (3.6) can be practically implemented. The binary character of the switching signal $s_i(t) \in [0, 1]$ can be compared to a Boolean response (i.e., [TRUE,FALSE] ~ [0, 1]). Using a system of logic gates a Boolean phase oscillator model can be realized as shown in Figure 3.3 [50]. In Figure 3.3, delay lines are represented by rectangles, multiplexers are trapezoids, inverters are denoted by triangles and XOR gates are shapes with ⊕ symbol. Using an analog to digital conversion module, the physical interactions of TCLs can be brought into digital environment by interfacing the measurements to a digital signal processing (DSP) or a field programmable gate array (FPGA) board, where the dynamics of the phase oscillator model can be programmed. The

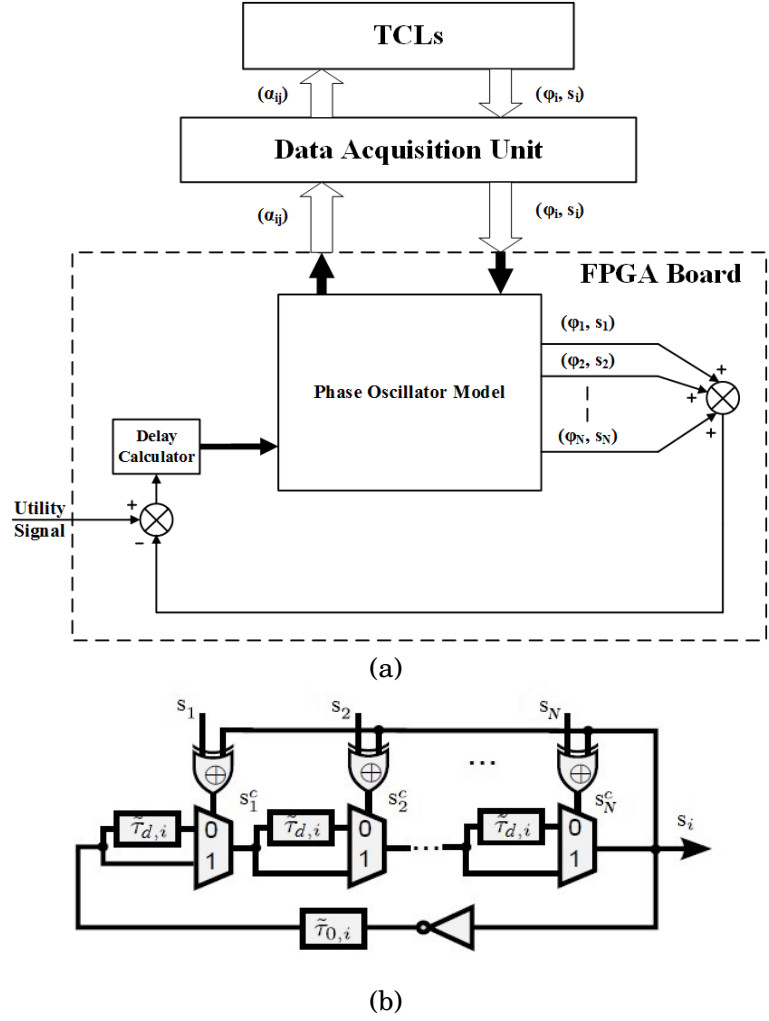

Figure 3.3: Block diagram representation of the phase oscillator model using Boolean fundamentals. (a) Closed-loop control of TCLs represented as a cyber-physical system. (b) Logic circuit diagram using XOR gates.

phase oscillator model calculates effective delay $\alpha_{ij}$ using DelayC and provides it to the TCLs using digital to analog conversions at the output of DSP/FPGA board. It can be noted that, right hand side of (3.3) can achieve both positive as well as negative values. Thus, in order to map outputs from (3.3) to a binary set and to maintain a practically feasible framework, an XOR type behavior can be justified. Hence, an addition of modulus in (3.6) helps achieve above mentioned objectives (i.e., $|a-b| \in [1,0]; \forall a, b \in [1,0]$). Additionally, a system of the form (3.6) can be realized by a set of XOR gates by choosing, $K = 2\tilde{\tau}_{d,i}\left(\frac{4\pi^2}{2\tau_{0,i}}\right)$ and $\tau_{0,i} = \tilde{\tau}_{0,i} + N\tilde{\tau}_{d,i}$, and can be rewritten in Boolean form as follows,

$$\dot{\phi}_i = \omega_i + K \sum_{j=1, j\neq i}^{N} s_j(\phi_j) \oplus s_i(\phi_i + \alpha_{ij}), \tag{3.7}$$

where $\oplus : \{0,1\} \times \{0,1\} \to \{0,1\}$ denotes the Boolean XOR function (using the analogy, $a \oplus b = |a-b|$), $\tau_{0,i}$ is the induced time lag/advance by the digital delays. Thus, it can be inferred

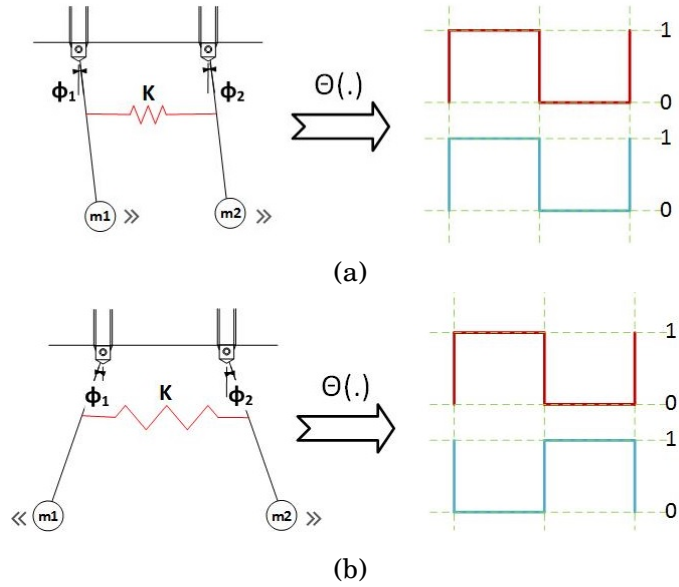

Figure 3.4: Pendulums coupled by a spring, oscillating in two equilibrium modes. (a) Oscillations in 'in-phase' mode. (b) Oscillations in 'anti-phase' mode.

that the proposed idea (3.6) can be implemented easily using Boolean fundamentals.

### 3.1.4 Coupled Pendulums

In order to understand the various modes of oscillations that emerge in Kuramoto type phase oscillator framework proposed in previous section and map them to the equivalent behavior seen in TCLs, I provide an analogy to pendulum behavior in this section. Consider two pendulums coupled with a spring as shown in Figure 3.4. For the sake of simplicity, assume both the pendulums to be identical and dissipative losses to be zero. The oscillations of these pendulums can be written using Kuramoto model and further can be converted into its equivalent boolean dynamics using Heaviside function as in (3.6). It can be easily noted that the Boolean dynamics obtained using (3.6) resemble to the switching characteristics $s(t)$ from (3.1) and can be adjusted to match its statistical parameters. Thus, for further proposal I would use TCLs and coupled pendulums interchangeably in this section.

Now as seen from Figure 3.4, pendulums starting from any initial conditions settle down in either of the two modes of oscillations at steady state, i.e., Figure 3.4(a)-'in-phase' or Figure 3.4(b)-'anti-phase' synchronization. Specifically, 'in-phase' oscillations are those in which difference between the phases is 0 radians and 'anti-phase' with phase difference of $\pi$ radians. While converting the phase dynamics into its equivalent Boolean dynamics, the cumulative sum of the amplitudes in the '$T_{ON}$' time is higher for 'in-phase' type synchronization than for 'anti-phase' synchronization.

For a system of TCLs aggregate power at any given time must be a combination of 'anti-phase' as well as 'in-phase' type synchronization. Hence, a phase lag of $\pi$ radians is evident and can be added to either of the oscillator dynamics in (3.6) or by adding $\pi/2$ radians in both the system dynamics. It has been additionally observed that for a system of $N$ agents, an equivalent of $\pi/N$ radians can be added to each TCL dynamics to achieve 'anti-phase' type synchronization.

## 3.2 Control of TCLs using Phase Oscillator Model

Preliminaries discussed in section 3.1 can be used to modify overall system dynamics of TCLs from individual to a coupled architecture using Kuramoto form. In order to better understand, I start by explaining the phase oscillator model from a multi-agent perspective in (3.6); i.e. 'Ensemble TCL' model. Later, these dynamics are extended for a 'Single TCL' model, nullifying the synchronization dynamics.

### 3.2.1 Ensemble TCL model

Integrating the phase oscillator dynamics in (3.6) with hybrid model TCL temperature dynamics in (3.1); the overall behavior in the coupled framework for a system of ensemble of TCLs can be written as follows,

$$
\begin{aligned}
\dot{T}_i &= -\frac{1}{[RC]_i}[T_i(t) - T_a + s_i(t)[PR]_i], \\
\dot{\phi}_i &= \omega_i + K \sum_{j=1, j \neq i}^{N} \left| \Theta\left[sin(\phi_j)\right] - \Theta\left[sin(\phi_i + \alpha_{ij})\right] \right|, \\
s_i(t) &= \Theta\left[sin(\phi_i) - s_{i,0}\right], \\
s_{i,0} &= sin\left[\frac{\pi - T_{ON,i}}{2}\right],
\end{aligned}
\tag{3.8}
$$

where $\omega_i$ is the switching frequency, $T_{ON,i}$ is the ON time of the $i$-th switching signal, $s_{i,0}$ is the bias to the Heaviside function used to imitate TCL characteristics in Kuramoto framework (shown in Figure 3.5). Bias $s_{i,0}$ can also serve as a parameter for maintaining heterogeneity in the operating set (and hence TCL behaviors) of the Heaviside function (Remark 3.1). For instance, as shown in Figure 3.5 scaling up or down of 'phase ON' time (i.e., angle equivalent to '$T_{ON,i}$', in radians) defines the overall duty cycle of the switching signal.

**Remark 3.1.** *It must be noted that the bias $s_{i,0}$ added in* (3.8) *helps retain duty cycle of the TCL and must not be mistaken for control.*

Conclusively, controlled de-synchronization of a set of '$N$' coupled TCLs in Kuramoto form can be achieved using the phase lag parameter $\alpha_{ij} \in \mathbb{R}; \alpha_{ij} \in [-\pi, \pi]$radians. Additionally, it is observed that an equidistant phase lag on a circle $\alpha_{ij} = \alpha_{ji} = 2\pi/N$, minimizes aggregate power consumed. Thus, any in between power demands can be attained using relevant time delay adjustments.

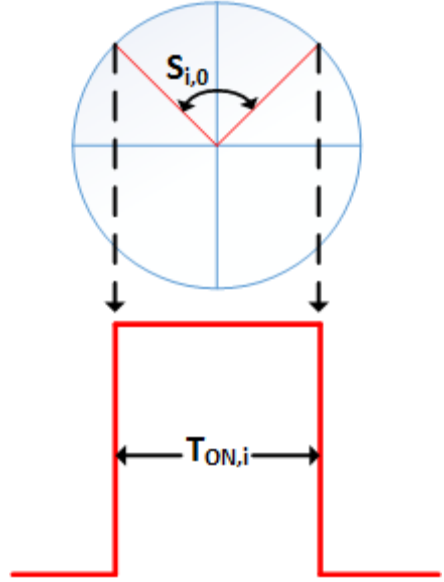

Figure 3.5: Controlling the thresholds of the Heaviside function using the bias $s_{i,0}$, thus adjusting the duty cycle.

### 3.2.2 Single TCL model

An individual TCL doesn't require synchronization dynamics and hence I modify 'Ensemble TCL' model by substituting $N = 1$ in (3.8), i.e., nullifying the synchronizing dynamics and re-writing the model as follows,

$$\begin{aligned}
\dot{T} &= -\frac{1}{[RC]}[T(t) - T_a + s(t)[PR]], \\
\dot{\phi} &= \omega, \\
s(t) &= \Theta\left[sin(\phi) - s_0\right], \\
s_0 &= sin\left[\frac{\pi - T_{ON}}{2}\right].
\end{aligned} \tag{3.9}$$

It is evident that the model shown in (3.9) can be derived from (3.6) by equating synchronization dynamics to 0 and its frequency is governed by $\omega$, i.e., the natural frequency of the TCL switching signal oscillations.

**Remark 3.2.** *It must be noted that, the information about the set point temperatures ($T_s$) is carried by switching frequencies ($\omega_i$) and dead band ($\delta$) of TCLs and a relation can be easily derived using* (3.1) *[49].*

## 3.3 Test Cases - Phase Oscillator Model

In this subsection, I use the system parameters mentioned in Table 3.1 to present a proof of concept for the phase oscillator model proposed in previous sections. With the models already proposed in section 3.2, I simulate the phase oscillator model of a single TCL thereby comparing the results with those obtained in Figure 3.1. Then, an ensemble TCL model of '$N = 4$' is considered for a minimum aggregate power consumption case and the results are verified against literature. Additionally, in order to verify that the user experience is not altered, a quick FFT analysis carried out on switching signals obtained from (3.6) against those attained using hybrid model in (3.1) provide validity to the proposed model.

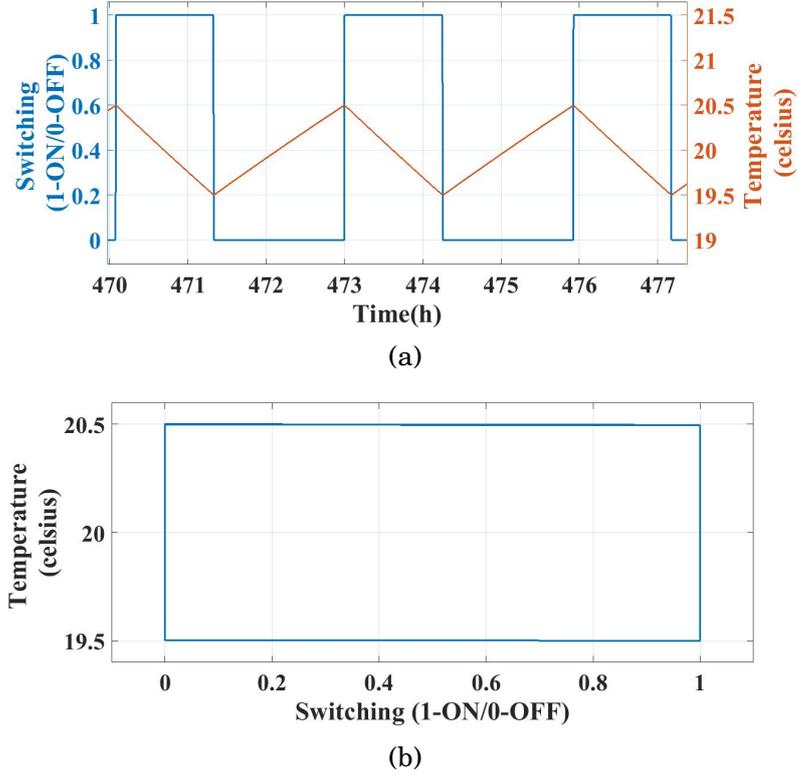

Figure 3.6: Dynamics of a single TCL. (a) Behavior of internal temperature $T$ and switching signal $s$ plotted against time. (b) Phase portrait of TCL; internal temperature $T$ plotted against switching signal $s$.

### 3.3.1 Single TCL

The system dynamics for single TCL proposed in (3.9) and data provided in Table 3.1 with $\omega$ = 2.15 rad/h (0.342 Hz) are used. Randomly, a duty cycle of 43% is chosen and the results obtained in Figure 3.6, replicate hybrid state dynamics shown in Figure 3.1. Thus, verifying the behaviour of an individual TCL.

### 3.3.2 Ensemble TCLs

Next, an ensemble model of 4 TCLs (i.e., '$N = 4$') has been considered. The results are verified for 50% and 43% duty cycles by tuning the bias $s_{i,0}$. Using $\alpha_{ij} = \alpha_{ji} = \alpha = (2\pi/N)$, i.e., $\alpha = [0; \pi/2; -\pi/2; -\pi/2]$ or $\alpha = [0; \pi/2; 3\pi/2; 2\pi]$ (distributed along its dimension) and ensemble model proposed in (3.8) following results are obtained.

#### 3.3.2.1 Homogeneous Case with 50% Duty cycle

Here a homogeneous set of TCLs with a duty cycle of 50% defined by the TCL design parameters is considered. A power transfer rate of $P = 12$KW and $\omega_i = 0.27$ rad/h (0.043 Hz) modulates overall duty cycle to 50%, keeping all the other parameters constant. As shown in Figure 3.7 the TCLs de-synchronize gradually with effective steady state phase lag given by $\alpha_{ij} = (2\pi/N)$. From Figure 3.7, aggregate power consumed by the TCLs (with $\eta = 1$, i.e., performance ratio chosen to be 100% for simplicity) appears to be minimum. A time series plot of $P_{agg} = \sum_{j=1}^{N} P_j, P_j = P\eta s_j$; at every iteration till a steady state phase difference is achieved is observed. On the other hand, a circle plot of $\phi_i(t)$ at steady state shown in Figure 3.7 can be used to visualize effective angular separation between TCLs.

#### 3.3.2.2 Homogeneous Case with 43% Duty cycle

Here a homogeneous set of TCLs with a duty cycle of 43% defined by the TCL design parameters is considered. A power transfer rate of $P = 14$KW and $\omega_i = 0.55$ rad/h (0.088 Hz) modulates overall duty cycle to 43%, keeping all the other parameters constant. As shown in Figure 3.8 the TCLs de-synchronize gradually with effective steady state phase lag given by $\alpha_{ij} = (2\pi/N)$. From Figure 3.8, aggregate power consumed by the TCLs (with $\eta = 1$, i.e., performance ratio chosen to be 100% for simplicity) appears to be minimum. A time series plot of $P_{agg} = \sum_{j=1}^{N} P_j, P_j = P\eta s_j$; at every iteration till a steady state phase difference is achieved is observed. On the other hand, a circle plot of $\phi_i(t)$ at steady state shown in Figure 3.8 can be used to visualize effective angular separation between TCLs. It is worth noticing that the bias $s_{i,0}$ helps retain duty cycle of regenerated switching signal to match with the one obtained from (3.1).

Further, to verify that the proposed model doesn't hamper user experience as well as retains its hybrid state model behavior, a quick frequency analysis is performed. A standard FFT of two switching signals: one obtained from (3.1) and other from (3.8) is computed. Figure 3.9 shows frequency spectrum of switching signal for one of the TCLs from the set of $N = 4$ against that obtained from the hybrid model. This provides validity to the proposed Kuramoto type model to achieve controlled de-synchronization thereby maintaining original system behavior.

### 3.3.3 Heterogeneous Ensemble TCLs

Parameters mentioned in Table 3.1 are subjected to variations dependent on factors like type of occupancy, separation medium, retention of closed system characteristics, etc. Thus, it is necessary to check the effectiveness of the proposed model to unknown uncertainties or known heterogeneities. For instance, all TCLs might not be of same make, construction or design and can't be considered homogeneous. Again, TCLs installed at public places (like an office or an institute) might perturb thermal conductivity of a closed system and happens to be a heterogeneous scenario. Heterogeneity can be captured through either natural frequency, thermal resistances, capacitances, etc., which are again inter-related. As seen from [49],

$$\omega_i = 2\pi / \left[ R_i C_i \ln\left( \frac{(T_a - T_{min})(T_{max} - T_a + PR_i)}{(T_a - T_{max})(T_{min} - T_a + PR_i)} \right) \right] \quad (3.10)$$

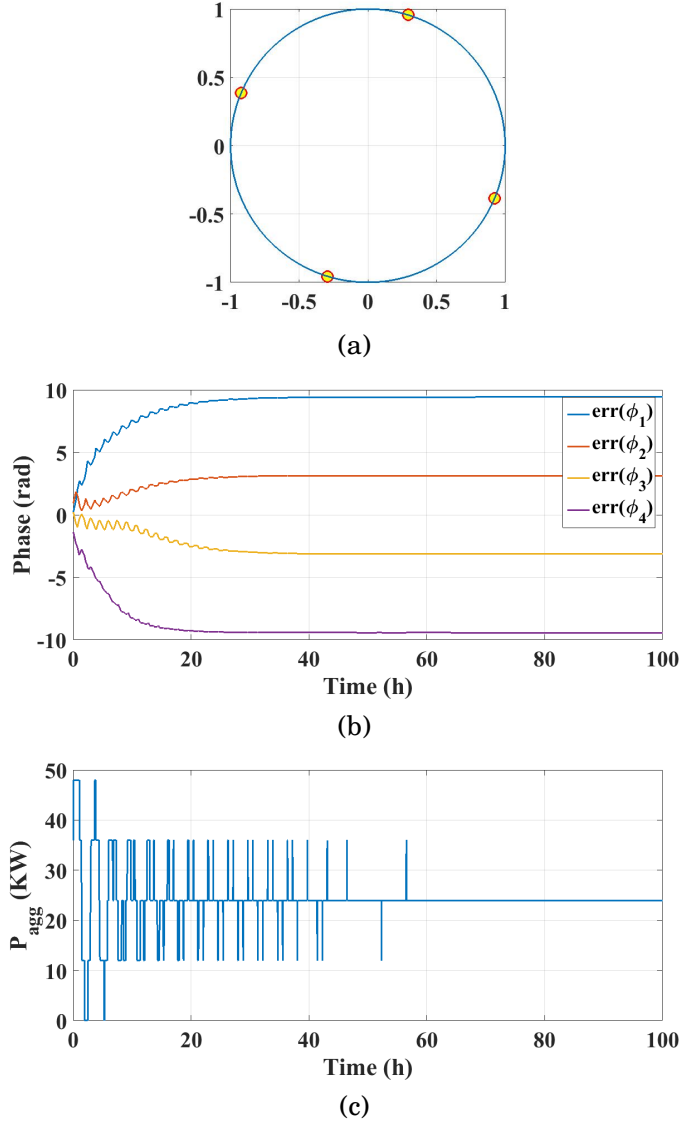

Figure 3.7: Dynamics of an ensemble of TCLs. (duty cycle = 50%) (a) Circle plot of a system of coupled TCLs showing relative angular separations. (b) Time series plot of cumulative relative angular separations symbolized as error induced due to time delay parameter, given by $err(\phi_i(t)) = \sum_{j=1, j\neq i}^{N} (\phi_j(t) - \phi_i(t))$. (c) $P_{agg}$ plot depicting effective aggregate power consumed against time.

To analyze effects of heterogeneity, I consider varying natural frequencies in (3.8) within 5% of its nominal value with a duty cycle of 50%. It can been observed (see Figure 3.10) that phase oscillator model for a heterogeneous case starts behaving as homogeneous set at steady state. This is evident due to inherent synchronization dynamics in Kuramoto equations and can be interpreted as a thermostatic equilibrium when the system parameters

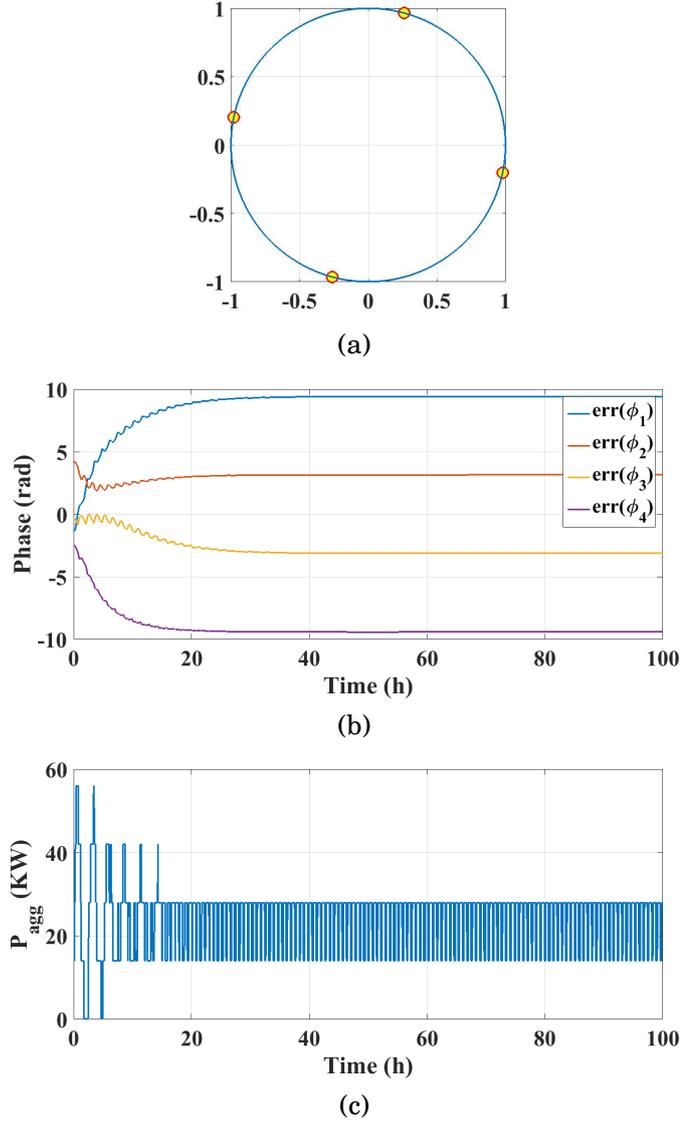

Figure 3.8: Dynamics of an ensemble of TCLs. (duty cycle = 43%) (a) Circle plot of a system of coupled TCLs showing relative angular separations. (b) Time series plot of cumulative relative angular separations symbolized as error induced due to time delay parameter, given by $err(\phi_i(t)) = \sum_{j=1, j \neq i}^{N} (\phi_j(t) - \phi_i(t))$. (c) $P_{agg}$ plot depicting effective aggregate power consumed against time.

synchronize.

### 3.3.4 A large Congregation of TCLs

Further, in a huge infrastructure number of TCLs would increase upto $N > 1000$ or $N > 10000$ or even higher. Hence, in this section I check the model behavior for a population of TCLs by modifying $N = 4$ to $N = 100$. As a matter of simplicity, it is assumed that the utility demands minimization of TCL loads while all the other system parameters are kept constant. A common observation to settle down to a minimum aggregate power of $P(N/2)$ KW is observed.

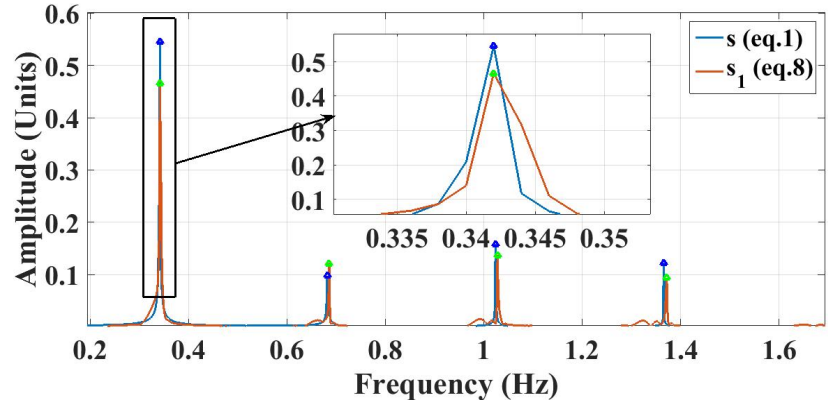

Figure 3.9: FFT of $s(t)$ from (3.1) versus (3.8).

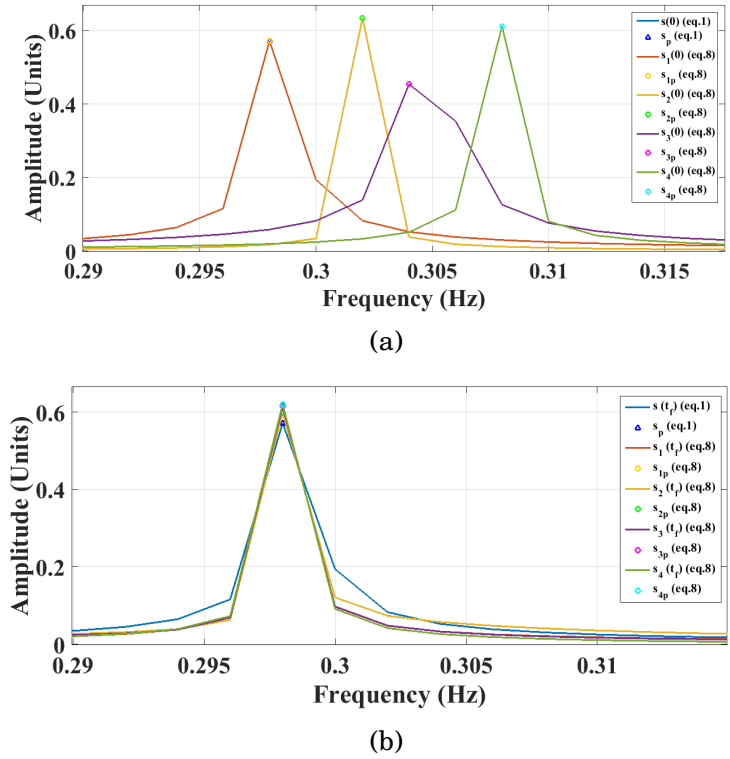

(a)

(b)

Figure 3.10: Frequency analysis of a system of heterogeneous TCLs in phase oscillator form. (a) FFT of $s(t)$ from (3.1) versus (3.8) at $t = 0$. (b) FFT of $s(t)$ from (3.1) versus (3.8) at $t = t_f$.

The TCLs either form clusters of equally distributed oscillators across a phase difference of $\pi$ radians or distribute themselves across the circle, thereby minimizing aggregate power consumed. This can be seen from Figure 3.11 where aggregate power is plotted against time.

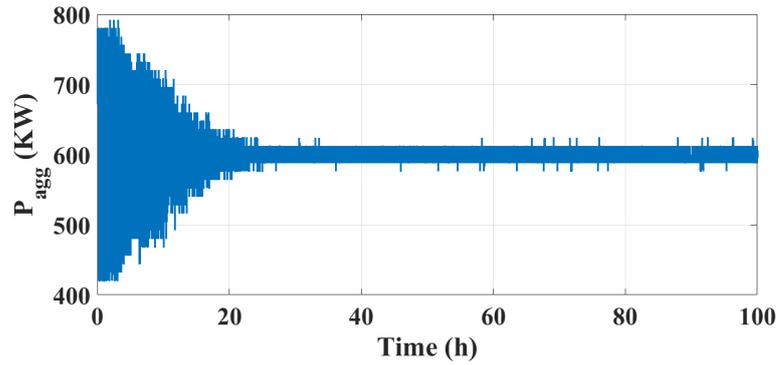

Figure 3.11: Effective aggregate power consumed by a population of TCLs plotted against time.

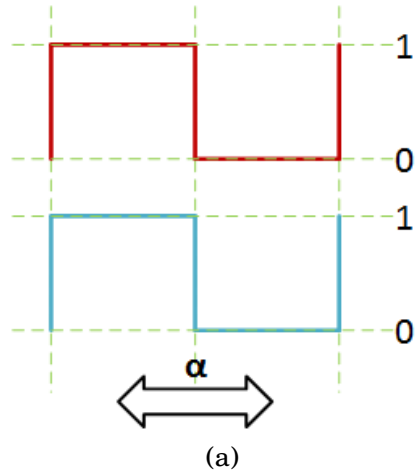

(a)

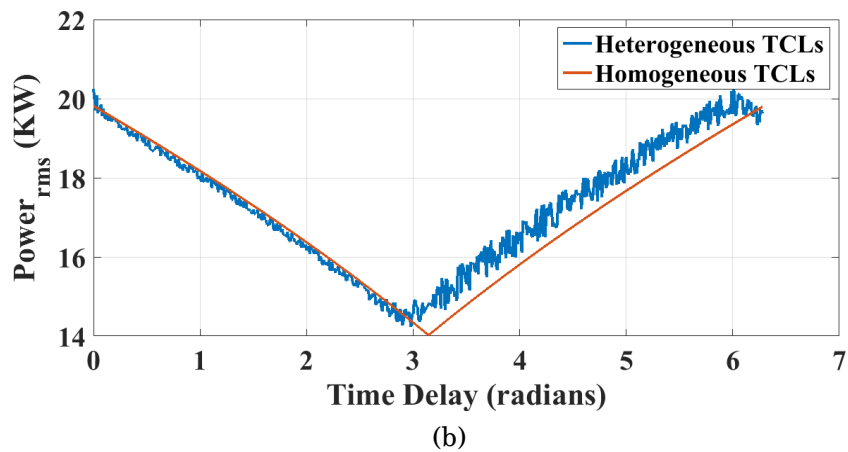

(b)

Figure 3.12: Delay calculator characteristics to map power reference set by the utility with delay $\alpha$ radians. (a) A pair of switching signals driving the TCLs delayed/advanced by $\alpha$ radians. (b) Computation of effective rms value of aggregate power consumed against $\alpha$ radians.

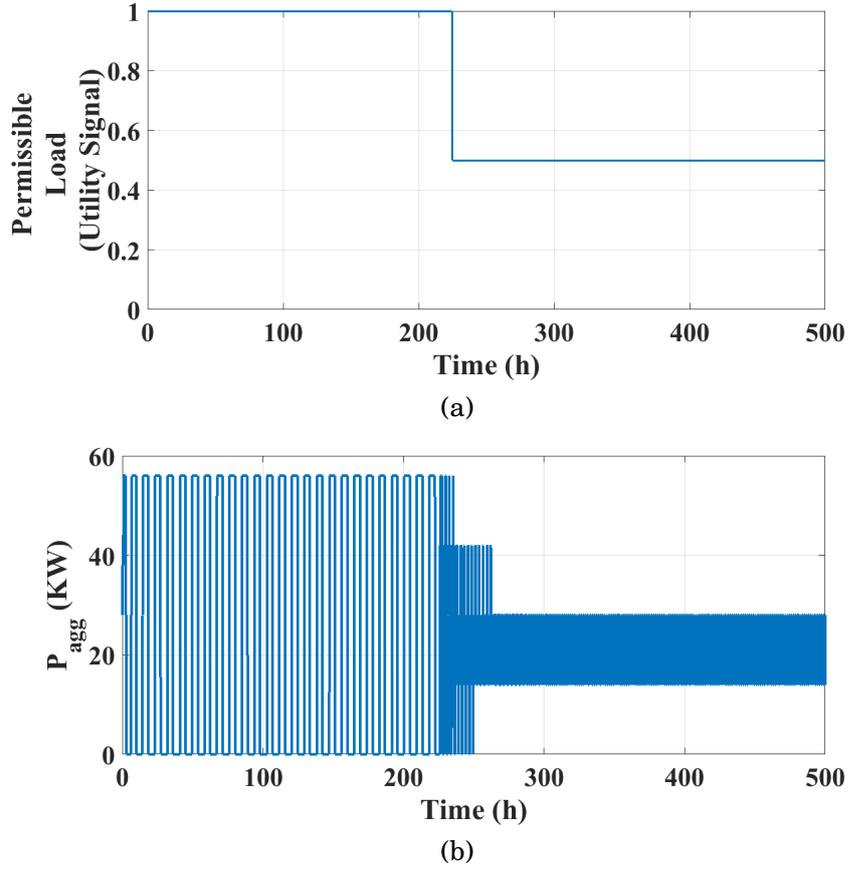

Figure 3.13: Load following scenario for an ensemble of TCLs driven by a utility defined signal. (a) Utility reference signal switching from 100% to 50% thereby demanding reduction in aggregate power consumed. (b) Impact of inducing time delay/advance parameter $\alpha$ so as to reduce effective aggregate power consumed and track utility demands.

### 3.3.5 Load following of TCLs as an AS

In order to provide valuable AS to the central power grid, I check the feasibility of the proposed algorithm to allow load following through utility guided command signal. As mentioned in previous sections, a Delay Calculator (DelayC) provides corresponding delay/advance signal to the TCLs in order to coalesce with the required power aggregate. For simplicity, a small set comprising of two TCLs is considered and delay $\alpha$ is varied to compute effective power aggregate. To maintain consistency and ease of validation, I calculate root mean square (rms) of the resultant aggregate power and the characteristics of the map is as shown in Figure 3.12. A duty cycle of 43% and a frequency of $\left(\frac{1}{2\pi}\right)$ Hz are considered and $\alpha$ is varied in increments of 0.01 radians. For heterogeneous case the frequency is varied in the range $\left[\left(\frac{1}{2\pi}\right) \pm 5\%\right]$ Hz and process is repeated. DelayC calculations show that any aggregate power reference can be achieved by choosing corresponding delay parameter as in Figure 3.12 and provide utility defined power reference.

The feedback loop shown in Figure 3.2 is used to examine load following scenario, whereby the utility demand signal is changed from 100% to 50% and effects are noted.

Keeping the system parameters constant, the overall responses of the system can be seen in Figure 3.13. It is observed that, the aggregated power follows utility demanded signal thereby verifying the proposed idea. It is worth noticing that, the effective dead band reduces in width but the set point temperatures are left unaffected. This makes sure that the proposed control action doesn't alter the user experience, which is an added advantage in reference to the existing approaches in the literature. Alternatively, the same can be achieved using clustering phenomena in Kuramoto framework and phase delay can be tuned from 0 radians to $\pi$ radians thereby allowing $P_{agg}$ to transit from maximum to minimum. In an autonomous frame, i.e., in case the communication from utility is lost, an internal signal maintaining minimization of loads can be programmed so as to keep the grid loading to its minimum at any given time.

**Remark 3.3.** *Frequency calculated in previous sections are scaled down by hours, i.e. $\omega_i$ is calculated as $1/T$ where, $T$ is in hours. In succeeding sections of the chapter time is scaled back to seconds.*

### 3.3.6 Shortcomings of Boolean phase oscillator model

Although, the Boolean phase oscillator model provides motivating results, some shortcomings in the model can be listed as follows,

1. It must be noted that the natural frequency $\omega_i$ used in (3.8) can be easily calculated using mathematical manipulations as follows,

$$\omega_i = \omega_{FFT} - K \sum_{j=1, j\neq i}^{N} \left| \Theta\left[sin(\phi_j)\right] - \Theta\left[sin(\phi_i + \alpha_{ij})\right] \right| \tag{3.11}$$

   Although, it is observed that as the number of TCLs increase, value of $\omega_i$ needs correction. For instance, $N = 16$ requires $\omega_i = 8.81$ Hz and $N = 100$ needs $\omega_i = 62.18$ Hz, to achieve $\omega_{FFT} = 0.298$ Hz. It is observed that (3.11) behaves in a nonlinear way and requires an additional correction factor $e$. Also, since the system under consideration consists of mechanical devices such as compressors and switches, such high frequencies are undesirable. Additionally, as shown in [93] the TCLs are frequency sensitive devices and drastic changes in the frequency must be avoided to prolong the product life. In a practical scenario, where system parameters (e.g., $N, \alpha_{ij}$) change dynamically, computation of $\omega_i$ in real-time would seem difficult. Thus, rendering the Boolean phase oscillator model difficult to be practically realized.

2. $K$ is a parameter that is computed using dynamical equations governing the system parameters [100]. As shown in section 3.1.3, value of $K$ can be calculated from the induced time delays in a programmable frame. Although, the inter-dependence of $K$ and $\omega_i$ in (3.8), adds complexity to its accurate computation.

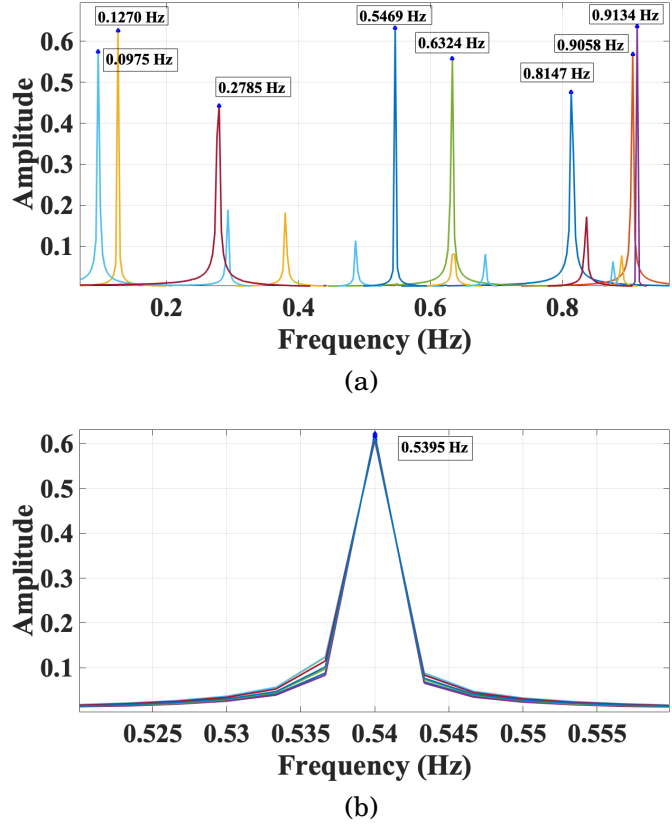

Figure 3.14: Frequency dynamics (a) FFT of switching signal at t = 0h. (b) FFT of switching signal at t = 15h.

## 3.4 Distributed Averaging Model for TCLs

Consider an undirected graph $G = \{V, E\}$ with set of nodes $N$ in $V$ and edge set $E$. Each edge of the graph $E \in G$, is an unordered pair of distinct nodes. A real scalar quantity $x_i(t)$ is associated with node $i$ at any given time $t$. The distributed averaging consensus algorithm computes the average $\frac{1}{N}\sum_{i=1}^{N} x_i(0)$ iteratively thus, facilitating a local communication link between the nodes. The nodes are updated continuously based on their and the states of all other neighbouring nodes. A widely accepted and used linear iterative algorithm can be stated as follows,

$$\dot{x}_i(t) = \sum_{j=1, j\neq i}^{N} W_{ij}\left(x_j(t) - x_i(t)\right) \tag{3.12}$$

for $i \in \{1, \ldots, n\}$ and $t \in \mathbb{R}^+$ and $W_{ij}$ being the weight associated with each edge $E_{ij}$. Since the nodes, in the presented case, are undirected, the weights associated with them are symmetric i.e., $W_{ij} = W_{ji}$.

$$W = w_{ij}\left(I_{N \times N} - diag(1_N)\right) \tag{3.13}$$

where $w_{ij}$ is the weight of edge $E_{ij}$ and $I_{N\times N}$ is identity matrix and $diag(1_N)$ is diagonal matrix. Thus, the matrix $W$ is the communication matrix that describes the strength of the link between two nodes. Extending this theory to system of TCLs, where the nodes are analogous to TCLs and the scalar quantity to the frequency of switching signal,

$$\dot{f}_i = W \sum_{j=1, j \neq i}^{N} (f_j - f_i) \qquad (3.14)$$

where $W$ is the tuning parameter of synchronization, left to the control designer. (3.14) synchronizes the frequency of the switching signal to their mean value as shown in Figure 3.14. Thus, (3.8) can be re-framed as follows,

$$\begin{aligned} \dot{T}_i &= -\frac{1}{[RC]_i}[T_i(t) - T_a + s_i(t)[PR]_i], \\ \dot{f}_i &= W \sum_{j=1, j \neq i}^{N} (f_j - f_i), \\ s_i(t) &= \Theta\left[sin(2\pi f_i t + \alpha_{ij}) - s_{i,0}\right], \\ s_{i,0} &= sin\left[\frac{\pi - T_{ON,i}}{2}\right], \end{aligned} \qquad (3.15)$$

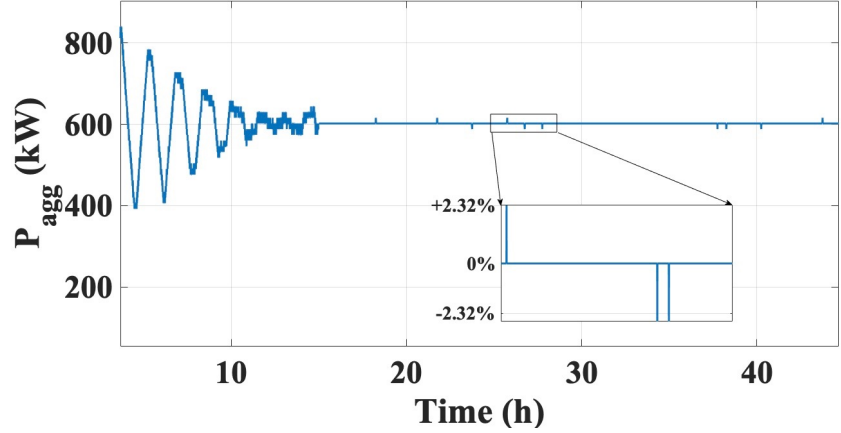

Figure 3.15: $P_{agg}$ vs time showing the aggregate power consumed at a given time for $N = 100$ heterogeneous TCLs.

It should be noted that unlike the Kuramoto based model where phase is used as the controlling parameter, (3.15) bifurcates the onus of controlling frequency as well as angular separation.

For the matter of comparison, a similar ensemble of heterogeneous TCLs, as in Figure 3.11 (natural frequencies in the same range as of ±5% of $\omega_i = 0.271$Hz), are taken. Additionally, the value of weight ($W$), in Figure 3.15, is taken as 0.06. By using (3.15), the system is de-synchronized and the results obtained are depicted by Figure 3.15. Similar results can be noted by comparing the two plots, however, as opposed to Kuramoto based model, in Figure 3.15, high frequency switching and computational hindrance have been avoided.

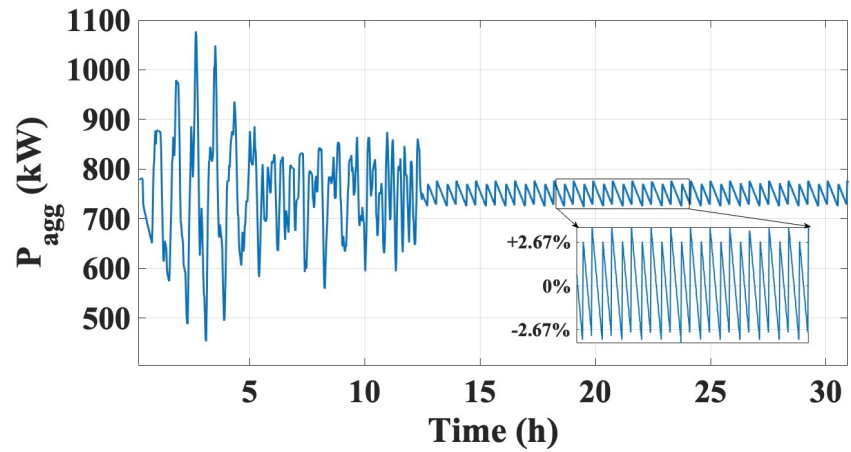

Figure 3.16: $P_{agg}$ vs time showing the aggregate power consumed at a given time for $N = 1000$ heterogeneous TCLs.

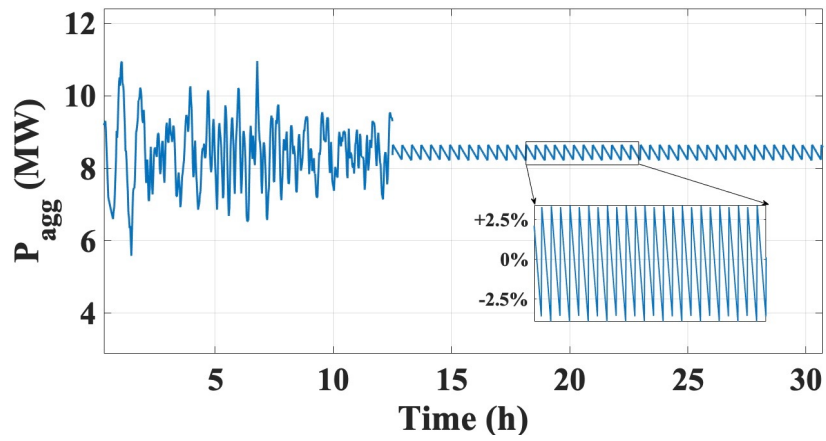

Figure 3.17: $P_{agg}$ vs time showing the aggregate power consumed at a given time for $N = 10000$ heterogeneous TCLs.

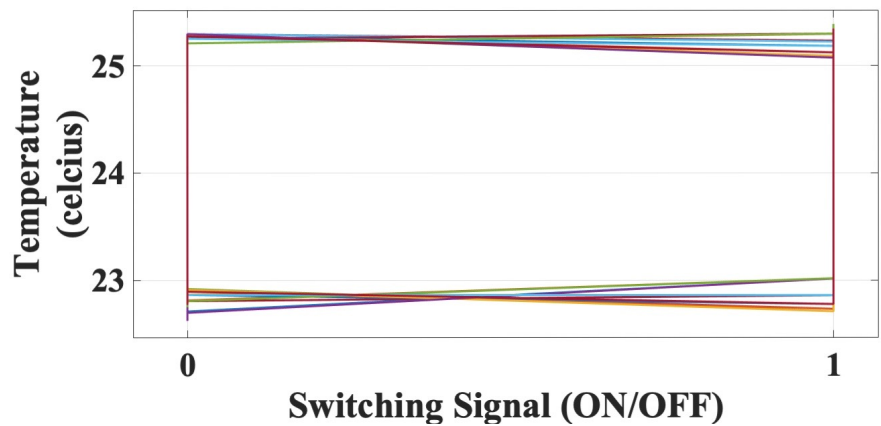

Figure 3.18: Phase portrait for $N = 10000$ TCLs.

## 3.5 Case study - Distributed Averaging Model

This section presents the application of the distributed averaging protocol in form of case study and later describes how it is practically implemented.

### 3.5.1 Software Simulation

To make the two cases put forth as realistic as possible, real time data of a TCL has been used to simulate the cases. For this, a TCL of a certain make was observed and the data listed in the Table 3.3 and Table 3.2 were collected. This TCL had a power rating of 1.66kW and capacity of 1.5tons.
Depending upon these values, following two cases have been chosen.

1. Case 1: $N = 1000$

   For $N = 1000$, heterogeneous TCLs are defined having a duty cycle in the range of 0.422 to 0.482, $\delta$ as 3.0°$C$ and $f$ between 0.0029Hz and 0.0033Hz respectively. The TCLs have been set to work at a set point temperature of 27°$C$.

   Following the distributed averaging model, TCLs settle at their respective phase difference, i.e., $\alpha_{ij} = \alpha = 2\pi/N$, which in this case is $\alpha = \pi/500$. As depicted in Figure 3.16, the power fluctuation for $N = 1000$ heterogeneous TCLs dampen over time and move towards steady state. The power at which the TCLs settle is 750.4472kW and fluctuations in the power are ±2.67%.

2. Case 2: $N = 10000$

   For $N = 10000$, heterogeneous TCLs are defined having a duty cycle in the range of 0.4812 to 0.5354, $\delta$ as 2.0°$C$ and $f$ between 0.0026Hz and 0.0036Hz respectively. The TCLs have been set to work at a set point temperature of 24°$C$.

   Following the distributed averaging model, TCLs settle at their respective phase difference, i.e., $\alpha_i = \alpha = 2\pi/N$, which in this case is $\alpha = \pi/5000$. As depicted in Figure 3.17, the power fluctuation for $N = 10000$ heterogeneous TCLs dampen over time and move towards steady state. The power at which the TCLs settle is 8.4453MW and fluctuations in the power are $\pm 2.5\%$. In Fig. 3.18, phase portraits of 100 TCLs chosen randomly from the population has been shown. It can be seen that the portrait dissimilarity is minimal although set point temperatures remain unaffected.

Table 3.2: List of Parameters (at set point temperature = 24°$C$)

| Parameter | Meaning | Value | Unit |
|---|---|---|---|
| $\omega$ | frequency of TCL | 0.0027 | Hz |
| $\delta$ | thermostat deadband | 2 | °$C$ |
| $d$ | duty cycle of TCL | 0.5083 | - |

Table 3.3: List of Parameters (at set point temperature = 27°$C$)

| Parameter | Meaning | Value | Unit |
|---|---|---|---|
| $\omega$ | frequency of TCL | 0.0031 | Hz |
| $\delta$ | thermostat deadband | 3 | °$C$ |
| $d$ | duty cycle of TCL | 0.452 | - |

### 3.5.2 Load Following as an AS

Next, I show how a load following case can be emulated providing a valuable AS to the utility. In [8], a function mapping RMS (root mean square) aggregate power of a population of TCLs with effective delay induced was proposed. Using the same idea, I choose different utility demands and corresponding delay $\alpha_{ij}$ is calculated. The effective power so obtained is normalized using total RMS power capacities; i.e.,

$$P_{norm}(\%) = \frac{(P_{rms,agg} - P_{rms,\alpha})}{P_{rms,agg}} \times 100 \tag{3.16}$$

where $P_{rms,agg}, P_{rms,\alpha}, P_{norm}$ are the rms of maximum aggregate power, rms of aggregate power at steady-state after application of $\alpha_{ij} = \alpha$, and normalized power of all the TCLs along a given time period respectively. These results can be seen from Fig. 3.19 where utility demands for random reduction/increase in the total allowable aggregate power and corresponding change in the delay induced causes controlled de-synchronization and hence required load following scenario is achieved. It can be inferred that the resultant aggregate power appears to be a significant AS to the centralized power grid.

### 3.5.3 Effects in a Power Grid

It should be noted that apart from control of power aggregation, the proposed algorithm benefits the grid by reducing the power system fluctuations. To compare/visualize the same, a small population of $N = 100$ TCLs is considered. In order to exhibit random behavior (depicting as-is characteristics without applying control) the frequency and phases of TCLs are chosen randomly from a distribution. Next, the same set of heterogeneous TCLs are acted upon by the distributed averaging protocol achieving effective de-synchronization. Both the cases are checked for the amount of peak to peak oscillations seen by the power grid. As shown in Fig. 3.20($a$), the system without application of distributed averaging strategy induce higher oscillations in the grid as compared to after application of the proposed scheme. In Fig. 3.20($b$) the quantum of reduction is observed to be $\sim 40\%$ of random case, showing a significant reduction where,

$$P_{red}(\%) = \frac{(P_{random} - P_{de-synchronized})}{P_{random}} \times 100 \tag{3.17}$$

Further, the results are extended for $N = 10000$ TCLs and similar analysis is performed. It can be observed, that the effective percentage reduction lowers to $\pm 2\%$ but is relatively comparable. Hence, it can be inferred that apart from providing load following objectives the proposed scheme can also help reduce amplitude of power system oscillations.

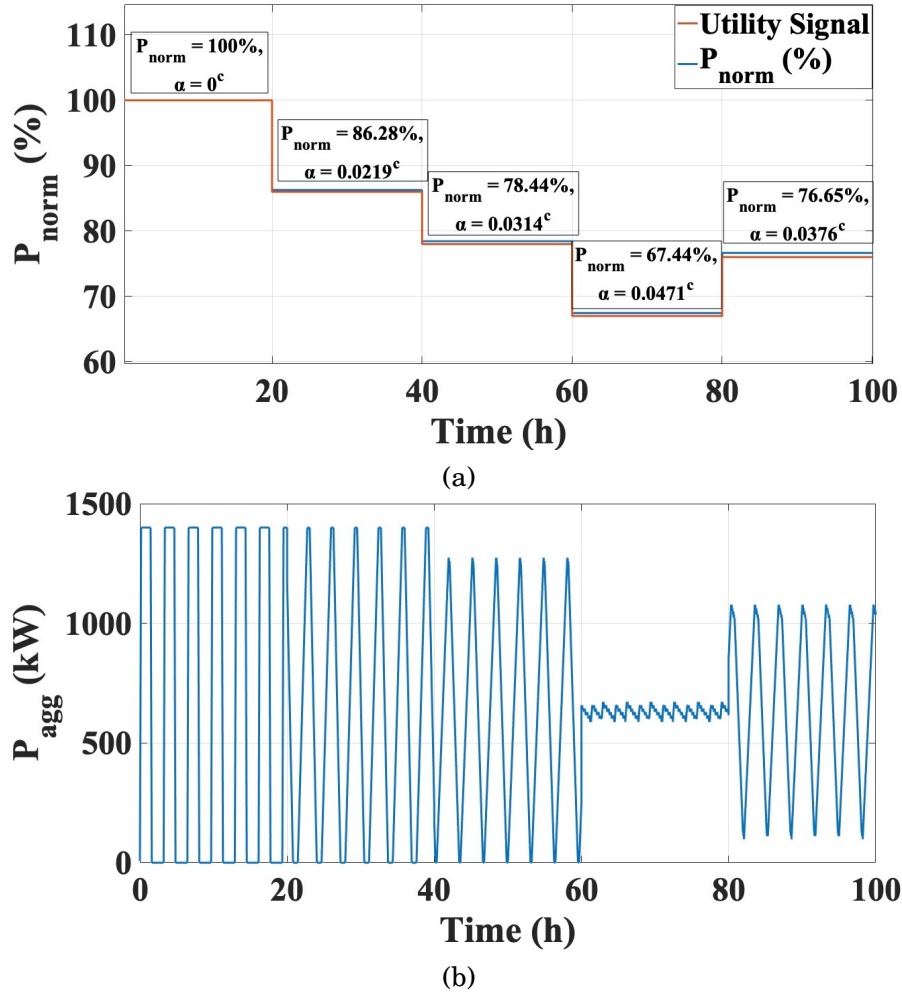

Figure 3.19: (a) $P_{norm}(\%)$ vs time showing variation of $P_{norm}(\%)$ with $\alpha$. (b) $P_{agg}$ vs time plot showing variation of $P_{agg}$ with $\alpha$.

### 3.5.4 Hardware Implementation

To validate the presented theory, I replicate the same model on a hardware and the results have been discussed later. Light emitting diodes (LEDs) provide close resemblance to a switching device, analogous to TCLs. Hence, for the demonstration purpose, LEDs have been used in place of actual TCLs. Since, the number of LEDs were limited to four, a microcontroller with basic computational capabilities and decent clock frequency was sufficient.

#### 3.5.4.1 Hardware

The components used for the implementation includes 4 LEDs, a breadboard, 5V DC supply, wires for connection and probes to display the output on the oscilloscope. The digital storage

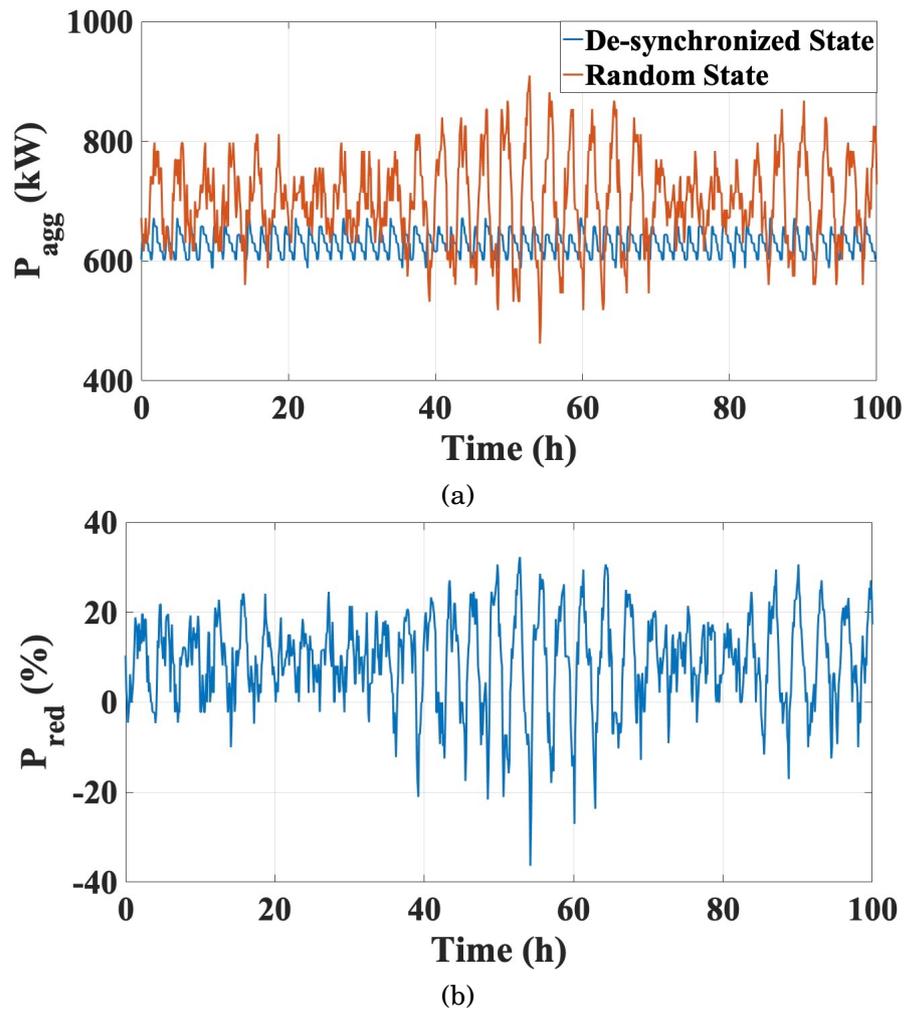

Figure 3.20: (a) $P_{agg}$ vs time plot for random and de-synchronized state. (b) $P_{red}(\%)$ vs time plot showing the reduction in fluctuations.

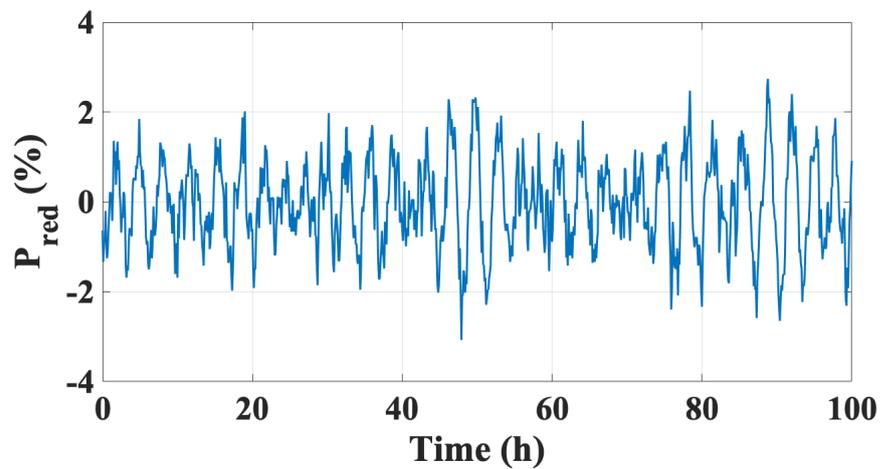

Figure 3.21: Calculation of $P_{red}(\%)$ for $N = 10000$ TCLs.

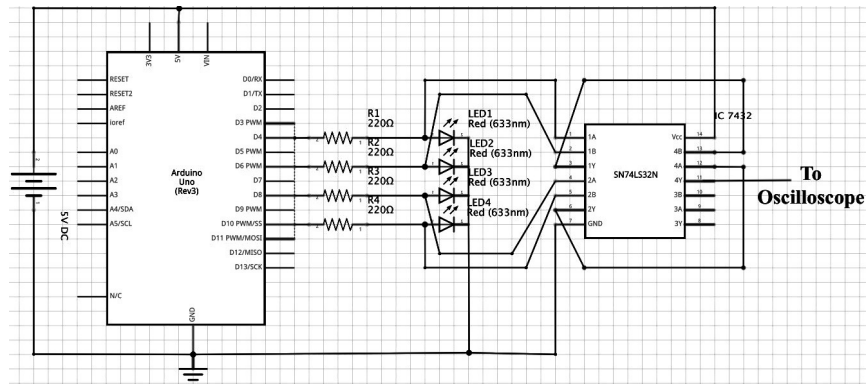

Figure 3.22: Circuit diagram used for practical implementation.

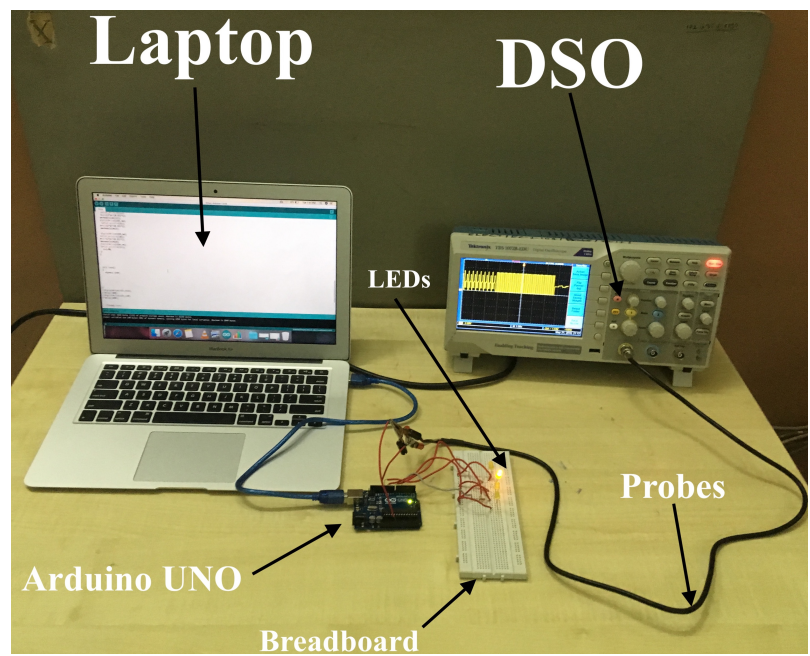

Figure 3.23: Practical setup used for implementation of distributed averaging model.

oscilloscope (DSO) used has the following specifications,

- Product Range : TBS1000B-EDU

- Scope Channel : 2

- Bandwidth : 70MHz

- Sampling Rate :1 GSPS

- Display Memory Depth : 2.5kpts

- Scope Display Type: WVGA LCD Colour

- Plug Type: EU, SWISS, UK

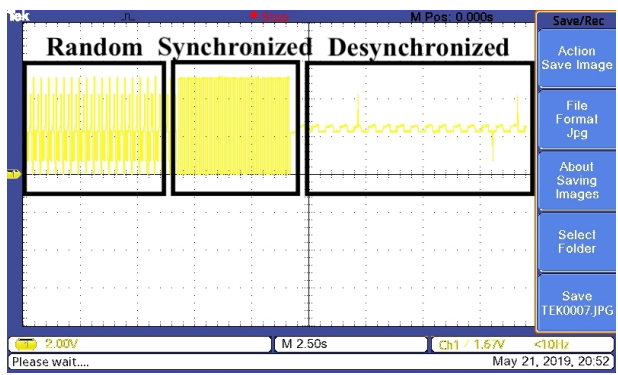

Figure 3.24: Oscilloscope output for $N = 4$ depicting random, synchronized and de-synchronized period of LEDs.

The micro-controller used is Arduino Uno with following specifications,

- Micro-controller: ATmega328
- Operating Voltage: 5V
- Digital I/O Pins: 14 (of which 6 are PWM pins)
- Analog Input Pins: 6
- DC Current per I/O Pin: 40 mA
- DC Current for 3.3V Pin: 50 mA
- Clock Speed: 16 MHz

### 3.5.4.2 Implementation

The setup for practical implementation of the distributed averaging model is shown in the Figure 3.23. Based on the distributed averaging model, as described by (3.15), the logic is coded and burned on Arduino. The Arduino generates switching signals which are then sent to the LEDs. The output of the LEDs are observed on the oscilloscope. As it can be seen, Figure 3.24 maps the output of the Figure 3.15.

In the beginning, i.e., between t = 0s to t = 6.25$s$, Figure 3.24, depicts the condition where the LEDs emulate random behaviour of the TCLs where no control is enforced. After t = 6.25$s$, the LEDs get synchronized and settle at their mean frequency. During this period, the LEDs have no phase difference between them. Figure 3.25 shows the synchronized state of the LEDs. It should be noted that the synchronized condition of the TCLs is a highly undesirable condition in a power system. Further, succeeding t = 12$s$, the LEDs induce phase difference of $\alpha = \pi/2$ between them. As seen from Figure 3.26, the output of the LEDs are minimized. Additionally, the fluctuations have also been minimized.

The results of the hardware implementation were in conjunction with the results of computer simulations. Thus, the results of the proposed theory have been validated.

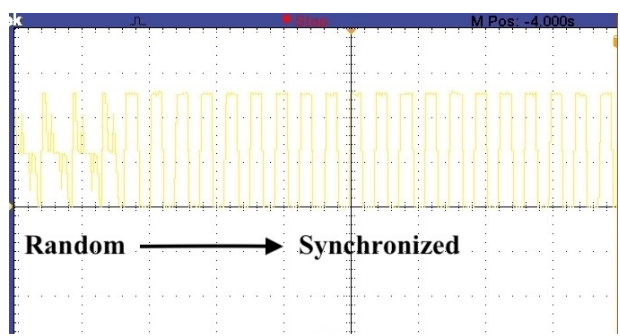

Figure 3.25: Transition of LEDs from the random period to synchronized state.

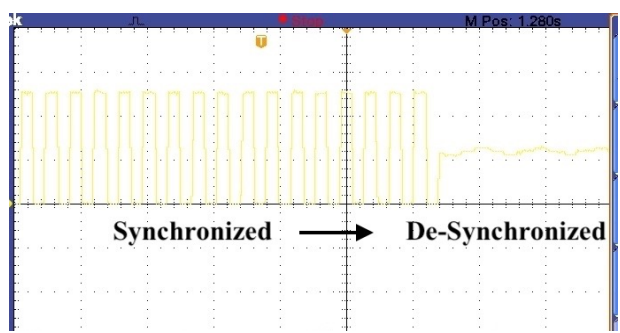

Figure 3.26: Transition of LEDs from the synchronized period to de-synchronized state.

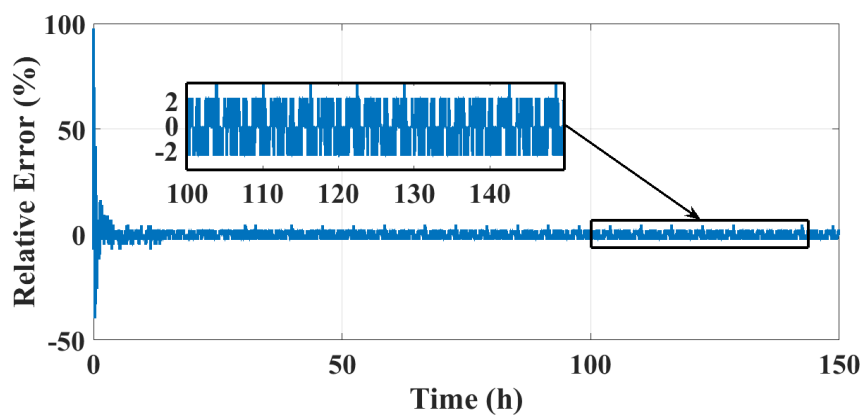

Figure 3.27: Plot of relative error versus time for a population of heterogeneous TCLs, showing effectiveness of the proposed model.

## 3.6 Competitive Benchmarking

In this section, I check the competencies of the proposed models vis-a-vis those available in the literature. A load following scenario discussed in the previous subsections is extended to analyse the efficiency, accuracy and reliability of the proposed models against existent techniques. A practical case of $N \in [100, 1000, 10000]$ of homogeneous as well as heterogeneous TCLs has been considered to be driven from maximum to a minimum aggregate power. The Kuramoto phase oscillator model is checked to achieve minimum aggregate power at $P = 14KW$ with $N = 100$. For the homogeneous case a steady state relative error ($err(t) = (P_{ref}(t) - P_{agg}(t))/P_{ref}(t) \times 100\%$) of 0% is observed, whereas heterogeneous set of TCLs attain a relative error in the range of ±2% as shown in Figure 3.27. Also, in order to benchmark the model with those available in the literature, I calculate effective root mean square error (RMSE %) defined by,

$$\text{RMSE \%} = \sqrt{\frac{1}{T}\frac{\int_0^T \left(P_{ref}(t) - P_{agg}(t)\right)^2 dt}{P_{base}^2}} \times 100 \qquad (3.18)$$

where $P_{base}$ is the baseline power or average power consumed by a population of TCLs. For the Kuramoto phase oscillator model the steady state RMSE % computed using (3.18) is found to be 6.3%. In reference to literature, the minimum control variance law (MVC) discussed in [101] achieves a relative error of < 5% with RMSE % of < 1%. On the other hand, from the inferences of [95] evaluating four different schemes, a wide range of 30%−50% relative errors and RMSE % of 1.18%−8.7% is observed. Further, a relative error in the range 4%−8% and a RMSE % of 2.31%−5.89% is registered in [49]. A novel feedback control proposed in [102] achieves an effective RMSE % of ≤ 2.27%. Thus, it can be inferred that the proposed model in the Kuramoto frame is competent against some widely studied schemes in the literature.

Table 3.4: Comparison with state-of-the-art-methods

| Parameter | Relative error(%) | RMSE (%) |
|---|---|---|
| [101] | 5% | 1% |
| [95] | 30 − 50% | 1.18 − 8% |
| [49] | 4 − 8% | 2.31 − 5.89% |
| [102] | - | 2.27% |
| Phase Oscillator Model | 2% | 6.3% |
| Distributed Averaging Model | 2.49% | 4.87% |

Next, in order to benchmark the distributed averaging scheme, a heterogeneous ensemble of $N \in [100, 1000, 10000]$ TCLs has been considered. Keeping the parameters of comparison constant as for the Kuramoto model, the relative error attained was < 2.4% and an RMSE % of 4.8685% was logged. It can be seen that both the models are effective in terms of peer-to-peer competencies and are comparable against state-of-the-art methods.

Table 3.4 shows the efficacy of phase oscillator model and distributed averaging model against several state-of-the-art methods. As compared to the Boolean phase oscillator model presented in section 3.2, the relative errors are similar. Thus, distributed averaging model achieves similar results while avoiding computational complexities.

## 3.7  Summary and Inferences

To infer the major results for the application considered in this chapter, following can be summarised,

1. In this study, heating and cooling loads are modelled as oscillators and a novel mathematical model for achieving dynamic dispatch using coupled and controlled de-synchronization of an ensemble of TCLs has been proposed.

2. An effective controlled de-synchronization of TCL phases has been achieved using Kuramoto framework a time delay/advance parameter keeping its set-point temperatures unaffected.

3. It has been shown that inclusion of a simple phase shift parameter can achieve required aggregate power thereby provide a significant AS to the central power grid.

4. In the next section, Boolean phase oscillator model so proposed is extended to a distributed averaging framework.

5. The results are simulated using MATLAB for both single as well as ensemble of TCLs (both homogeneous and heterogeneous cases) and have been validated against those available in the literature.

6. Further, the results are extended for a congregation of TCLs thereby calculating reduction in oscillations induced by TCLs to the central grid.

7. A load following scenario proves control of TCLs to be a significant AS to the central power grid.

8. On a futuristic note, it would be interesting to analyze various fascinating behaviors studied in Kuramoto framework (e.g., chimera states, etc.) for TCL type systems.

9. A distributed averaging model helps overcome some of the shortcomings in the phase oscillator model and have been simulated as well as implemented using a hardware, providing validity to the proposed idea.

# CHAPTER 4

# PHASE SYNCHRONIZATION IN POWER SYSTEMS NETWORK

**Table of Contents**

Electrical power systems are complex engineering networks which are crucial to the present day infrastructure. A power network comprises of blocks consisting of numerous interconnected sub-systems, making them challenging to analyze and understand. With continuous rise in electricity demands and the trend for more interconnections, an issue of concern is the mitigation and analysis of low-frequency interarea oscillations. In literature, oscillations associated with individual generators in a power plant are called local mode oscillations typically ranging from 0.7-2.0Hz [103]-[104]. The stability of these oscillations characterized as intraarea (same area) and interarea (across areas) has been extensively studied [103]-[104]. These oscillations between the generators which are inherent to power systems require appropriate mathematical models and techniques for their analysis.

Kuramoto-type models have been widely used to study the dynamics of a power system network through swing equations [105]. It must be noted though that a power system network has an added second order term due to generator inertia and are only similar to Kuramoto model. Power dissipation terms that arise in the swing equation model are absent in conventional Kuramoto model, which can be shown existent by few mathematical adjustments. In power systems, globally coupled phase oscillators of Kuramoto form have been viewed as electromechanical generators mutually coupled to deliver load power. A conventional second order Kuramoto-type oscillator can be written as follows,

$$J_i\ddot{\delta}_i + d_i\dot{\delta}_i = \omega_i + \sum_{j\neq i, j=1}^{n} k_{ij}sin(\delta_j - \delta_i), \quad i \in \{1,\ldots,n\}, \tag{4.1}$$

where $\delta_i$ is angular position of rotor with respect to the synchronously rotating reference frame (for consistency, all the angles throughout the work are in radians), $J_i$ inertia in kgm$^2$, $d_i$ damping and $\omega_i$ is a natural frequency chosen from an appropriate distribution $g(\omega)$ of $i$-th oscillator. $[k_{ij}]; i,j \in \{1,\ldots,n\}$ is the matrix of coupling constants and $n$ defines the number of oscillators. The standard Kuramoto-type equation assumes value of coupling constants $[k_{ij}]$ to be always positive and symmetric. In this work, I explore the mapping between a power grid and Kuramoto oscillators.

Interarea oscillations emerge when two areas having independent sets of power generators experience supply-demand imbalance. The generators in individual areas are observed to beat against each other with frequencies ranging from 0.1-0.8Hz, classified as low frequency interarea oscillations in a power grid. These oscillations can be visualized as two large generators trying to desynchronize each other in the event of supply-demand balance being achieved in each individual area. The above phenomena is analyzed using small-signal or modal analysis [104], however additionally it would be advantageous to have a nonlinear (large-signal) model to capture the different behaviors and effects of these oscillations. I propose a novel 'conformist-contrarian' [48] second order Kuramoto-type model (henceforth, referred to as CC-Kuramoto) which captures the in-phase (intraarea) and the anti-phase (interarea) oscillations in a power system.

The motivation behind developing such a model is to address some of the challenges related to modeling low frequency oscillations in power systems [106]. Conventionally, small-signal analysis and damping control is used by power system engineers to assure system stability at planning stage and thereby execution. It has to be noted though, that over the years software packages on these design/analysis have become computationally efficient in terms of execution time, but still bears significant computation cost for near real-time implementation. Some key challenges related to models for power systems, identified from the literature are as follows:

1. The major problems related to power system oscillations are of perturbed damping of overall system which are regulated conventionally using power system stabilizers. These oscillations are identified using eigenvalue analysis which are computationally costly.

2. In cases when power transfer needs to be increased or decreased, the groups of individual generators in the source and sink sides are dispatched in order of their sensitivities to the critical modes with respect to the output of these generators. This assist in increased power levels without adding any further damping control actions. Computation of these critical modes and calculation of generator sensitivities to it in real-time is difficult.

3. Modeling such behaviors is not an easy task due to observed in-coherency from planning to actual implementations in the past. Details of major equipment and inclusion of newer loads like those of induction motors is still not simple in small-signal models.

4. On the similar lines, [107] shows using a − sync behavior in system characteristics to allow mitigation of homogeneous power oscillations in an interarea setup. [108] describes the behavior and analysis of interarea oscillations in a nonlinear form by adding periodic disturbances to the major parameters that have significant impact. The model must be capable to reproduce these through simple modifications.

Apart from fabricating a perfect model that can overcome above mentioned complexities, a power system engineer looks for a model that can provide significant inferences. With increasing vulnerability of modern power systems due to inclusion of various ancillary services (ASs) it is important to study the settings that might lead to partial stability or instability. It must be noted though, that power grids are not simple physical network of transmission lines and are deeply impacted by its structural as well as dynamical interactions. Thus, a dynamic redesign/modification of existent power network is not possible, as it can be a major limiting factor in optimizing synchronization. With these constraints as reference, I show occurrence of various stable, partially stable and unstable states via tuning of system parameters, and avoid fiddling with the structure. It is observed that these parameters beyond a certain threshold lead to randomization of steady state equilibrium points thereby existence of a chaotic behaviour. The same power grid setup (and some other complex systems in nature) shows a state of partial stability by clustering themselves into islands of synchronised and de-synchronised oscillators, commonly referred to as chimera in literature [109]. Hence, I emulate the existence of these chimera state behaviors and correlate them with blackouts with islanding commonly seen in power grids [110].

Nonlinear modes associated with instabilities have been analysed and discussed in [111–113] related to power grid synchronization. These provide an informative decomposition of nonlinear oscillations when the network loses synchrony. [100, 114] establishes sufficient conditions for synchronization, obtained via perturbation analysis for non-uniform Kuramoto oscillators. It must be noted though, that the conditions attained in previous studies maintain homogeneity in system parameters, whereas I obtain conditions on various power grid parameters and hence heterogeneity. On the similar lines, [115] show effects of heterogeneity on power grid networks but it would be of great use if these can be formulated in Kuramoto-type framework and extended to practical blackout scenarios. Thus, in this work I take a standard example from power systems to show that chimera behaviors can lead to blackouts and can be correlated in a distributed grid. To summarize, major contributions of this work are as follows. I propose a nonlinear model for analysis of power systems in simplistic form to avoid discussed computational complexities. A practical example from power systems is used to correlate and showcase advantages of the proposed model. It has been shown, that the model not only provides information about the nominal states but also existence of chimera behaviors in power systems. This could help site engineers to take actions apriori or equip with necessary tools at the right time.

The work in this chapter is divided into two parts. *Part-I*: I first start with modeling a standard power grid in Kuramoto form. Then, gradually move towards addressing complexities discussed before and how it can be easily incorporated in a large scale (nonlinear model) using analogy to a standard physics example. Next, a standard power systems network to study interarea oscillations is considered and results are verified using computer simulations. *Part-II*: A detailed bifurcation analysis is performed on the proposed model parameters in order to analyse the system stability. Finally, I emulate chimera behavior [109] commonly referred to in the literature and discuss its implications in a power network.

## 4.1 Part - I: Power Network and Kuramoto Oscillators

### 4.1.1 Mathematical model of Power Grid

The fundamental units of a power grid consists of active generators and passive machines/loads. The generator generates electrical power by converting any tangible source of energy connected to it, produced by the prime mover of the generator with the frequency close to the standard or natural frequency $\Omega$ of an electrical system. All generators in a power grid can be looked upon as set of synchronous machines rotating at synchronous frequency $\Omega$, with the stator windings of the generator delivering electrical power to the grid. Any power generator in a power system is described by a power balance equation of the form,

$$P_{accumulated} + P_{dissipated} = P_{source} - P_{transmitted}, \qquad (4.2)$$

where $P_{source}$ is the rate at which the energy is fed into the generator at frequency $\Omega$ (i.e., $2\pi\Omega = 50$Hz). Hence, the phase angle $\theta_i$ at the output of the $i$-th generator in stationary frame is then given by,

$$\theta_i = \Omega t + \delta_i. \qquad (4.3)$$

$P_{accumulated}$ is the rate at which kinetic energy is accumulated by the generator:

$$P_{accumulated} = \frac{1}{2}J_i \frac{d}{dt}(\dot{\theta}_i)^2, \qquad (4.4)$$

where $J_i$ is the moment of inertia of the $i$-th generator in kgm$^2$. For the sake of simplicity, I assume identical machines (i.e., $J_i = J$). $P_{transmitted}$ is the power transmitted from generator $i$ to $j$ with phase difference, $\Delta\theta_{ij} = \theta_j - \theta_i \neq 0$.

$$P_{transmitted} = -P_{max}sin(\Delta\theta_{ij}). \qquad (4.5)$$

$P_{max}$ being maximum electrical power input in watts. The dissipated power ($P_{dissipated}$) with $K_D$ the dissipation constant of the prime mover in Ws$^2$/rad$^2$, can be expressed as:

$$P_{dissipated} = K_D(\dot{\theta}_i)^2. \qquad (4.6)$$

Since, all the generators share common frequencies $\Omega$, $\Delta\theta_{ij} = \Delta\delta_{ij} = \Phi_{ij}$. Substituting (4.4), (4.5) and (4.6) in (4.2), following can be computed,

$$P_{source} = J\ddot{\theta}_i\dot{\theta}_i + K_D(\dot{\theta}_i)^2 - P_{max}sin(\Phi_{ij}). \tag{4.7}$$

Differentiating (4.3) with respect to time and further double differentiating it; thereby assuming perturbations around the synchronous frequency being very small, $i.e., \dot{\delta}_i \ll \Omega$, (4.7) can be approximated as,

$$P_{source} \cong J\Omega\ddot{\delta}_i + [J\ddot{\delta}_i + 2K_D\Omega]\dot{\delta}_i + K_D\Omega^2 - P_{max}sin(\Phi_{ij}). \tag{4.8}$$

Under practically relevant assumptions, the coefficient of first derivative is constant and neglecting acceleration terms, as well as knowing that the rate at which the energy is stored in kinetic term is much lower as compared to rate at which energy is dissipated in friction, (4.8) is reduced to,

$$J\Omega\ddot{\delta}_i = P_{source} - K_D\Omega^2 - 2K_D\Omega\dot{\delta}_i + P_{max}sin(\Phi_{ij}). \tag{4.9}$$

Now, using the fact that, $P_{max} = E_i E_j |Y_{ij}|$; $E_i$ being internal voltage of $i$-th generator, $Y_{ij}$ the Kron reduced admittance matrix denoting maximum power transferred between generators [100] and choosing $P_{m,i} = P_{source} - K_D\Omega^2$, where $P_{m,i}$ is the mechanical power input,

$$J\Omega\ddot{\delta}_i = P_{m,i} - E_i^2\Re(Y_{ii}) - 2K_D\Omega\dot{\delta}_i + \sum_{j\neq i, j=1}^{n} E_i E_j |Y_{ij}| sin(\Phi_{ij}). \tag{4.10}$$

Dividing both sides by $J\Omega$,

$$\ddot{\delta}_i = \left[\frac{P_{m,i}}{J\Omega} - \frac{E_i^2\Re(Y_{ii})}{J\Omega}\right] - \frac{2K_D}{J}\dot{\delta}_i + \sum_{j\neq i, j=1}^{n} \frac{E_i E_j |Y_{ij}|}{J\Omega} sin(\Phi_{ij}). \tag{4.11}$$

(4.11) can be rewritten as follows,

$$\ddot{\delta}_i = \omega_i - \alpha\dot{\delta}_i + \sum_{j\neq i, j=1}^{n} k_{ij} sin(\delta_j - \delta_i), \tag{4.12}$$

where $\alpha = \frac{2K_D}{J}$ is the dissipation constant, coupling constant $[k_{ij}] = \frac{E_i E_j |Y_{ij}|}{J\Omega}$ and natural frequency $\omega_i = \left[\frac{P_{m,i}}{J\Omega} - \frac{E_i^2\Re(Y_{ii})}{J\Omega}\right]$. From a graph theoretic viewpoint $[k_{ij}]$ can be seen as a weighted laplacian matrix with $k_{ij} = 0$ when generators are not connected to each other and $k_{ij} \geq \frac{(\omega_{max} - \omega_{min})n}{(2(n-1))}$ otherwise (i.e., assumed to be greater than critical coupling, to ensure steady state synchronization [100]). It can be observed that, (4.12) has the same form as a second order Kuramoto oscillator model [100].

### 4.1.2 CC-Kuramoto Model for Interarea Oscillations

Further, I extend (4.12) to spring coupled oscillators. In Kuramoto oscillators or spring coupled pendulums for that sake; the coupling term introduces restoring forces on the oscillators. Considering the special case, when there is no energy transfer between oscillators, either of in-phase or anti-phase steady state oscillations may exist (as observed in a spring coupled pendulum - Figure 4.1). For the in-phase oscillations, the restoring forces are zero implying absence of the coupling term, which is not a tangible explanation for a coupled system in practice because there would always be some sort of restoring forces present in a coupled system. On the other hand, in the case of anti-phase oscillations the spring keeps contributing restoring forces whilst the energy transfer is zero [116]. This also explains Huygens observations [117] and is a valid template for modeling interarea oscillations in power systems.

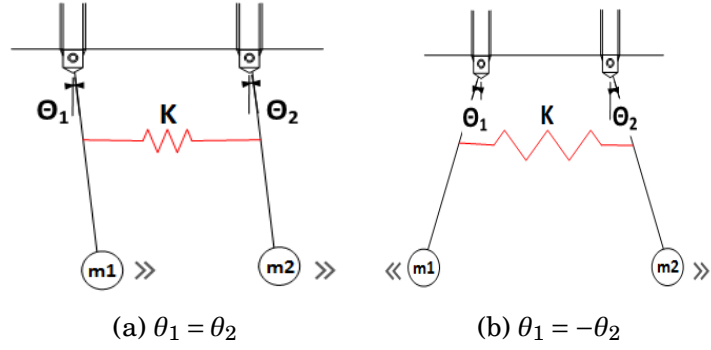

(a) $\theta_1 = \theta_2$      (b) $\theta_1 = -\theta_2$

Figure 4.1: Pendulums coupled by a spring, oscillating in two equilibrium modes. (a) Oscillations in 'in-phase' mode. (b) Oscillations in 'anti-phase' mode.

Any oscillator of the form given in (4.12), assuming damping/dissipation constant to be zero, with identical natural frequencies ($\omega_i = \omega$) and $H(\Phi_{ij}) = \sum_{j \neq i, j=1}^{n} k_{ij} sin(\Phi_{ij})$ and $\Phi_{ij} = (\delta_j - \delta_i) = (\theta_j - \theta_i)$, can be written as,

$$\ddot{\delta}_i = \omega + H(\Phi_{ij}), \qquad (4.13)$$

and thereby, following can be deduced,

$$\ddot{\Phi}_{ij} = H(-\Phi_{ij}) - H(\Phi_{ij}) = -2H(\Phi_{ij}). \qquad (4.14)$$

The above (4.14) has fixed points $\Phi_{ij} = \beta(2\pi)$ or $\Phi_{ij} = (2\beta - 1)\pi; \forall \beta \in \mathbb{Z}$ which are respectively the in-phase and anti-phase modes of the oscillator. Linearizing (4.14) about its fixed points,

$$\begin{aligned}\ddot{\Phi}_{ij} &\approx \left[-2\frac{\partial H(\Phi_{ij})}{\partial \Phi_{ij}}\bigg|_{H(\Phi_{ij})=0}\right]\Phi_{ij}, \\ &\approx \left[-2k_{ij}cos(\Phi_{ij})\big|_{H(\Phi_{ij})=0}\right]\Phi_{ij}.\end{aligned} \qquad (4.15)$$

Thus, from (4.15) it can be deduced that in-phase solution $\Phi_{ij} = 0$ is synchronizing and stable, whereas anti-phase solution $\Phi_{ij} = \pi$ is desynchronizing and unstable. These results are similar to small-signal stability analysis performed by linearizing the nonlinear power system dynamics [104]. Hence, next I integrate these equilibrium/critical modes directly in the nonlinear dynamics of Kuramoto model (4.12). The in-phase or 'conformist' model of

Kuramoto oscillators can be given as follows,

$$\ddot{\delta}_i = \omega_i - \alpha_i \dot{\delta}_i + \sum_{j\neq i, j=1}^{n} k_{ij} sin(\delta_j - \delta_i), \qquad (4.16)$$

whereas, an anti-phase or 'contrarian' Kuramoto model can be obtained by replacing $H(\Phi_{ij})$ with $-H(\Phi_{ij})$ in (4.13) to give,

$$\ddot{\delta}_i = \omega_i - \alpha_i \dot{\delta}_i - \sum_{j\neq i, j=1}^{n} k_{ij} sin(\delta_j - \delta_i). \qquad (4.17)$$

Linearisation of the 'contrarian' model (4.17) on the lines of (4.15) yields $\ddot{\Phi}_{ij} \leq 0$ for anti-phase modes and otherwise for in-phase modes. Thus, in the 'contrarian' model the anti-phase mode is stable and in-phase mode is unstable. All oscillations in physical systems in general and power systems in particular are weighted sum of in-phase and anti-phase modes. Hence, to study the oscillations in power systems, I propose the CC-Kuramoto model of coupled oscillators given as,

$$\begin{aligned}\ddot{\delta}_i^{a_1} &= \omega_i - \alpha_i \dot{\delta}_i^{a_1} + \sum_{j\neq i, j=1}^{p} k_{ij} sin(\delta_j^{a_1} - \delta_i^{a_1}) \\ &\quad - \sum_{j=p+1}^{n} k_{ij} sin(\delta_j^{a_2} - \delta_i^{a_1}), \\ \ddot{\delta}_i^{a_2} &= \omega_i - \alpha_i \dot{\delta}_i^{a_2} + \sum_{j\neq i, j=p+1}^{n} k_{ij} sin(\delta_j^{a_2} - \delta_i^{a_2}) \\ &\quad - \sum_{j=1}^{p} k_{ij} sin(\delta_j^{a_1} - \delta_i^{a_2}),\end{aligned} \qquad (4.18)$$

where without loss of generality I assume $\delta_i^{a_c} \in \left[\delta_1^{a_1}, \delta_2^{a_1}, \delta_3^{a_2}, \delta_4^{a_2}\right]$, $p$ set of generators in area 1 ($a_1$) and ($n-p$) generators in area 2 ($a_2$).

Further, a CC-Kuramoto model settles into one of the three type of states, depending upon the system parameters and initial conditions: Incoherent state - a state of complete desynchronization, $\pi$-state - when two groups of coherent oscillators are separated by phase difference of $\pi$ radians and the Travelling wave state - where the two coherent groups are apart by a phase difference less than $\pi$ radians. Not only do the above states exhibit rich dynamical behavior but other interesting outcomes arise in the process of transition between these states. The direct relation between CC-Kuramoto and the power system network facilitates the study of complex dynamics arising in power networks.

### 4.1.3 Case study: Classical two-area four-machine system

In order to validate the proposed model, I use a classical two-area four-machine power system developed in [104] for interarea oscillation analysis. The system is symmetric; consisting of two identical areas connected through a relatively weak tie ($J_i = J = 0.4 \text{kgm}^2$, $\alpha_i = \alpha = 0.125$). Each area includes two synchronous generators with equal power output. The single line diagram of the system considered is as shown in Figure 4.2,

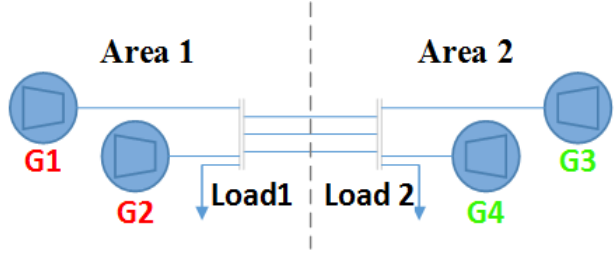

Figure 4.2: Single line diagram of a two-area four-machine power system.

All loads are constituted to behave as constant impedances. The number of tie-circuits in service were varied in order to vary tie-line impedances. Power transfer between two areas is emulated, either by an uneven splitting of system loads or by an uneven distribution of generation between the areas. The combinations for tie-line power flow are given in Table 4.1.

Table 4.1: Load and Tie-Line Power of Test System

|  | Generation/Load (MW) | | Power flow from Area 1 to Area 2 (MW) |
|---|---|---|---|
|  | Area 1 | Area 2 |  |
| Case 1 | 1400/1367 | 1400/1367 | 0 |
| Case 2 | 1400/967 | 1450/1767 | 400 |

The number of tie-line in service is two and the transfer level along the tie-line of two areas varies from 0 MW to 400 MW in accordance to the variation in load levels. *Case 1* relates to no power transfer between two areas. On the other hand, the event of power transfer between areas has been designated as *Case 2*.

$$K_{Case1} = \begin{bmatrix} 0 & 1.9689 & 0.1766 & 0.1782 \\ 1.9689 & 0 & 0.1782 & 0.1801 \\ 0.1766 & 0.1782 & 0 & 1.9363 \\ 0.1782 & 0.1801 & 1.9363 & 0 \end{bmatrix} \quad (4.19)$$

$$K_{Case2} = \begin{bmatrix} 0 & 2.5960 & 0.2130 & 0.2151 \\ 2.5960 & 0 & 0.2151 & 0.2171 \\ 0.2130 & 0.2151 & 0 & 1.7214 \\ 0.2151 & 0.2171 & 1.7214 & 0 \end{bmatrix} \quad (4.20)$$

The coupling matrix $[k_{ij}]$ is represented as (4.19) for *Case 1* and (4.20) for *Case 2*. Elements of coupling matrix $[k_{ij}]$ are derived from $k_{ij} = \frac{E_i E_j |Y_{ij}|}{J\Omega}$. Natural frequencies are calculated using $\omega_i = \left[\frac{P_{m,i}}{J\Omega} - \frac{E_i^2 \Re(Y_{ii})}{J\Omega}\right]$ and are shown in Table 4.2. For simplicity, I assume $\Omega$ = 1Hz.

The model proposed in (4.18) was solved using MATLAB and following observations were made.

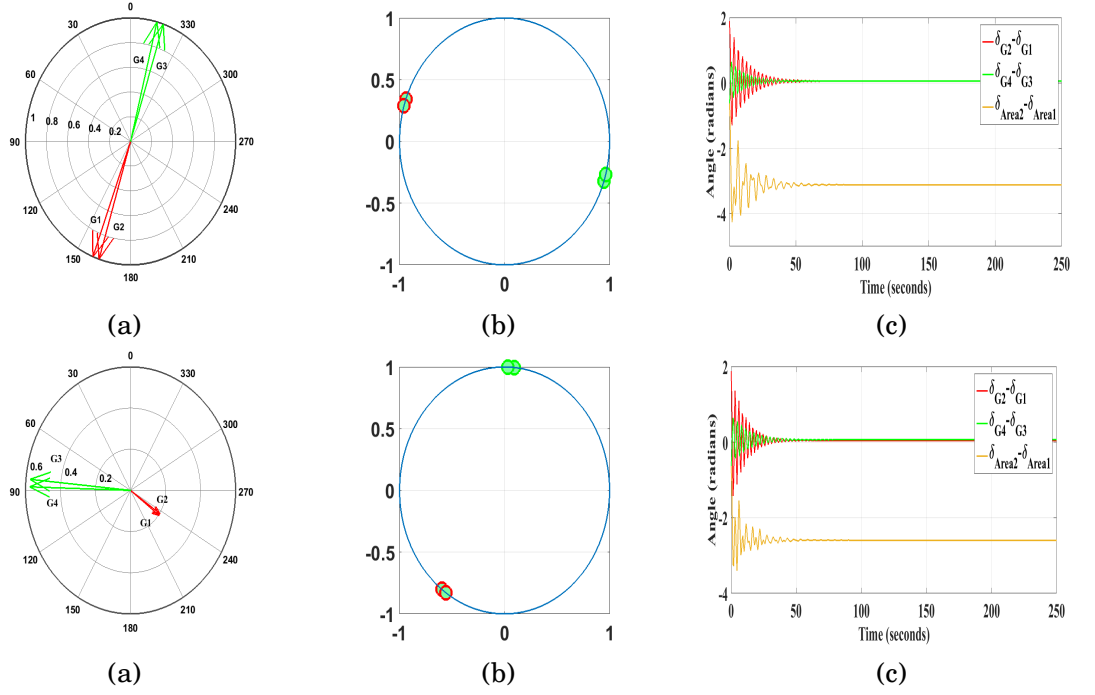

Figure 4.3: *Case 1* (first row from the top) - Dynamics of the proposed model for interarea oscillations, when no power is transferred. (a) Compass plot. (b) Circle plot. (c) Time series plot. *Case 2* (second row from the top) - Dynamics of the proposed model for interarea oscillations, when power is delivered from area 1 to area 2. (a) Compass plot. (b) Circle plot. (c) Time series plot.

- *Case 1*: The generators oscillate anti-phase ($-3.12 \approx -\pi$ radians) in interarea and in-phase ($0.06 \approx 0$ radians) intraarea. To validate the results, I provide compass plots of normalized eigen modes, circle plot as well as time-domain plots of generators as shown in Figure 4.3. The compass plots were obtained by using the steady state vectors: $\vec{c}_i(t) = (\delta_j - \delta_i)_{rms} \angle \delta_i(t) \sim (\delta_j - \delta_i)_{rms} \left( \sum_{i=1}^{n} e^{\lambda_i t} u_i v_i^T \delta_i(0) \right)$, where $u_i$ is the normalized left eigen-vector, $v_i$ is the normalized right eigen-vector and $\lambda_i$ are the eigen-values of the linearized system. It can be seen that the compass plot of steady state vectors show behavior similar to normalized eigen modes obtained by small-signal analysis performed traditionally.

- *Case 2*: In the case when power is transferred between areas, the phase difference between interarea generators were observed to be $-2.6$ radians (i.e., $\neq -\pi$) and $0.05$ radians in intraarea, as shown in Figure 4.3. The results are comparable with [104], providing validity to the proposed model.

Table 4.2: Natural frequencies (in rad/s)

|  | $\omega_1$ | $\omega_2$ | $\omega_3$ | $\omega_4$ |
| --- | --- | --- | --- | --- |
| Case 1 | 17.5290 | 17.7923 | 17.5640 | 17.8285 |
| Case 2 | 16.8882 | 17.1532 | 17.7931 | 18.0629 |

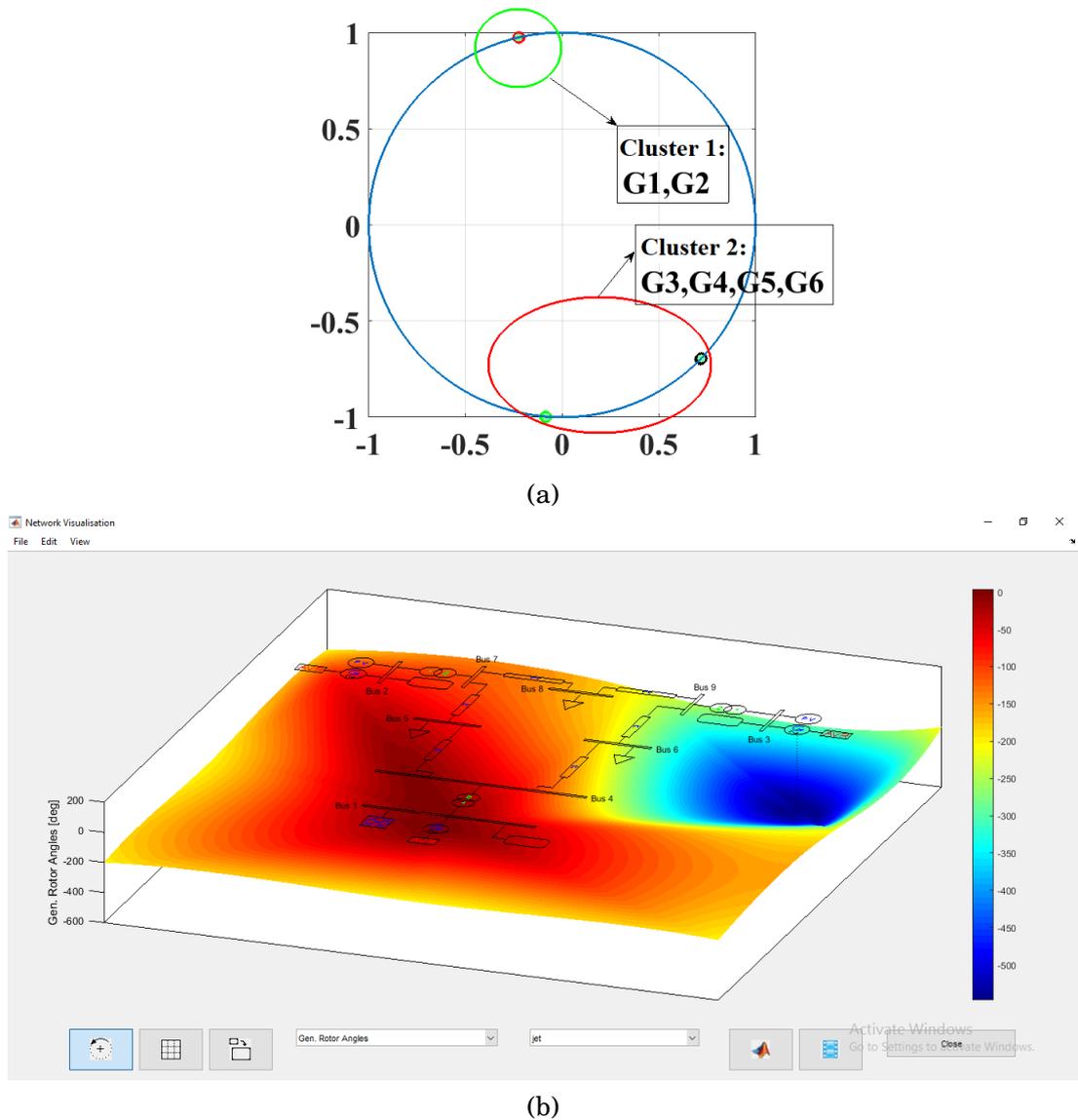

(a)

(b)

Figure 4.4: Validation of CC-Kuramoto model against PSAT for a three area power network. (a) Circle plot using CC-Kuramoto model. (b) Heat map of phase angles on PSAT interface.

Further, I extend the idea to a multi-area framework by extending $N = 4 \rightarrow 6$ and keeping the number of oscillators in any of the three areas equally distributed. The CC-Kuramoto model is then tested for the same and the outcomes analysed. It has been observed, that these areas form clusters amongst themselves joining the group that is relatively similar. As shown in Figure 4.4, the phase angles obtained from CC-Kuramoto are analysed using circle plot. The results are almost identical to those obtained using a similar three area model provided in Power System Analysis Toolbox (PSAT) [118], commonly referred to by power systems community. The results so obtained provide validity to the proposed idea in a multi-area frame thereby standing competent to the existing tools in the literature.

In order to check the similarity of the proposed model to the conformist-contrarian characteristics in the literature [48], following observations were made. For a population of

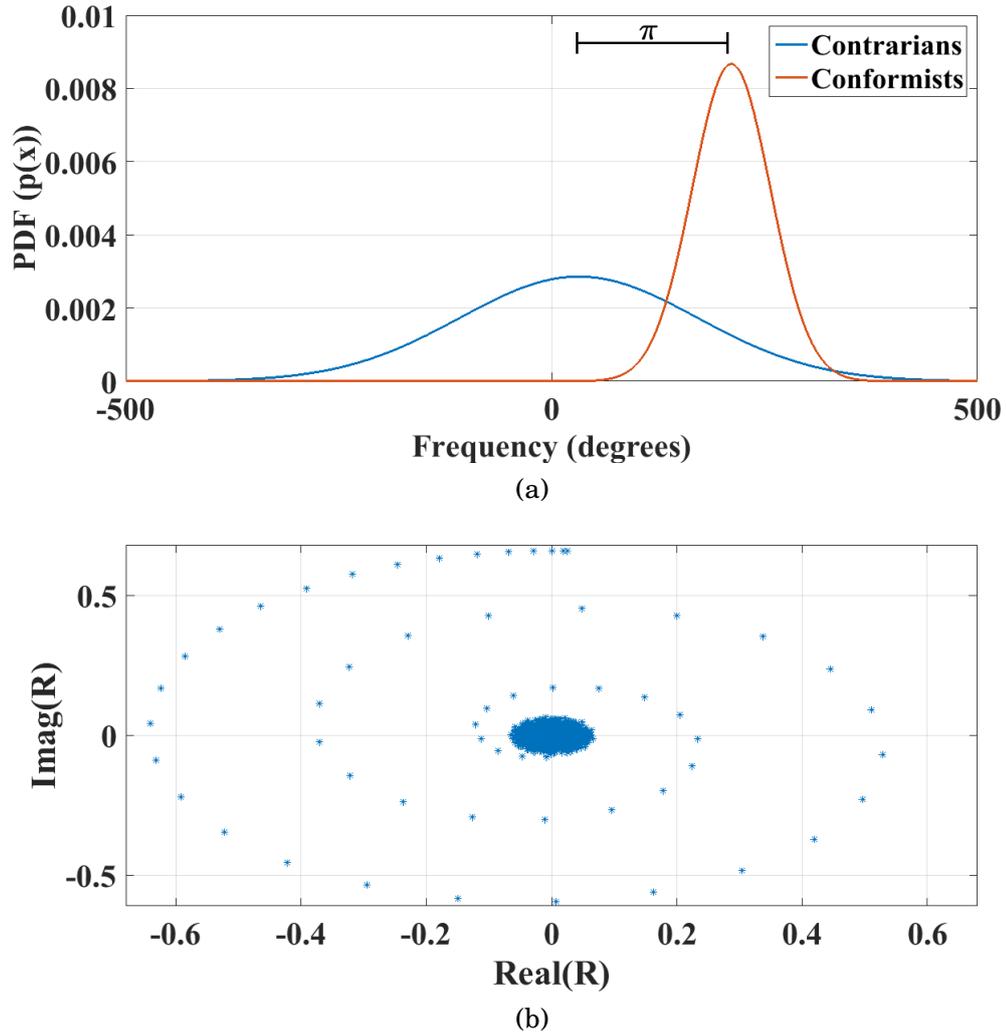

Figure 4.5: $\pi$-state for a set of 500 generators in CC-Kuramoto form: (a) Probability distribution function (PDF). (b) Order parameter $\left(R = \frac{1}{N}\sum_{j=1}^{N} e^{i\theta_j}\right)$ plot.

'$N = 500$' generators taken at once, at any given time the phase of these are seen to transit between '$\pi$' state and transmitted wave state defined before. As shown in Figure 4.5, the set of '$N$' generators in CC-Kuramoto form have tendency to form clusters which are $\pi$ radians apart called as the $\pi$-state. Consider, the case when power load changes resulting in tie-line experiencing travelling wave effect (Figure 4.6); i.e., where the phase of clustered generators are $< \pi$ radians apart. Thus, at any time the power line experiences load transitions, and hence a to and fro switch between these two states is observed until it stabilizes in $\pi$-state form. On the contrary, in the case of tie line breakage the effective coupling constants are zero resulting in loss of synchrony between coupled generators leading to power system instability. In addition, I show order parameter plots to understand the rate of synchronization which in both the cases are similar showing effective consensus states. Thus, it can be inferred that the proposed CC-Kuramoto model verifies with the discussed characteristics in power systems as well as physics applications.

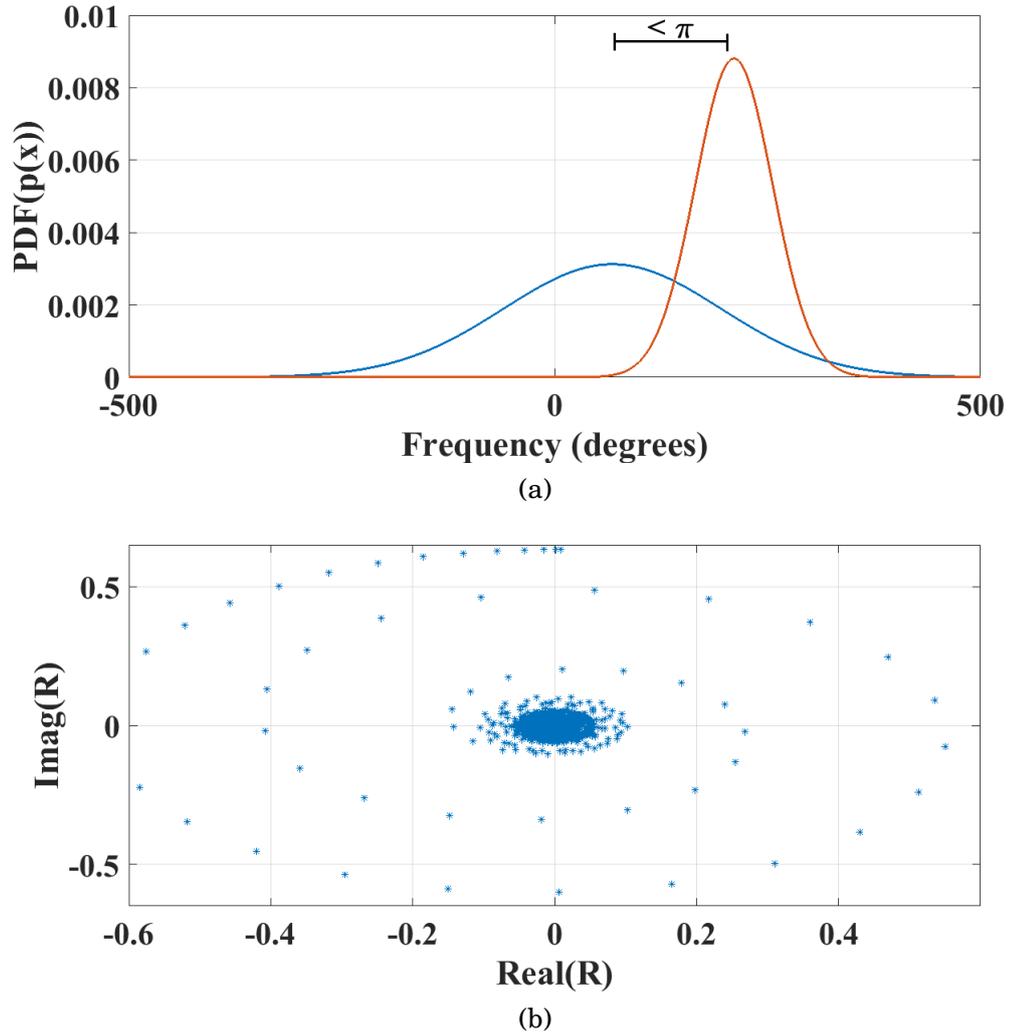

Figure 4.6: Travelling wave state a set of 500 generators in CC-Kuramoto form: (a) Probability distribution function (PDF). (b) Order parameter $\left(R = \frac{1}{N}\sum_{j=1}^{N} e^{i\theta_j}\right)$ plot.

## 4.2 Part - II: Partial Stability in Power Systems

In this section, I study the bifurcation analysis of the proposed CC-Kuramoto model in order to understand the effect of the design parameters on the stability of a power network. In order to do so, I first analyse some of the characteristics nodes in power systems.

### 4.2.1 Equal Area Criteria in Power Systems

In power systems, the equal area criterion is a "graphical technique used to examine the transient stability of the machine systems (one or more than one) with an infinite bus". The areas under the curve of a power angle diagram are equated across to calculate effective acceleration/deceleration thereby comment on the stability of the system. For instance, consider (4.18), rewriting in terms of mechanical and electrical power interactions,

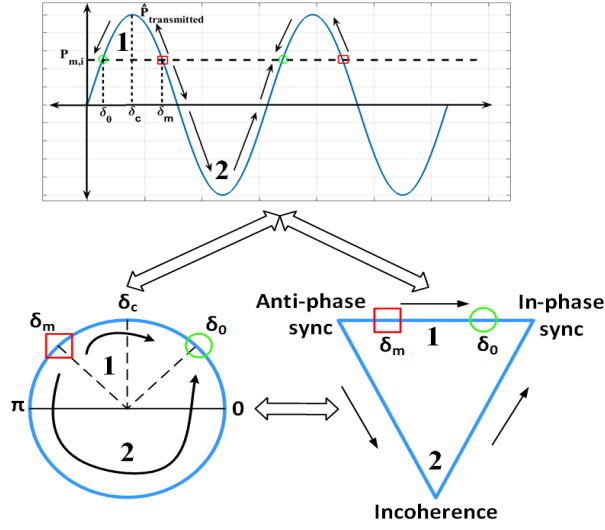

Figure 4.7: Equivalence of equal area criterion to equilibrium spaces. Green circles denote 'sink' nodes where system achieves equilibrium. Red squares denote 'source' node where system attains acceleration and hence increase in accumulated power.

$$\ddot{\delta}_i = P_{m,i} - \hat{P}_{transmitted}, \tag{4.21}$$

where $\hat{P}_{transmitted} = \alpha_i \dot{\delta}_i - P_{max} sin(\Delta \delta_{ij}) = P_{transmitted} - P_{dissipated}$. It can be seen that the collective acceleration of generators is dependent on the difference of mechanical and electrical power inputs. As shown in Figure 4.7, difference in electrical-mechanical inputs either accelerate or decelerate the generators to achieve equilibrium. The generators accelerate when mechanical power is higher than the transmitted electrical power (i.e., $P_{m,i} > P_{transmitted}$) and decelerate when electrical power is higher (i.e., $P_{m,i} < P_{transmitted}$). This is due to the fact that, the difference in the power gives rise to the rate of change of accumulated power ($P_{accumulated}$) in the rotor masses. The change in accumulated power and generator inertia results in effective change in rotor angles, thereby acceleration/deceleration and vice-versa. The coupled set of generators happen to achieve steady state, when mechanical power of the generator from turbines match the transmitted electrical power. This can be visualized as creation of a 'sink' node, where the system tries to drive itself in order to achieve stability. Thus, any generator starting from a rotor angle $\in [\delta_0, \delta_c, \delta_m]$, will try to move towards $\delta_0$ (or stability). On the other hand, $\delta_m$ happens to be critically stable and a small perturbation towards $\pi$ radians can render increase in rotor angles due to effective acceleration of generators. In such case, a 'source' node is formed at $\pi$ radians, as shown in Figure 4.7.

Generators are mechanical devices that exhibit high inertia and hence require time to achieve stability, once perturbed from its equilibrium. The natural speed (angular speed) of rotation $\omega_i$ and generator inertia $J$ along-with accumulated power $P_{accumulated}$ in turn defines the rate at which stability is achieved or if system becomes unstable. As shown in

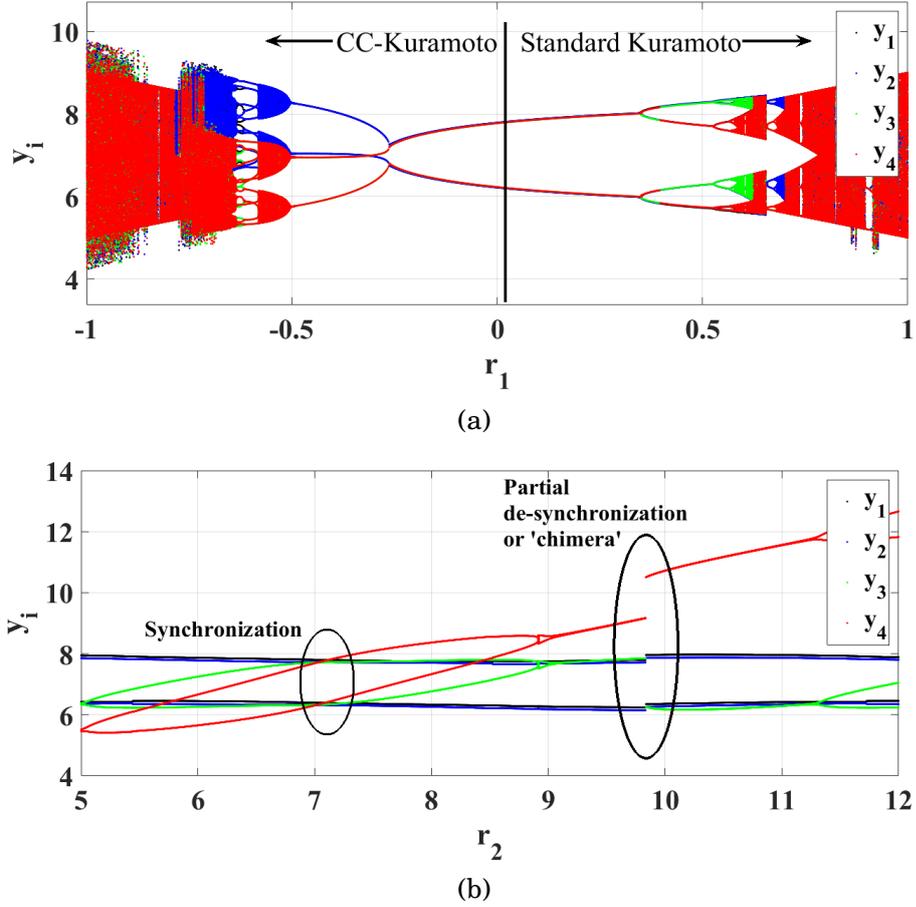

Figure 4.8: Bifurcation analysis of CC-Kuramoto model. (a) Analysis of system stability with variation in interarea coupling constants. (b) Stability analysis of the model by varying the natural frequencies in certain areas.

Figure 4.7, although $\omega_i$ and $J$ are constants if generator achieves stability via route 2 has higher $P_{accumulated}$ and is highly susceptible to instability as compared to route 1. I show this using circle plot and thereby its relevance to '$\pi$', 'transmitted' and 'incoherence' state as mentioned in previous sections. Thus, it can be inferred that a 'source' node imparts instability whereas 'sink' node stabilizes a power setup. These inferences can be easily made using CC-Kuramoto model as these nodes are pretty evident. In the next section, I perform bifurcation analysis on CC-Kuramoto model in order to understand the effect of system parameters on system stability.

### 4.2.2 Case Study: Bifurcation Analysis

#### 4.2.2.1 Using Equilibrium Points

Next, I rewrite (4.18) in order to study the bifurcation in power system using CC-Kuramoto model. Let $\delta_i^{a_c} = x_i^{a_c}, \dot{\delta}_i^{a_c} = y_i$, and $r_1$ be the parameter of bifurcation on interarea coupling.

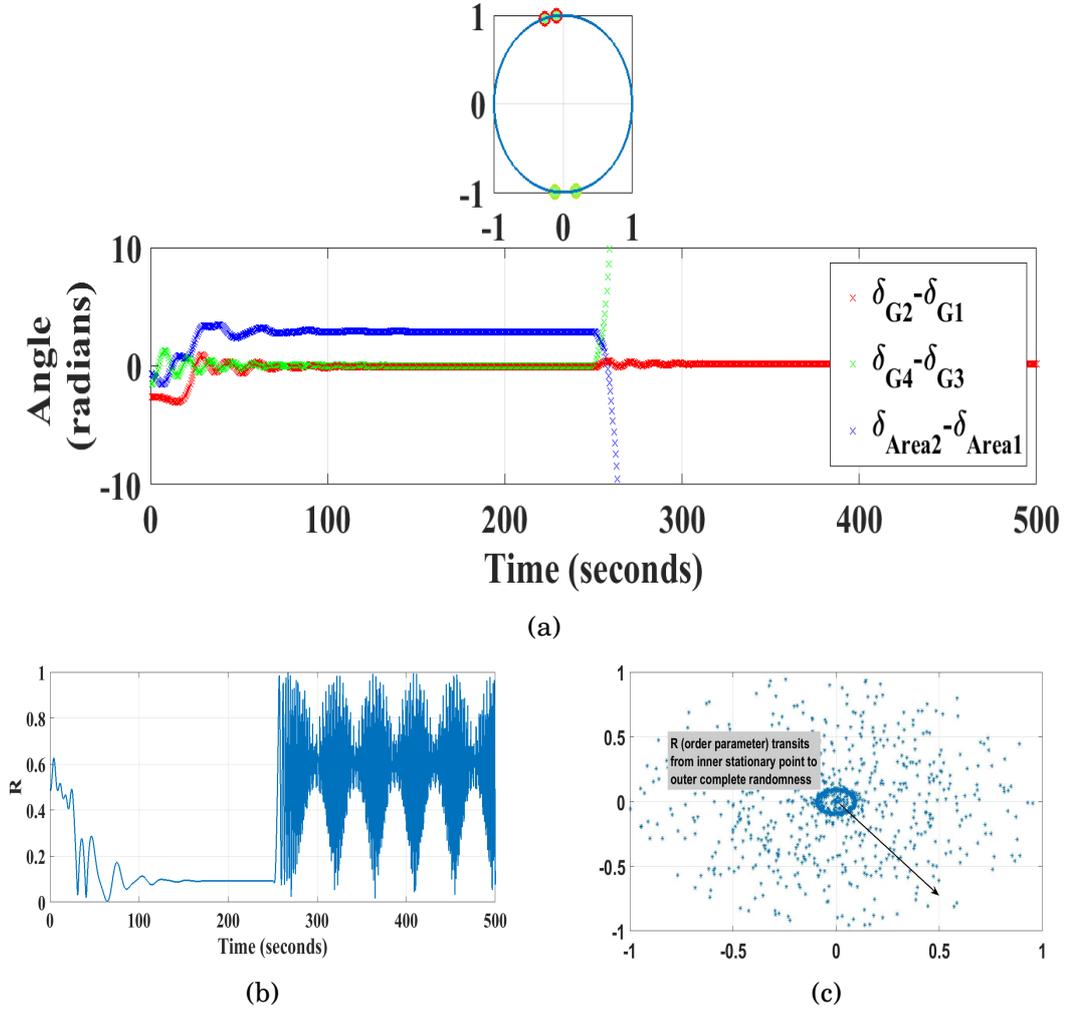

Figure 4.9: Simulation of 'chimera' behavior in a system of coupled generators. (a) Circle plot against Time series plot showing loss of synchronicity at time ∼ 250 seconds. (b) Order parameter ($R = \frac{1}{n}\sum_{j=1}^{n} e^{i\delta_j}$) plot against time. (c) Polar plot of order parameter.

$$\begin{aligned}
0 &= \omega_i - \alpha_i y_i + \sum_{j\neq i, j=1}^{p} k_{ij} \sin(x_j^{a_1} - x_i^{a_1}) \\
&\quad - \sum_{j=p+1}^{n} r_1 \sin(x_j^{a_2} - x_i^{a_1}), \\
0 &= \omega_i - \alpha_i y_i + \sum_{j\neq i, j=p+1}^{n} k_{ij} \sin(x_j^{a_2} - x_i^{a_2}) \\
&\quad - \sum_{j=1}^{p} r_1 \sin(x_j^{a_1} - x_i^{a_2}).
\end{aligned} \quad (4.22)$$

With the same setup of two-area four-machine system, parameter $r_1$ (i.e., interarea coupling) is varied in the range $r_1 \in [-1, 1]$ with homogeneous natural frequencies (i.e., $\omega_i = \omega$). Under this setting, I solve for equilibrium points of $\dot{x}_i^{a_c} = \hat{y}_i = f(x_i^{a_c}, y_i) = 0$ and thereby fixed points of the dynamical equations (4.22). The results were obtained using MATLAB and are shown in Figure 4.8 (a). It can be seen that interarea coupling of $0.5 \geq r_1 \geq -0.5$ achieves period doubling, although the solutions are constant. Beyond $-0.5 > r_1 > 0.5$ CC-Kuramoto model achieves chaotic behavior. This can be visualised as complete loss of synchronism and thereby chaos.

For the next case, $\omega_4^{a_2}$ is varied in a range $\omega_4^{a_2} = r_2 \in [5, 12]$rad/s keeping coupling parameter constant, showing synchronization at $r_2 = 7$rad/s and leaves synchronicity at $r_2 = 10$rad/s showing chimera behavior. I solve for

$$\begin{aligned} 0 &= \hat{\omega}_i - \alpha_i y_i + \sum_{j \neq i, j=1}^{p} k_{ij} sin(x_j^{a_1} - x_i^{a_1}) \\ &\quad - \sum_{j=p+1}^{n} k_{ij} \, sin(x_j^{a_2} - x_i^{a_1}), \\ 0 &= \hat{\omega}_i - \alpha_i y_i + \sum_{j \neq i, j=p+1}^{n} k_{ij} sin(x_j^{a_2} - x_i^{a_2}) \\ &\quad - \sum_{j=1}^{p} k_{ij} \, sin(x_j^{a_1} - x_i^{a_2}), \end{aligned} \quad (4.23)$$

where $\hat{\omega}_i = [\omega_1^{a_1}, \omega_2^{a_1}, \omega_3^{a_2}, r_2]^T$. This scenario can be interpreted as a gradual overload of one of the generators from two areas leading to de-synchronization in one area, whereas other area remains synchronized (refer Figure 4.8 (b)). Further, using circle and time-series plots for angular separations (as shown in Figure 4.9 (a)), the same partial de-synchronization is observed. These can be inferred as islanding of power network through circuit breakers to avoid the impact of excessive overloading of generators in a neighbouring area (and hence blackouts or cascaded failures [119]). To summarize, the heterogeneity was introduced by increasing frequencies $\omega_i$ incrementally in one area, while keeping parameters of other area constant (i.e., $\omega_1^{a_1}, \omega_2^{a_1}, \omega_3^{a_2} \in g_1(\omega); \omega_4^{a_2} \in r_2 g_1(\omega) = g_2(\omega), r_2 \in \mathbb{R}$; $g_1, g_2$ being frequency distributions). As seen from Figure 4.9 area experiencing incremental perturbations lose synchronicity whereas other area remains unaffected, emulating blackout conditions with islanding. In order to measure loss of synchronicity, I make use of order parameters $R = \frac{1}{N} \sum_{j=1}^{N} e^{i\theta_j}$ as shown in Figure 4.9 (b), (c).

### 4.2.2.2 Using Eigen value analysis

In this subsection, I provide an eigenvalue based justification for bifurcation phenomena observed in previous subsection. For instance, consider (4.15) and let $\lambda_{ap}, \lambda_{ip}, \lambda_{inc}$ be eigenvalues of anti-phase, in-phase and incoherent dynamics respectively. Where

$$\lambda = \begin{cases} \lambda_{ap} < 0 & \text{if } \Phi_{ij} = m(\pi) \\ \lambda_{ip} > 0 & \text{if } \Phi_{ij} = 0, m(2\pi) \\ \lambda_{inc} = 0 & \text{if } \Phi_{ij} = m(\pi/2), \end{cases} \quad (4.24)$$

$m \in \mathbb{Z}$. Now, since $\lambda_{ap} = -\lambda_{ip}$, these two nodes exchange stability through $\lambda_{inc}$ and hence following can be concluded. (i) anti-phase and in-phase modes have converse stabilities and are never stable simultaneously, (ii) these critical modes swap stabilities at incoherence state and (iii) if either of anti-phase or in-phase states are stable incoherence state must be unstable and vice-versa. Particularly, in power systems these states rest in anti-phase (unstable), in-phase (stable) and chimera (partially stable) states.

## 4.3 Summary and Inferences

To summarize the outcomes of the research in this chapter, following can be inferred:

1. In this study, a mathematical model for interarea oscillations is proposed using Kuramoto-type framework with its analogy in power grids.

2. It is shown how these oscillations can be visualized in a 'conformist-contrarian' form to better understand interarea oscillations.

3. Validity of the choices has been justified using analogy of spring coupled pendulums.

4. In order to verify the model, a standard four generator power system was considered from the literature. Simulations were performed in MATLAB and results were verified and validated.

5. The proposed model is used to investigate various phenomena like spatial/temporal chimera [109] and spontaneous failures in power systems [110].

# CHAPTER 5

## INFERENCES AND FUTURE WORK

In this work, I provide a deeper insight to some of the practical applications existing in nature that can be controlled using ideas from synchronization perspective. The major highlights of the work integrated in this thesis can be enlisted as follows:

1. A brief introduction describing the examples from nature that helped motivate me to define a problem statement has been discussed in the pilot section. This section provides examples as well as background of inception of synchronization and helps reader to motivate themselves in the direction.

2. On the basis of the motivation and the literature studied, frequency synchronization problem in micro grids was studied and a novel control law for the same was devised. The new approach suggested provides outcomes that are better in terms of the control laws already existing in the literature and forms the first contribution of the thesis.

3. From the same framework of synchronization, I move to achieve controlled de-synchronization in TCLs to attain required power aggregate defined by the utility. A load following objective is achieved and the results are verified for homogeneous, heterogeneous and population of TCLs. It has been shown how a Kuramoto model can be implemented to achieve effective de-synchronization in TCLs and thereby desired objectives. The effectiveness of the model is tested against various models available in the literature and the efficacies are tabulated. These help achieve sustainability objectives in a world of constant demand-supply mismatch.

4. In the fourth chapter, various power grid analysis software existing in literature have been studied and in order to overcome its complexities a novel nonlinear model for power system analysis in Kuramoto form has been proposed. This model is verified against standard results available in literature and a bifurcation analysis for the same has been provided. It has been shown how a 'chimera' behavior commonly referred to in oscillator theory can be mapped to blackout scenarios in power systems.

In terms of proposing future endeavours and ideas that this dissertation might open up, following can be noted:

1. Voltage synchronization is one such problem already persisting in an AC micro grid and needs robust controllers similar to the ones proposed in this work. Apart from individual control it would be of great advantage if these parameters can be controlled simultaneously in an uncertain environment.

2. A plug-n-play device that can be installed in residential homes with minimum change in the existing infrastructure and allowing control of TCLs to reduce load on power grid would be of great advantage.

3. The models proposed for power grid analysis in concluding chapter can be directly implemented into the existing supervisory control and data acquisition (SCADA) systems and blackout scenarios can be studied to better learn the model for discrepancies.

**Major Contributor**

1. **Bajaria, Pratik**, and N. M. Singh. "Dynamic dispatch of thermostatically controlled loads using phase oscillator model." Sustainable Energy, Grids and Networks 18 (2019): 100220.

2. **Bajaria, Pratik**, "Phase Oscillator Model for De-synchronization of Thermostatically Controlled Loads", 38th Chinese Control Conference, 2019.

3. **Bajaria, Pratik** and Singh, Navdeep, "Frequency synchronisation of islanded microgrids using robust adaptation over Luré forms." 2018 Indian Control Conference (ICC). IEEE, 2018.

4. Singh, Khsitij and **Bajaria, Pratik**, "A Distributed Algorithm for Dynamic Dispatch of Thermostatically Controlled Loads", NAPS 2019, IEEE-PES. [*Accepted*]

5. Khan, Afzal and **Bajaria, Pratik**, "Hybrid Energy Storage Using Battery and Supercapacitor for Electric Vehicle", ICPES 2019, IEEE-PES. [*Accepted*]

6. **P. Bajaria**, S. R. Wagh and N. M. Singh, "Perspectives on Interarea Oscillations in Power Systems", *AIP Advances*, AIP. [*Submitted*]

7. K. Singh and **P. Bajaria**, "Demand Response Based De-synchronization of Thermostatically Controlled Loads", *Sustainable Energy, Grids and Networks*, Elsevier. [*Submitted*]

**Supportive Contributions**

1. Bhopale, P., **Bajaria, P.**, Singh, N., Kazi, F. (2017, June). Enhancing reduced order model predictive control for autonomous underwater vehicle. In International Conference on Computer Science, Applied Mathematics and Applications (pp. 60-71). Springer, Cham.

2. Bhopale, P. S., **Bajaria, P. K.**, Kazi, F. S., Singh, N. M. (2016, December). LMI based depth control for autonomous underwater vehicle. In 2016 International Conference on Control, Instrumentation, Communication and Computational Technologies (ICCICCT) (pp. 477-481). IEEE.

3. Bansode, P., Deshpande, A., Bahadure, S., **Bajaria, P.**, Kazi, F., Singh, N. (2017, January). A Stackelberg game theoretic approach to resilient control of an islanded microgrid. In 2017 Indian Control Conference (ICC) (pp. 212-218). IEEE.